\documentclass[prd,aps,floats,floatfix,superscriptaddress,preprintnumbers,showpacs,eqsecnum,nofootinbib,twocolumn]{revtex4}
\usepackage{amsfonts,amsmath,epsfig,amssymb,latexsym,color,graphicx,array,subfigure,bm,float }

\definecolor{red}{rgb}{1,0,0}
\definecolor{blue }{rgb}{0,0,1}
\definecolor{green}{rgb}{0,1,0}


\newcommand{\bea}{\begin{eqnarray}}
\newcommand{\ena}{\end{eqnarray}}
\newcommand{\beann}{\begin{eqnarray*}}
\newcommand{\enann}{\end{eqnarray*}}

\newcommand{\dsl}{\pa \kern-0.5em /}

\newcommand{\pa}{\partial}

\newcommand{\nn}{\nonumber\\}

\newcommand{\vect}[1]{\!\!\mbox{ \boldmath $#1$}}
\newcommand{\gsim}{\, \mbox{\raisebox{-1.ex}
{$\stackrel{\textstyle>}{\textstyle\sim}$}}\,}
\newcommand{\lsim}{\, \mbox{\raisebox{-1.ex}
{$\stackrel{\textstyle<}{\textstyle\sim}$}}\,}

\begin{document}

\date{\today}

\title{Dynamics of Binary System  around Supermassive Black Hole}

\author{Kei-ichi Maeda}
\affiliation{Department of Physics, Waseda University,  Shinjuku, Tokyo 169-8555, Japan}
\affiliation{Center for Gravitational Physics and Quantum Information, Yukawa Institute
for Theoretical Physics, Kyoto University, 606-8502, Kyoto, Japan}

\author{Priti Gupta}
\affiliation{Department of Physics, Indian Institute of Science, Bangalore 560012, India}

\author{Hirotada Okawa}
\affiliation{Waseda Institute for Advanced Study {\rm (WIAS)}, 1-21-1 Nishi Waseda, Shinjuku-ku, Tokyo 169-0051, Japan }

\begin{abstract}
We discuss motion of a binary system around a supermassive black hole.
Using Fermi-Walker transport, we construct a
local inertial reference frame and set up a Newtonian binary system.
Assuming a circular geodesic observer around a Schwarzschild black hole, 
we write down the equations of motion of a binary.
Introducing a small acceleration of the observer, we
 remove the interaction terms between the center of mass (CM) of a binary and its relative coordinates.
The CM follows the observer's orbit, but its motion deviates from an exact 
circular geodesic.
We first solve the relative motion of a binary system, and  then 
find the motion of the CM by the perturbation equations with the small acceleration. 

We show that there appears the Kozai-Lidov (KL) oscillations when a binary is compact and the initial inclination is larger than a critical angle. In a hard binary system, KL oscillations are regular, whereas in a soft binary system, oscillations are irregular both in period and in amplitude, although stable. We find an orbital flip when the initial inclination is large. As for the motion of the CM, the radial deviations from a circular orbit become stable oscillations with very small amplitude.
\end{abstract}

\maketitle





\section{Introduction}
\label{Introduction}

After the remarkable success of the LIGO-Virgo-KAGRA Collaboration~\cite{Abbott_2020,Abbott_2021}, the study of gravitational wave (GW) emission has received a significant boost.
The analysis of data (obtained through the first three observational runs) produced over a hundred confident detections~\cite{Abbott_2019,o3population} from binary black holes (BH), a binary neutron star (NS) and BH-NS systems, with more to follow in the next decade. The scientific insights emerging from the detections have significantly revolutionized our understanding of the sources. For instance, some notable events revealed heavier stellar-mass BHs~\cite{GW150914}, and their origin is still under discussion. Using the electromagnetic counterpart, we found that the speed of GWs is very close to the speed of light as predicted by the general theory of relativity~\cite{GW170817}. With the increase in detections, we can probe more fundamental questions like testing theories of gravity in strong field regimes, finding the redshift distribution of BHs and their environment, and so on~\cite{test1,test2,test3,test4,test5}.

Unlocking the scientific potential of GWs depends on our theoretical knowledge. In order to filter the GW signal from the detector noise, it is necessary to model the predicted waveform accurately. Current observations are from isolated binary systems. It is possible, however, that nature will provide us with more exotic sources. This paper will examine a three-body system as one of the likely sources. The environment near supermassive black holes (SMBHs) in galactic nuclei comprises many stars and compact objects. A binary system could emerge in these surroundings, which composes a natural hierarchical triple system~\cite{Heggie1975,Hut1993,Samsing2014,Riddle2015,Fabio2016,stephan2019}. Recent LIGO events suggest the hierarchical systems as a possible dynamical formation channel of the heavy merging binary BHs~\cite{sym13091678,Gayathri_2020,Gerosa_2021}.

In a hierarchical triple, the distance between two bodies (forming an ‘inner’ binary) is much less than the distance to the third body. In 1962, Kozai and Lidov independently explored the dynamics of restricted hierarchical triples~\cite{Kozai62,Lidov62}, revealing a remarkable phenomenon (known as Kozai-Lidov (KL) resonance) — when the two orbits are inclined relative to each other, there is a periodic exchange between orbital eccentricity and relative inclination in secular timescale~\cite{Shevchenko17}. The orbital eccentricity can reach extreme values leading to large emission of GWs~\cite{Lisa2019,hoang18,Gupta_2020}. 

There has been extensive work on dynamics of such systems based on Newtonian or post-Newtonian approximation~\cite{naoz13b,Naoz12,Naoz16,Naoz2020,tey13,Li15,Will14a,Will14b}. Recently we also find many work focusing on gravitational waves from such systems~\cite{Amaro-Seoane2010, Antonini2012,hoang18,Antonini2016,Meiron2017,Robson2018,Lisa2018,Lisa2019,Hoang2019,Loeb2019,Gupta_2020,kuntz2022transverse}. It has been shown that KL resonance leaves an imprint on the waveform and lies in the detectable range of future space-based detectors like LISA and DECIGO. Indirect observation of GW from a triple system is also studied by analyzing the cumulative shift of periastron time of a binary pulsar undergoing KL oscillations~\cite{Haruka2019,Suzuki:2020zbg}. 

In this paper, when we discuss on a binary system near SMBH, we follow another approach, i.e., a binary system is treated as perturbations of SMBH spacetime. In the case of a single object in SMBH spacetime, it can be treated a test particle. But in  the case of a binary system, it is not the case because the self-gravitational mutual interaction is much stronger than the gravitational tidal force by SMBH. In order to analyze such a hierarchical system, we first prepare a local inertial frame and set up a binary in this frame. When a binary is tightly bounded but the mutual gravitational interaction is not so strong, the binary motion in this frame can be discussed by Newtonian gravitational dynamics.

Using Fermi normal coordinate system or Fermi-Walker transport, we can construct a local inertial frame~\cite{1963JMP.....4..735M,Nesterov_1999,Delva:2011abw}. Using such a technique, there are several discussions on a tidal force acting on stars near SMBH~\cite{Banerjee_2019,PhysRevD.71.044017,Cheng_2013,Kuntz_2021}, but only a few works on a binary system have been discussed~\cite{Gorbatsievich_Bobrik,Chen_Zhang,camilloni2023tidal}. In this paper, we analyze such a system in detail. Assuming an observer is moving along a circular geodesic around a Schwarzschild SMBH, we construct a local inertial frame, and set up a binary system. We then discuss motion of a binary, showing existence of the KL oscillations when a binary is compact and the initial inclination angle is larger than a critical value.

The paper is organized as follows: We review 
how to construct a local inertial proper reference frame by use of 
Fermi-Walker transport in Sec. II A. In Sec. II B, we perform post-Newtonian expansion for a test particle motion in this frame. In Sec. III, we 
set up a self-gravitating system in the proper reference frame and derive the Lagrangian in the Newtonian limit.
In Sec. IV, assuming an observer moving along a circular geodesic in Schwarzschild 
black hole, we derive the equations of motion for a binary system. 
We also discuss the interaction terms between the center of mass (CM) of a binary and its relative coordinates.  Introducing small acceleration of an observer, we remove 
the interaction terms, finding the equations of motion for the CM, which gives 
small deviations from a circular geodesic.
We then analyze twelve models numerically and show the properties of binary motions
such as the KL oscillations, chaotic features, and orbital flips in Sec. V. 
We also discuss motions of the CM of a binary. Concluding remarks follow in Sec. VI.
In the Appendix A, we provide some numerical and 
analytic solutions in the coplanar case.
We also present the Lagrange planetary equations of the model 
and write down the equations for the orbital parameters of a binary 
taking averages over inner and outer binary cycles in Appendix B.
We show that this simplified method recovers numerical results obtained by 
direct integration of the equations of motion in the case of a hard binary.
It also provides the KL oscillation time scale and the maximum and minimum values of eccentricity. 

\textit{Notation used}: Greek letters range from 0 to 3, while Roman letters run from 1 to 3; Hatted indices denote tetrad components in a proper reference frame rotating along an observer; Bar over symbols correspond to quantities in a static tetrad frame.

\section{Proper reference frame}
\label{proper reference frame}
\subsection{Proper reference frame of an arbitrary observer
in a curved spacetime}

We first discuss how to set up a local inertial frame in a curved spacetime
\cite{Misner1973,1963JMP.....4..735M,Gorbatsievich_Bobrik}. The spacetime  metric is given by
\bea
d\bar s^2=\bar g_{\mu\nu}dx^\mu dx^\nu
\label{background_metric}
\,.
\ena
We then consider  an observer, 
whose orbit  is given by 
a world line $ \gamma$ described by 
\beann
x^\mu=z^\mu(\tau)
\,,
\enann
where $\tau$ is a proper time of the observer. 
The 4-velocity is given by
\beann
u^\mu(\tau)\equiv {dz^\mu\over d\tau}
\,.
\enann

We now set up an orthonormal tetrad system $\{e_{\hat \alpha}^\mu\}$ along the observer's world line $\gamma$, which is defined  by
the conditions such that
\beann
e_{\hat \alpha}^{~\mu} e_{\hat \beta \mu}=\eta_{\hat \alpha \hat \beta}\,,~~e_{\hat 0}^{~\mu}=u^\mu\,,
\enann
where $\eta_{\hat \alpha \hat \beta}$ is 
Minkowski spacetime metric.

For a given 4-velocity $u^\mu$,
 this tetrad system is determined up to three-dimensional rotations.
The tetrad $e_{\hat \alpha}^\mu$  is transported  along the observer's world line  $ \gamma$ 
as
\beann
{D e_{\hat i}^{~\mu}\over d\tau}=-\Omega^{\mu\nu}\,e_{\hat i \,\nu}
\,,
\enann
where 
\beann
\Omega^{\mu\nu}\equiv a^\mu  u^\nu-u^\mu a^\nu  +u_\alpha \omega_\beta \epsilon^{\alpha\beta\mu\nu}
\,.
\enann
Here,
\beann
a^\mu&\equiv& {Du^\mu\over d\tau},
\enann
and 
\beann
\omega_\mu&\equiv&{1\over 2}
u^\alpha\epsilon_{\alpha\mu\rho\sigma}\Omega^{\rho\sigma}={1\over 2}\stackrel{(3)}{\epsilon}_{\mu\rho\sigma}\Omega^{\rho\sigma},
\enann
are  the acceleration of the observer
and  the angular velocity of a rotating 
spatial basis vector $e_{(a)}^\mu$, respectively. A non-rotating tetrad frame for which $\omega^\mu=0$ is called the Fermi-Walker transport.
If the orbit is a geodesic ($a^\mu=0$ and $\omega^{\hat k}=0$), we find 
$\displaystyle{{D e_{\hat a}^\mu\over d\tau}=0}$, which is just a parallel transport.

Next, we construct a local coordinate system (the observer's proper reference system) 
near the world line $\gamma$, which is described as
\beann
(x^{\hat\mu})=(c\tau, x^{\hat a})\,,
\enann
where the spatial components $x^{\hat a}$ is measured from the point at $\tau$ on the world line $\gamma$ along the spatial hypersurface $\Sigma(\tau)$ perpendicular to $\gamma$.

\begin{widetext}

We find that the metric form of this proper reference frame 
up to the second order of $x^{\hat a}$ 
is given by
\bea
g_{\hat \mu\hat \nu}=\eta_{\hat \mu \hat \nu}+\varepsilon_{\hat \mu \hat \nu}
+O(|x^{\hat k}|^3),
\label{metric_reference_frame}
\ena 
where
\bea
\varepsilon_{\hat 0 \hat 0}&=&-{1\over c^2}\left[2a_{\hat k}
x^{\hat k}
+\left(c^2
\bar{\cal R}_{\hat 0 \hat k \hat 0 \hat \ell }-
\omega_{\hat j\hat k} \omega^{\hat j}_{~\hat \ell} \right)
x^{\hat k}  x^{\hat \ell} +{\left(a_{\hat k} x^{\hat k}\right)^2\over c^2}   \right],
\label{epsilon00}\\
\varepsilon_{\hat 0 \hat j}&=&-{1\over c^2}\left[c\, \omega_{\hat j\hat k} x^{\hat k} 
 +{2\over 3}c^2 \bar{\cal R}_{\hat 0\hat k \hat j  \hat \ell }x^{\hat k}x^{\hat \ell}\right],
\label{epsilon0j}\\
\varepsilon_{\hat i \hat j}&=&-{1\over c^2}\left[{1\over 3}c^2\bar{\cal R}_{\hat i \hat k \hat j\hat \ell}x^{\hat k}x^{\hat \ell}
\right]
\,.
\label{epsilonij}
\ena
where
$\bar{\cal R}_{\hat \mu\hat\nu\hat\rho\hat \sigma}$ is the tetrad component of the Riemann curvature of a background spacetime
and
$
\omega_{\hat j\hat k}\equiv \epsilon_{\hat j \hat k \hat \ell}\omega^{\hat \ell}
$~\cite{Misner1973,1963JMP.....4..735M,Gorbatsievich_Bobrik}.
\end{widetext}

The acceleration and angular frequency in the proper reference frame are defined by
\beann
a^{\hat j}&\equiv & e^{\hat j}_{~\mu} {Du^\mu\over d\tau},
\\
\omega^{\hat j}&\equiv &{1\over 2}\stackrel{(3)}{\epsilon}^{\hat j\hat k\hat \ell}e_{\hat \ell \mu}{De_{\hat k}^{~\mu}\over d\tau}
\,.
\enann

For the Fermi-Walker transport, $\omega^{\hat k}=0$. If it is 
 the geodesic  ($a^{\hat j}=0$ and $\omega^{\hat k}=0$),
we recover the Fermi normal coordinates.
The tetrad is parallely transformed along the world line.

\subsection{Test particle motion in a proper reference frame}
\label{particle}
First, we consider motion of a test particle in the above proper reference frame.
The action for a test particle with mass $m$ in a given spacetime is given by
\beann
S=-m c \int \sqrt{-ds^2}
\,.
\enann
For a test particle in the proper reference frame, 
since the world interval $ds^2$ is 
given by the metric $g_{\hat \mu \hat \nu}$,
we  find the action for a test particle as
\beann
S=\int d\tau {\cal L}
\,,
\enann
where 
\beann{\cal L}&\equiv& 
-mc \sqrt{-\left(\eta_{\hat \mu\hat \nu}+\varepsilon_{\hat \mu\hat \nu}\right)
{dx^{\hat\mu}\over d\tau}{dx^{\hat\nu}\over d\tau}},
\enann
is Lagrangian of the test particle and $\varepsilon_{\hat \mu\hat \nu}$ is a small deviation from the Minkowski spacetime 
since $|\varepsilon_{\hat \mu\hat \nu}| \ll 1$.

\begin{widetext}
Assuming that the test particle  moves slowly
in the proper reference frame, 
 we perform the post-Newtonian expansion in terms of $v^{\hat j}/c$ as
follows:
First, we expand the square root term in the Lagrangian as
\beann
{\cal L}
&=&-mc^2 \sqrt{
1-{\vect{v}^2\over c^2}
-\varepsilon_{\hat 0\hat 0}-2\varepsilon_{\hat 0\hat j}{v^{\hat j}\over c}-
\varepsilon_{\hat i\hat j}
{v^{\hat i}v^{\hat j}\over c^2}
}
\\
&=&
-mc^2\Big\{1-{\vect{v}^2\over 2 c^2}-{\varepsilon_{\hat 0\hat 0}\over 2}-\varepsilon_{\hat 0\hat j}{v^{\hat j}\over c}-{1\over 2}
\varepsilon_{\hat i\hat j}
{v^{\hat i}v^{\hat j}\over c^2}
-{1\over 8}\left[{\vect{v}^2\over c^2}
+\varepsilon_{\hat 0\hat 0}+2\varepsilon_{\hat 0\hat j}{v^{\hat j}\over c}+
\varepsilon_{\hat i\hat j}
{v^{\hat i}v^{\hat j}\over c^2}\right]^2+\cdots
\Big\}.
\enann
Note that $x^{\hat 0}=c\tau$.
Inserting Eqs. (\ref{epsilon00}), (\ref{epsilon0j}), and (\ref{epsilonij}),
and expanding the above Lagrangian in terms of $v^{\hat j}/c$, 
we obtain
\beann
{\cal L}=
{\cal L}_0+{\cal L}_{1/2}+
{\cal L}_1,
\enann
where
\beann
{\cal L}_0&=&{1\over 2}m\vect{v}^2-ma_{\hat k}x^{\hat k}-{1\over 2}m \omega_{\hat j\hat k} \omega^{\hat j}_{~\hat \ell}
x^{\hat k}  x^{\hat \ell} 
-m\omega_{\hat j\hat k}v^{\hat j}x^{\hat k}
-{1\over 2}mc^2\bar{\cal R}_{\hat 0 \hat k \hat 0 \hat \ell }x^{\hat k}  x^{\hat \ell},
\\
{\cal L}_{1/2}&=&
-{2\over 3}mc^2\bar{\cal R}_{\hat 0\hat k \hat j  \hat \ell }
x^{\hat k}x^{\hat \ell}\, {v^{\hat j}\over c},
\\
{\cal L}_1&=&-{m\over 2}{(a_{\hat k}x^{\hat k})^2\over c^2}
-{1\over 6}
mc^2\bar{\cal R}_{\hat i \hat k \hat j\hat \ell}x^{\hat k}x^{\hat \ell}{v^{\hat i}v^{\hat j}\over c^2}
+{m\over 8c^2}\left[\vect{v}^2-
2a_{\hat k}x^{\hat k}-\omega_{\hat j\hat k}\omega^{\hat j}_{~\hat l}
x^{\hat k} x^{\hat \ell}-2\omega_{\hat j\hat k}x^{\hat k}v^{\hat j}
-c^2\bar{\cal R}_{\hat 0 \hat k \hat 0 \hat \ell }x^{\hat k} x^{\hat \ell}
\right]^2
\,,
\enann 
 are Newtonian, 0.5 PN and 1PN Lagrangian, respectively.

In this expansion, we find 0.5 PN term formally, but it can be
 an apparent term which comes from a choice of an observer's coordinates. 
In fact, if we choose an appropriate observer's acceleration, which 
appears as 0.5 PN term, 
we can remove the above ${\cal L}_{1/2}$.
However, if we have multi-particle system as we will discuss later, 
this adjustment can be used 
only for one particle or the center of mass of the system
(See \S. \ref{binary}). Hence, we keep the ${\cal L}_{1/2}$  term
and discuss ``Newtonian" dynamics including such a term.
\\
\end{widetext}

\section{Self-gravitating Newtonian system in a curved spacetime}
\label{binary}

\subsection{Self-gravitating $N$-body system}
\label{N_body}
Now we discuss self-gravitating $N$-body system in a fixed curved background spacetime, which is given by the metric in Eq.~\eqref{background_metric}~\cite{Gorbatsievich_Bobrik}. 
We are interested in the case where Newtonian dynamics is valid 
 in the observer's proper reference frame. 
 The necessary condition is 
that the typical scale  $\ell_{N\text{-body}}$  of $N$-body system should satisfy 
\beann
\ell_{N\text{-body}}\ll 
{\rm min} \left[{1\over |a^{\hat j}|}, {1\over |\omega^{\hat j}|}, \ell_{\bar{\cal R}}\right],
\enann
where $\ell_{\bar{\cal R}}$ is the minimum curvature radius defined by
\beann
\ell_{\bar{\cal R}} \equiv {\rm min} \left[|\bar{\cal R}_{\hat \mu\hat \nu\hat \rho\hat \sigma}|^{-{1\over 2}}, |\bar{\cal R}_{\hat \mu\hat \nu\hat \rho\hat \sigma ; \hat \alpha}|^{-{1\over 3}}, |\bar{\cal R}_{\hat \mu\hat \nu\hat \rho\hat \sigma ; \hat \alpha;\hat \beta}|^{-{1\over 4}}
\right]
\,.
\enann

In order to find the metric contributions from $N$-body system, we first focus on the motion of the $I$th particle, which is gravitating with the other $(N-1)$ particles. The metric contribution from the those $(N-1)$ particles is given by
\beann
\varphi^I_{\hat 0\hat 0}=-{2\Phi_I\over c^2}\,,
\enann
where $\Phi_I$ is the Newtonian potential produced by the $(N-1)$ particles, which is given by
\beann
\Phi_I(x^{\hat i})=-G\sum_{J\neq I}^N{m_J\over |x^{\hat i}-x_J^{\hat i}|}
\,.
\enann
We assume that the other components of $\varphi^I_{\hat \mu\hat \nu}$ vanish
because we are interested in Newtonian dynamics in the proper reference frame.

We then obtain the metric of the observer's proper reference frame for the $I$th particle as
\bea
g^I_{\hat \mu \hat \nu}=\eta_{\hat \mu \hat \nu}+h^I_{\hat \mu \hat \nu},
\label{metric_g+h}
\ena
where
\bea
h^I_{\hat \mu \hat \nu}=\varepsilon_{\hat \mu \hat \nu}+\varphi^I_{\hat \mu \hat \nu}
\,.
\label{metric_e+p}
\ena

The equation of motion for the $I$th particle 
can be derived by the variation of the action
\beann
{\cal S}_I=\int d\tau{\cal L}_I,
\enann
where
\beann
{\cal L}_I\equiv -m_I c\sqrt{-g^I_{\hat \mu\hat \nu}{dx_I^{\hat \mu}\over d\tau}{dx_I^{\hat \nu}\over d\tau}}
\,.
\enann
\begin{widetext}

By use of  the metric form (\ref{metric_g+h}),
we can expand the above Lagrangian up to 0.5 PN order as
\beann
{\cal L}_I&=&
{1\over 2}  m_I \left({d\vect{x}_I\over d\tau}\right)^2
- m_I  \Phi (x_I) -m_I a_{\hat k}x_I ^{\hat k}-{1\over 2}m_I  \omega_{\hat j\hat k} \omega^{\hat j}_{~\hat \ell}
x_I ^{\hat k}  x_I ^{\hat \ell} 
-m_I \omega_{\hat j\hat k}v_I ^{\hat j}x_I ^{\hat k}
\\
&&
-{1\over 2}m_I c^2\bar{\cal R}_{\hat 0 \hat k \hat 0 \hat \ell }
x_I ^{\hat k}  x_I ^{\hat \ell}-{2\over 3}
m_I c^2\bar{\cal R}_{\hat 0\hat k \hat j  \hat \ell }
x_I^{\hat k}x_I^{\hat \ell}\, {v_I^{\hat j}\over c}.
\enann

The total action of $N$-body system and its Lagrangian are given by summing up 
each Lagrangian ${\cal L}_I$.
We finally  obtain
\beann
{\cal S}_{N\mathchar`-{\rm  body}}=\int d\tau {\cal L}_{N\mathchar`-{\rm body}},
\enann
where
\beann
{\cal L}_{N\mathchar`-{\rm body}}={1\over 2}  \sum_{I} m_I \left({d\vect{x}_I\over d\tau}\right)^2
+\sum^N_I\sum^N_{J\neq I}
{G m_I m_J\over 2|x_I^{\hat i}-x_J^{\hat i}|}
+{\cal L}_a+{\cal L}_\omega+{\cal L}_{\bar{\cal R}},
\enann
with
\beann
{\cal L}_a&=&-\sum_I^N m_I a_{\hat k}x_I ^{\hat k},
\\
{\cal L}_\omega&=&-{1\over 2}\sum_I^N m_I  \left[\omega_{\hat j\hat k} \omega^{\hat j}_{~\hat \ell}
x_I ^{\hat k}  x_I ^{\hat \ell} 
+2\omega_{\hat j\hat k}v_I ^{\hat j}x_I ^{\hat k}\right],
\\
{\cal L}_{\bar{\cal R}}&=&-{1\over 2}\sum_I^N  m_I c^2\bar{\cal R}_{\hat 0 \hat k \hat 0 \hat \ell }
x_I ^{\hat k}  x_I ^{\hat \ell}-{2\over 3}\sum_I
m_I c^2\bar{\cal R}_{\hat 0\hat k \hat j  \hat \ell }
x_I^{\hat k}x_I^{\hat \ell}\, {v_I^{\hat j}\over c}
\,.
\enann

Note that ${\cal L}_a$  comes from the inertial force of the accelerated observer, and ${\cal L}_\omega$ originates in 
the rotation of the observer (the centrifugal force and the Coriolis force). ${\cal L}_{\bar{\cal R}}$ describes the 
curvature effect of the third body (the tidal force).

\subsection{Binary system in a curved spacetime}
\label{binary_in_curved_ST}
Next, we discuss a binary system in a fixed curved background.
A binary consists of two point particles with the masses $m_1$ and $m_2$.
The Lagrangian up to 0.5 PN order is given by
\bea
{\cal L}_{\rm binary}={\cal L}_{\rm N}+{\cal L}_{1/2},
\label{Lagrangian_binary}
\ena
where
\bea
{\cal L}_{\rm N}&\equiv& {1\over 2} \left[ m_1 \left({d\vect{x}_1\over d\tau}\right)^2+m_2 \left({d\vect{x}_2\over d\tau}\right)^2\right]
+  {G m_1m_2\over |\vect{x}_1-\vect{x}_2|}
+{\cal L}_{a}+{\cal L}_{\omega}+{\cal L}_{\bar{\cal R}}
\label{Lagrangian_N},
\ena
with
\beann
{\cal L}_{a}
&=&
-\sum_{I=1}^2 m_I a_{\hat k}x_I ^{\hat k},
\\
{\cal L}_{\omega}
&=&
-\sum_{I=1}^2 m_I \left[\epsilon_{\hat j\hat k\hat \ell}\omega^{\hat \ell}
x_I^{\hat k}{dx_I^{\hat j}\over d\tau}-
{1\over 2} \left(\vect{\omega}^2 \vect{x}_I^2 -(\vect{\omega} \cdot \vect{x}_I)^2
\right)\right],
\\
{\cal L}_{\bar{\cal R}}
&=&
-{1\over 2}  \sum_{I=1}^2 m_I \bar{\cal R}_{\hat 0\hat k\hat 0\hat \ell}x_I^{\hat k} x_I^{\hat \ell}
\,,
\enann
and 
\bea
{\cal L}_{1/2}&\equiv &-{2\over 3}\sum_{I=1}^2
m_I c^2\bar{\cal R}_{\hat 0\hat k \hat j  \hat \ell }
x_I^{\hat k}x_I^{\hat \ell}\, {v_I^{\hat j}\over c}
\,.
\label{Lagrangian_0.5PN}
\ena

Introducing the center of mass coordinates and the relative coordinates by
\beann
\vect{R}&=&{m_1\vect{x}_1+m_2\vect{x}_2\over m_1+m_2},
\\
\vect{r}&=& \vect{x}_2-\vect{x}_1
\,,
\enann
we find the Newtonian Lagrangian (Eq.~\eqref{Lagrangian_N}) in terms of  $\vect{R}$  and $\vect{r}$ as
\bea
{\cal L}_{\rm N}={\cal L}_{\rm CM}(\vect{R}, \dot{\vect{R}})+
{\cal L}_{\rm rel}(\vect{r}, \dot{\vect{r}})
\label{Lagrangian_N_Rr}
\,,
\ena
where
\beann
{\cal L}_{\rm CM}(\vect{R},\dot{\vect{R}})
&=&{1\over 2} (m_1+m_2)\dot{\vect{R}}^2
+{\cal L}_{{\rm CM}\mathchar`-a}(\vect{R},\dot{\vect{R}})
+{\cal L}_{{\rm CM}\mathchar`-\omega}(\vect{R},\dot{\vect{R}})
+{\cal L}_{{\rm CM}\mathchar`-\bar{\cal R}}(\vect{R},\dot{\vect{R}}),
\enann
with 
\beann
{\cal L}_{{\rm CM}\mathchar`-a}
&=&-(m_1+m_2)a_{\hat k}x_I ^{\hat k},
\\
{\cal L}_{{\rm CM}\mathchar`-\omega}
&=&-(m_1+m_2)\left[
\epsilon_{\hat j\hat k\hat \ell}\omega^{\hat \ell}
R^{\hat k}\dot R^{\hat j}-{1\over 2}\left(\vect{\omega}^2 \vect{R}^2
-\left(\vect{\omega}\cdot \vect{R}\right)^2\right)\right],
\\{\cal L}_{{\rm CM}\mathchar`-\bar{\cal R}}
&=&
-{1\over 2}(m_1+m_2)
\bar{\cal R}_{\hat 0\hat k \hat 0 \hat \ell}R^{\hat k}R^{\hat \ell},
\enann
and
 \beann
{\cal L}_{\rm rel}(\vect{r},\dot{\vect{r}})&=&
{1\over 2}\mu \dot{\vect{r}}^2+  {G m_1m_2\over r}
+{\cal L}_{{\rm rel}\mathchar`-\omega}(\vect{r},\dot{\vect{r}})
+{\cal L}_{{\rm rel}\mathchar`- \bar{\cal R}}(\vect{r},\dot{\vect{r}}),
\enann
with 
\beann
{\cal L}_{{\rm rel}\mathchar`-\omega}
&=&-\mu \left[
\epsilon_{\hat j\hat k\hat \ell}\omega^{\hat \ell}
r^{\hat k}\dot r^{\hat j}-{1\over 2}\left(\vect{\omega}^2 \vect{r}^2
-\left(\vect{\omega}\cdot \vect{r}\right)^2\right)\right]\,,
\\
{\cal L}_{{\rm rel}\mathchar`-\bar{\cal R}}
&=&
-{1\over 2}\mu 
\bar{\cal R}_{\hat 0\hat k \hat 0 \hat \ell}r^{\hat k}r^{\hat \ell}
\,.
\enann
Here, $\mu = m_1 m_2/(m_1+m_2)$ is the reduced mass. When we consider only ${\cal L}_{\rm N}$, 
we can separate the variables $\vect{R}$ and $\vect{r}$.
 In particular, when the observer follows the geodesic ($\vect{a}=0$ and $\vect{\omega}=0$),
the orbit of $\vect{R}=0$ is a solution of the equation for $\vect{R}$.
It means that the center of mass (CM) follows the observer's geodesic.
We have only the equation for the relative coordinate $\vect{r}$. However, when we include the 0.5 PN term, it is not the case.
The 0.5PN Lagrangian ${\cal L}_{1/2}$  is rewritten by use of  $\vect{R}$  and    $\vect{r}$ as
\bea
{\cal L}_{1/2}={\cal L}_{1/2\mathchar`-{\rm CM}}(\vect{R}, \dot{\vect{R}})+
{\cal L}_{1/2\mathchar`-{\rm rel}}(\vect{r}, \dot{\vect{r}})
+{\cal L}_{1/2\mathchar`- {\rm int}}(\vect{R}, \dot{\vect{R},}\vect{r}, \dot{\vect{r}}),
\label{Lagrangian_1/2_Rr}
\ena
where
\bea
{\cal L}_{1/2\mathchar`-{\rm CM}}(\vect{R}, \dot{\vect{R}})&=&-{2\over 3}(m_1+m_2)R_{\hat 0\hat k \hat j \hat \ell}
R^{\hat k}R^{\hat \ell}\dot R^{\hat j},
\nn
{\cal L}_{1/2\mathchar`-{\rm rel}}(\vect{r}, \dot{\vect{r}})&=&- {2\over 3} \mu{(m_1-m_2)\over (m_1+m_2)}R_{\hat 0\hat k \hat j \hat \ell}r^{\hat k}r^{\hat \ell}\dot r^{\hat j},
\nn
{\cal L}_{1/2\mathchar`- {\rm int}}(\vect{R}, \dot{\vect{R},}\vect{r}, \dot{\vect{r}})&=&
- {2\over 3} \mu R_{\hat 0\hat k \hat j \hat \ell}
\left[r^{\hat k}r^{\hat \ell}\dot R^{\hat j}
+\left(R^{\hat k}r^{\hat \ell}+r^{\hat k}R^{\hat \ell}
\right)\dot r^{\hat j}\right].
\label{interaction_term}
\ena
Due of the interaction term ${\cal L}_{1/2\mathchar`- {\rm int}}$, the orbit of $\vect{R}=0$ is no longer a solution even if the acceleration vanishes. The motion of  the CM ($\vect{R}(\tau$)) couples with the relative motion $(\vect{r}(\tau))$. As a result, not only the orbit of a binary but also the motion of the CM will become complicated even if the observer's orbit is a geodesic.

However, if we introduce an appropriate acceleration $\vect{a}$
in 0.5PN order 
 to cancel the interaction terms, $\vect{R}=0$ will become a solution, i.e., 
 the CM can follow the observer's motion as follows:
Integrating by parts the interaction term, we find
 \beann
{\cal L}_{1/2\mathchar`- {\rm int}}(\vect{R}, \dot{\vect{R},}\vect{r}, \dot{\vect{r}})&=&
-{2\over 3}\mu \bar{\cal R}_{\hat 0\hat k \hat j \hat \ell}
\left[\dot R^{\hat j} r^{\hat k}r^{\hat \ell}
+\dot r^{\hat j}\left(R^{\hat k}r^{\hat \ell}+r^{\hat k}R^{\hat \ell}
\right)\right]
\\
&\approx&
2\mu\left[{1\over 3}{d\bar{\cal R}_{\hat 0\hat k \hat j \hat \ell}\over d\tau}
r^{\hat k}r^{\hat \ell} 
+ \bar{\cal R}_{\hat 0\hat k \hat j \hat \ell}
r^{\hat k}\dot r^{\hat \ell} \right]R^{\hat j}~~{\rm (integration~by~part)}
\,,
\enann
where the time derivative of the curvature is evaluated along the observer's orbit.

If we define the acceleration by
 \beann
 a_{\hat j}={2\mu \over m_1+m_2}\left[{1\over 3}{d\bar{\cal R}_{\hat 0\hat k \hat j \hat \ell}\over d\tau}
r^{\hat k}r^{\hat \ell} 
+ \bar{\cal R}_{\hat 0\hat k \hat j \hat \ell}
r^{\hat k}\dot r^{\hat \ell} \right]
\label{0.5PN_acceleration}
\,,
 \enann
two terms ${\cal L}_{1/2\mathchar`- {\rm int}}$ and ${\cal L}_{{\rm CM}\mathchar`-a}$ cancel each other.
As a result, the Lagrangians for $\vect{R}$ and $\vect{r}$ are decoupled,
and  $\vect{R}=0$ becomes an exact solution of the equation for $\vect{R}$, which is
derived from the Lagrangian
(${\cal L}_{\rm CM}+{\cal L}_{1/2\mathchar`-{\rm CM}}$).
The CM follows the observer's orbit and therefore, we obtain the decoupled equation for the relative coordinate $\vect{r}$.
 
In order to obtain the proper observer's orbit, 
which is not a geodesic but may be close to the geodesic, 
we have to solve the equation of motion including small acceleration such that
 \bea
 {Du_{\rm CM}^\mu\over d\tau}=a^\mu=e^{\mu\hat j } a_{\hat j}=
{2\mu \over m_1+m_2}e^{\mu\hat j }\left[{1\over 3}{d\bar{\cal R}_{\hat 0\hat k \hat j \hat \ell}\over d\tau}
r^{\hat k}r^{\hat \ell} 
+ \bar{\cal R}_{\hat 0\hat k \hat j \hat \ell}
r^{\hat k}\dot r^{\hat \ell} \right]
\,.
\label{eq_CM}
 \ena
 
As a result, we first solve the equation for the relative coordinate $\vect{r}$, which is 
obtained only by the Lagrangian ${\cal L}_{\rm rel}(\vect{r})
 +{\cal L}_{1/2\mathchar`-{\rm rel}}(\vect{r})$.
 Note that when $m_1=m_2$, we have only Newtonian 
 Lagrangian ${\cal L}_{\rm rel}$ because ${\cal L}_{1/2\mathchar`-{\rm rel}}=0$ vanishes.
After obtaining the solution of $\vect{r}(\tau)$, 
we find the  motion for the CM
 (or the observer)  in the background spacetime
 by solving Eq.~\eqref{eq_CM}. Using  the relative motion $\vect{r}(\tau)$ with the solution of the CM motion 
 ($x_{\rm CM}^\mu(\tau)$), 
 we will obtain a binary motion in a given curved background spacetime ($x_1^\mu(\tau)\,,x_2^\mu(\tau)$).
 \\

\end{widetext}

\section{A binary system in a Schwarzschild spacetime}
\label{EOM_Schwarzschild}

We consider a spherically symmetric supermassive black hole.
The background spacetime is given by Schwarzschild solution as
\bea
d\bar s^2=-f d\mathfrak{t}^2+{1\over f}d\mathfrak{r}^2+\mathfrak{r}^2d\Omega^2
\,,
\ena
where
\bea
f = 1-{\mathfrak{r}_g\over \mathfrak{r}}.
\ena
The gravitational radius $\mathfrak{r}_g$ is given by 
\beann
\mathfrak{r}_g\equiv {2GM\over c^2},
\enann
where $M$ is a gravitational mass of the supermassive black hole. In what follows, we set $G=1$ and $c=1$ for brevity unless specified otherwise. 

\subsection{Proper reference frame}
\begin{figure}[h]
\begin{center}
\includegraphics[width=4.5cm]{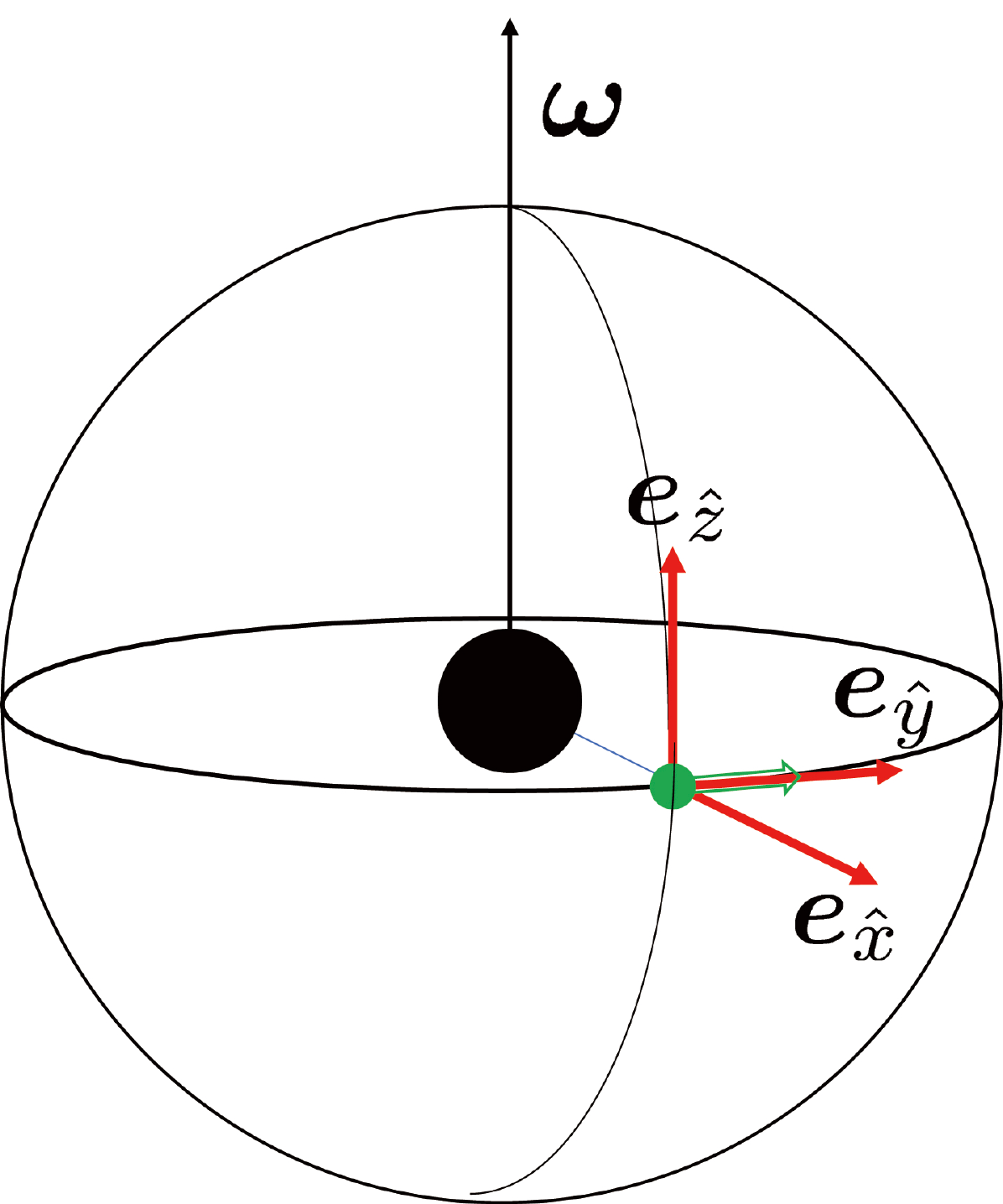}
\caption{A  local inertial  tetrad system $\{
\vect{e}_{\hat x}\,, \vect{e}_{\hat y}\,, \vect{e}_{\hat z} \}$ rotating with angular frequency
 $\mathfrak{w}_{\rm R}$ 
along a circular orbit.}
\label{tetrad}
\end{center}
\end{figure}

We first discuss Newtonian dynamics for which Lagrangian ${\cal L}_{\rm N}$ is given by Eq.~\eqref{Lagrangian_N}.
In this case, the CM of a binary system follows a geodesic observer.
We consider an observer,  which moves along a
circular geodesic with the radius $\mathfrak{r}=\mathfrak{r}_0$.
The orbit is assumed to be 
 on the equatorial plane without loss of generality.
We introduce a rotating tetrad system along this geodesic such that
\beann
e_{\hat 0}^{~\mu}&\equiv& u^\mu=\left({\sqrt{1+\mathfrak{r}_0^2\mathfrak{w}_0^2}\over \sqrt{f(\mathfrak{r}_0)}}\,, 0\,, 0\,,\mathfrak{w}_0\right),
\\
e_{\hat 1}^{~\mu}&=& 
\left(0\,,\sqrt{f(\mathfrak{r}_0)}\,, 0\,,0\right),
\\
e_{\hat 2}^{~\mu}&=&
\left( {\mathfrak{r}_0 \mathfrak{w}_0\over  \sqrt{f(\mathfrak{r}_0)}}\,, 0\,, 0\,,{\sqrt{1+\mathfrak{r}_0^2\mathfrak{w}_0^2}
\over \mathfrak{r}_0}\right),
\\
e_{\hat 3}^{~\mu}&=& \left(0\,, 0\,, -{1\over \mathfrak{r}_0}\,,0\right)
\,,
\enann
where
the angular frequency $\mathfrak{w}_0$  is defined by
\bea
\mathfrak{w}_0=u^3={d\varphi\over d\tau}
={1\over  \mathfrak{r}_0(\mathfrak{r}_0/M-3)^{1/2}}
\,.
\label{omega_0}
\ena
We then calculate the angular velocity $\omega^{\hat j}$ of this rotating tetrad system as
\beann
\omega^{\hat j}={1\over 2}\stackrel{(3)}{\epsilon}^{\hat j\hat k\hat \ell}e_{\hat \ell \mu}{De_{\hat k}^{~\mu}\over d\tau}=\mathfrak{w}_{\rm R}  \delta^{\hat j}_{~\hat 3}\,,
\enann
where
\bea
\mathfrak{w}_{\rm R} ={1\over \mathfrak{r}_0(\mathfrak{r}_0/M)^{1/2}},
\label{omega_R}
\ena
is the angular frequency of the rotating tetrad frame 
(See Fig.~\ref{tetrad}).
Note that this angular frequency is different from 
the observer's angular frequency $\mathfrak{w}_0$.
The difference between two angular frequencies $\mathfrak{w}_0$ and $\mathfrak{w}_R$, i.e., 
\bea
\mathfrak{w}_{\rm dS}\equiv \mathfrak{w}_0-\mathfrak{w}_{\rm R} 
\label{omega_dS}
\ena
describes the rotation of non-rotating inertial frame moving along a circular orbit, which may cause de Sitter precession~\cite{desitter1916}.

Next, we calculate the Riemann curvature in the above tetrad system.
In the static tetrad system 
$\{e^{\bar \alpha}_{~\mu}\}$ with 
\beann
e^{\bar 0}_{~0}=\sqrt{f(\mathfrak{r}_0)}\,,~e^{\bar r}_{~r}={1\over \sqrt{f(\mathfrak{r}_0)}}\,,~
e^{\bar \theta}_{~\theta}=\mathfrak{r}_0\,,~e^{\bar \phi}_{~\phi}=\mathfrak{r}_0\,,
\enann
 the non-trivial components of  the Riemann curvature of 
 the Schwarzschild spacetime are given by
\beann
&&
\bar{\cal R}_{\bar 0\bar r\bar 0\bar r}=-{\mathfrak{r}_g\over\mathfrak{r}_0^3}
\,,~
\bar{\cal R}_{\bar 0\bar \theta\bar 0\bar \theta}=\bar{\cal R}_{\bar 0\bar \phi\bar 0\bar \phi}={\mathfrak{r}_g\over 2\mathfrak{r}_0^3},
\\
&&
\bar{\cal R}_{\bar \theta\bar \phi\bar \theta\bar \phi}={\mathfrak{r}_g\over \mathfrak{r}_0^3}
\,,~
\bar{\cal R}_{\bar r\bar \theta\bar r\bar \theta}=R_{\bar r\bar \phi\bar r\bar \phi}=-{\mathfrak{r}_g\over 2\mathfrak{r}_0^3}
\,.
\enann
We then introduce the Descartes coordinates $(\bar x, \bar y, \bar z)$ 
such that 
$\bar x$-direction is the same as the $r$-direction, but the $\bar y$ and $\bar z$-directions are 
$\phi$ and $-\theta$-directions, respectively, as shown in Fig. \ref{tetrad}.
Since the tetrad system in this coordinates is given by
\beann
e^{\bar 0}_{~0}=\sqrt{f(\mathfrak{r}_0)}\,,~e^{\bar x}_{~r}={1\over \sqrt{f(\mathfrak{r}_0)}}\,,~e^{\bar y}_{~\phi}=\mathfrak{r}_0 \,,~
e^{\bar z}_{~\theta}=-\mathfrak{r}_0\,,
\enann
we find the curvature components in this tetrad system as
\beann
&&
\bar{\cal R}_{\bar 0\bar x\bar 0\bar x}=-{\mathfrak{r}_g\over\mathfrak{r}_0^3}
\,,~
\bar{\cal R}_{\bar 0\bar y\bar 0\bar y}=\bar{\cal R}_{\bar 0\bar z\bar 0\bar z}={\mathfrak{r}_g\over 2\mathfrak{r}_0^3},
\\
&&
\bar{\cal R}_{\bar y\bar z\bar y\bar z}={\mathfrak{r}_g\over \mathfrak{r}_0^3}
\,,~
R_{\bar x\bar y\bar x\bar y}=\bar{\cal R}_{\bar z\bar x\bar z\bar x}=-{\mathfrak{r}_g\over 2\mathfrak{r}_0^3}
\,.
\enann

The transformation matrix between the observer's tetrad and static tetrad ($\bar x,\bar y, \bar z$) is given by
\beann
\Lambda_{\hat 0}^{~\bar \alpha}&=&\left(\sqrt{1+\mathfrak{r}_0^2\mathfrak{w}_0^2}\,, 0\,,\mathfrak{r}_0\mathfrak{w}_0 \,, 0\right),
\\
\Lambda_{\hat 1}^{~\bar \alpha}&=& \left(0\,, 1\,, 0\,, 0\right),
\\
\Lambda_{\hat 2}^{~\bar \alpha}&=& \left(\mathfrak{r}_0\mathfrak{w}_0\,, 0\,, \sqrt{1+\mathfrak{r}_0^2\mathfrak{w}_0^2}\,, 0\right),
\\
\Lambda_{\hat 3}^{~\bar \alpha}&=&\left(0\,, 0\,, 0\,, 1\right).
\enann
It is just the Lorentz boost with velocity 
\beann
\mathfrak{v}={\mathfrak{r}_0\mathfrak{w}_0\over \sqrt{1+\mathfrak{r}_0^2\mathfrak{w}_0^2}},
\enann
in $0\,\mathchar`-\,y$ plane.
\begin{widetext}
The components in the observer's proper reference frame are given by
\beann
&&
\bar{\cal R}_{\hat 0\hat x\hat 0\hat x}=-\bar{\cal R}_{\hat y\hat z\hat y\hat z}= -\left(1+{3\over 2}\mathfrak{r}_0^2\mathfrak{w}_0^2\right){\mathfrak{r}_g\over  \mathfrak{r}_0^3}
\,,~
\\
&&
\bar{\cal R}_{\hat 0\hat y\hat 0\hat y}=-\bar{\cal R}_{\hat z\hat x\hat z\hat x}={1\over 2}{\mathfrak{r}_g\over \mathfrak{r}_0^3}
\,,~
\\
&&
\bar{\cal R}_{\hat 0\hat z\hat 0\hat z}=-\bar{\cal R}_{\hat x\hat y\hat x\hat y}
=\left({1+3\mathfrak{r}_0^2\mathfrak{w}_0^2\over 2}\right){\mathfrak{r}_g\over  \mathfrak{r}_0^3},
\\
 &&
\bar{\cal R}_{\hat 0\hat x\hat y\hat x}=-\bar{\cal R}_{\hat 0\hat z\hat y\hat z}=-{3\over 2}\epsilon_1
\mathfrak{r}_0\mathfrak{w}_0\sqrt{1+\mathfrak{r}_0^2\mathfrak{w}_0^2}
\, {\mathfrak{r}_g\over \mathfrak{r}_0^3}
\,.
\enann

\subsection{Equations of motion of a binary}
Since the CM of a binary follows the observer's circular geodesic ($\vect{R}=0$), we have to solve only the equations of motion for the relative coordinate $\vect{r}$.
Using $x=r^{\hat 1}\,,y=r^{\hat 2}\,,z=r^{\hat 3}$,
the relative motion of a binary is
given by the Lagrangian 
\bea
{\cal L}_{\rm rel}(\vect{r},\dot{\vect{r}})&=&
{1\over 2}\mu \dot{\vect{r}}^2+  {G m_1m_2\over r}
+{\cal L}_{{\rm rel} \mathchar`-\mathfrak{w}}(\vect{r},\dot{\vect{r}})
+{\cal L}_{{\rm rel}\mathchar`- \bar{\cal R}}(\vect{r},\dot{\vect{r}}),
\label{EoM_relative_coordinate}
\ena
with
\beann
{\cal L}_{{\rm rel} \mathchar`-\mathfrak{w}}
&=&-\mu\left[\mathfrak{w}_0\left(x\dot{y}
-y\dot{x}\right)
-{\mathfrak{w}_0^2\over 2}\left(x^2+y^2\right)\right],
\\{\cal L}_{{\rm rel} \mathchar`-\bar{\cal R}}
&=&-{\mu\over 2} 
\left(
\bar{\cal R}_{\hat 0\hat x\hat 0\hat x}x^2+\bar{\cal R}_{\hat 0\hat y\hat 0\hat y}y^2+\bar{\cal R}_{\hat 0\hat z\hat 0\hat z}z^2\right)
\\
&=&-{\mu\mathfrak{r}_g\over 4\mathfrak{r}_0^3} 
\left[y^2-2x^2+z^2+3\mathfrak{r}_0^2\mathfrak{w}_0^2(-x^2+z^2)\right]
\,.
\enann

The first and second terms in ${\cal L}_{{\rm rel}\mathchar`-\mathfrak{w}}$ describe the Coriolis force and 
the centrifugal force, respectively.
The first half terms in ${\cal L}_{{\rm rel}\mathchar`- \bar{\cal R}}$ are the same 
as those in Newtonian hierarchical triple  system under quadrupole approximation, 
while the last half terms are relativistic corrections.
Note that in the present approach (approximation up to the second order of $x^{\hat a}$),
we cannot go beyond quadrupole approximation.

In order to analyze the relative motion of a binary,
it may be better to go to a non-rotating initial reference frame.
Since the angular frequency of a rotating tetrad frame is $\mathfrak{w}_{\rm R}$, 
the position $(x, y, z)$ in the rotating frame can be replaced by the position $(\mathsf{x},\mathsf{y},\mathsf{z})$  in a non-rotating Descartes' coordinate system by use of the following transformation; 
\beann
x&=&\mathsf{x}\cos\mathfrak{w}_{\rm R}   \tau -\mathsf{y} \sin \mathfrak{w}_{\rm R}  \tau,
\\
y&=&
\mathsf{x} \sin\mathfrak{w}_{\rm R}   \tau +\mathsf{y} \cos \mathfrak{w}_{\rm R}    \tau,
 \\
z&=&\mathsf{z}
\,.
\enann

The Lagrangian ${\cal L}_{\rm rel}$ in a non-rotating proper reference frame 
is given by
\bea
{\cal L}_{\rm rel}={1\over 2}\mu \left({ d {\vect{\mathsf{r} }}\over d\tau}\right)^2+  {G m_1m_2\over \mathsf{r}}
+{\cal L}_{\rm rel\mathchar`-dS}(\vect{\mathsf{r}},\dot{\vect{\mathsf{r}}})
+{\cal L}_{{\rm rel}\mathchar`-\bar{\cal R}}(\vect{\mathsf{r}},\tau)
\,,
\label{EoM_relative_non-rotating}
\ena
where
\beann
{\cal L}_{\rm rel\mathchar`-dS}(\vect{\mathsf{r}},\dot{\vect{\mathsf{r}}})&=&\mu \mathfrak{w}_{\rm dS}\left(\dot{\mathsf{x}}\mathsf{y}
-\dot{\mathsf{y}}\mathsf{x}\right)+{\mu\over 2} \mathfrak{w}_{\rm dS}^2\left(\mathsf{x}^2+\mathsf{y}^2\right),
\\
{\cal L}_{{\rm rel}\mathchar`-\bar{\cal R}}(\vect{\mathsf{r}},\tau)
&=&-{\mu\mathfrak{r}_g\over 4\mathfrak{r}_0^3} 
\left[\mathsf{x}^2+\mathsf{y}^2+\mathsf{z}^2-3\left(1+\mathfrak{r}_0^2\mathfrak{w}_0^2\right)\left(\mathsf{x}\cos\mathfrak{w}_{\rm R}   \tau -\mathsf{y} \sin \mathfrak{w}_{\rm R}  \tau\right)^2
+3\mathfrak{r}_0^2\mathfrak{w}_0^2 \mathsf{z}^2 \right].
\enann
Since the momentum is defined by
\beann
\mathsf{p}_{\mathsf{x}}&\equiv &{\partial {\cal L}\over \partial \dot{\mathsf{x}}}=\mu \dot{\mathsf{x}}+\mu\mathfrak{w}_{\rm dS} \mathsf{y},
\\
\mathsf{p}_{\mathsf{y}}&\equiv &{\partial {\cal L}\over \partial \dot{\mathsf{y}}}=\mu \dot{\mathsf{y}}-\mu\mathfrak{w}_{\rm dS} \mathsf{x},
\\
\mathsf{p}_{\mathsf{z}}&\equiv &{\partial {\cal L}\over \partial \dot{\mathsf{z}}}=\mu \dot{\mathsf{z}}
\,,
\enann
we obtain the Hamiltonian as
\bea
{\cal H}_{\rm rel}={\cal H}_0+{\cal H}_1
\label{Hamiltonian_non-rotating}
\,,
\ena
where
\beann
{\cal H}_0
&=&{1\over 2\mu} {\vect{\mathsf{p}}}^2- {G m_1m_2\over \mathsf{r}},
\\
{\cal H}_1
&=&{\cal H}_{1\mathchar`-{\rm dS}}+{\cal H}_{1\mathchar`-\bar{\cal R}},
\enann
with
\beann
{\cal H}_{1\mathchar`-{\rm dS}}&=&
\mathfrak{w}_{\rm dS}\left(\mathsf{p}_{\mathsf{y}}\mathsf{x}-\mathsf{p}_{\mathsf{x}}\mathsf{y}\right),
\\
{\cal H}_{1\mathchar`-\bar{\cal R}}&=&{\mu\mathfrak{r}_g\over 4\mathfrak{r}_0^3} 
\left[\mathsf{x}^2+\mathsf{y}^2+\mathsf{z}^2-3\left(1+\mathfrak{r}_0^2\mathfrak{w}_0^2\right)\left(\mathsf{x}\cos\mathfrak{w}_{\rm R}   \tau -\mathsf{y} \sin \mathfrak{w}_{\rm R}  \tau\right)^2
+3\mathfrak{r}_0^2\mathfrak{w}_0^2 \mathsf{z}^2 \right].
\enann

The  term ${\cal H}_{1\mathchar`-{\rm dS}}$ gives the so-called 
de Sitter precession as follows:
Let us consider the model with ${\cal H}'={\cal H}_0+{\cal H}_{1\mathchar`-{\rm dS}}$.
The Hamilton equations are given as
\beann
\dot{\mathsf{x}}&=&{\partial{\cal H}'\over \partial\mathsf{p}_{\mathsf{x}}}
={\mathsf{p}_{\mathsf{x}}\over \mu}-\mathfrak{w}_{\rm dS}\mathsf{y},
\\
\dot{\mathsf{y}}&=&{\partial{\cal H}'\over \partial\mathsf{p}_{\mathsf{y}}}
={\mathsf{p}_{\mathsf{y}}\over \mu}+\mathfrak{w}_{\rm dS}\mathsf{x},
\\
\dot{\mathsf{z}}&=&{\partial{\cal H}'\over \partial\mathsf{p}_{\mathsf{z}}}
={\mathsf{p}_z\over \mu},
\enann
and 
\beann
\dot{\mathsf{p}}_{\mathsf{x}}&=&-{\partial{\cal H}'\over \partial\mathsf{x}}=-{Gm_1m_2\mathsf{x}\over \mathsf{r}^3}-\mathfrak{w}_{\rm dS}\mathsf{p}_{\mathsf{y}},
\\
\dot{\mathsf{p}}_{\mathsf{y}}&=&-{\partial{\cal H}'\over \partial\mathsf{y}}=-{Gm_1m_2\mathsf{y}\over \mathsf{r}^3}+\mathfrak{w}_{\rm dS}\mathsf{p}_{\mathsf{x}},
\\
\dot{\mathsf{p}}_{\mathsf{z}}&=&-{\partial{\cal H}'\over \partial\mathsf{z}}=-{Gm_1m_2\mathsf{z}\over \mathsf{r}^3}.
\enann
We then calculate the time evolution of the angular momentum $\vect{L}=(L_{\mathsf{x}}, L_{\mathsf{y}}, L_{\mathsf{z}})$.
Using the equations of motion, we find 
\beann
\dot{L}_{\mathsf{x}}&=&{d\over d\tau }(\mathsf{y}\mathsf{p}_{\mathsf{z}}-\mathsf{z}\mathsf{p}_{\mathsf{y}})=\dot{\mathsf{y}}\mathsf{p}_{\mathsf{z}}+\mathsf{y}\dot{\mathsf{p}}_{\mathsf{z}}-\dot{\mathsf{z}}\mathsf{p}_{\mathsf{y}}-\mathsf{z}\dot{\mathsf{p}}_{\mathsf{y}}=-\mathfrak{w}_{\rm dS}L_{\mathsf{y}},
\\
\dot{L}_{\mathsf{y}}&=&{d\over d\tau }(\mathsf{z}\mathsf{p}_{\mathsf{x}}-\mathsf{x}\mathsf{p}_{\mathsf{z}})=\dot{\mathsf{z}}\mathsf{p}_{\mathsf{x}}+\mathsf{z}\dot{\mathsf{p}}_{\mathsf{x}}-\dot{\mathsf{x}}\mathsf{p}_{\mathsf{z}}-\mathsf{x}\dot{\mathsf{p}}_{\mathsf{z}}=\mathfrak{w}_{\rm dS}L_{\mathsf{x}},
\\
\dot{L}_{\mathsf{z}}&=&{d\over d\tau }(\mathsf{x}\mathsf{p}_{\mathsf{y}}-\mathsf{y}\mathsf{p}_{\mathsf{x}})=\dot{\mathsf{x}}\mathsf{p}_{\mathsf{y}}+\mathsf{x}\dot{\mathsf{p}}_{\mathsf{y}}-\dot{\mathsf{y}}\mathsf{p}_{\mathsf{x}}-\mathsf{y}\dot{\mathsf{p}}_{\mathsf{x}}=0.
\enann
From these equations, we find that the $\mathsf{z}$-component of the angular momentum is conserved, and $(L_{\mathsf{x}}, L_{\mathsf{y}})$ rotates around the $\mathsf{z}$-axis with the constant angular frequency
$\mathfrak{w}_{\rm dS}$ which is just the de Sitter precession.
The full equations are given as
\bea
\dot{\mathsf{x}}&=&{\partial{\cal H}\over \partial\mathsf{p}_{\mathsf{x}}}
={\mathsf{p}_{\mathsf{x}}\over \mu}-\mathfrak{w}_{\rm dS}\mathsf{y},
\label{px}
\\
\dot{\mathsf{y}}&=&{\partial{\cal H}\over \partial\mathsf{p}_{\mathsf{y}}}
={\mathsf{p}_{\mathsf{y}}\over \mu}+\mathfrak{w}_{\rm dS}\mathsf{x},
\label{py}
\\
\dot{\mathsf{z}}&=&{\partial{\cal H}\over \partial\mathsf{p}_{\mathsf{z}}}
={\mathsf{p}_{\mathsf{z}}\over \mu},
\label{pz}
\ena
and 
\bea
\dot{\mathsf{p}}_{\mathsf{x}}&=&-{\partial{\cal H}\over \partial\mathsf{x}}=-{Gm_1m_2\over \mathsf{r}^3}\mathsf{x}-\mathfrak{w}_{\rm dS}\mathsf{p}_{\mathsf{y}}
-{\mu\mathfrak{r}_g\over 2\mathfrak{r}_0^3}
\left\{\mathsf{x}-3(1+\mathfrak{r}_0^2\mathfrak{w}_0^2)(\mathsf{x}\cos\mathfrak{w}_{\rm R}   \tau -\mathsf{y} \sin \mathfrak{w}_{\rm R}  \tau)\cos\mathfrak{w}_{\rm R}   \tau\right\},
\label{eq_x}
\\
\dot{\mathsf{p}}_{\mathsf{y}}&=&-{\partial{\cal H}\over \partial\mathsf{y}}=-{Gm_1m_2\over \mathsf{r}^3}\mathsf{y}+\mathfrak{w}_{\rm dS}\mathsf{p}_{\mathsf{x}}
-{\mu\mathfrak{r}_g\over 2\mathfrak{r}_0^3}
\left\{\mathsf{y}+3(1+\mathfrak{r}_0^2\mathfrak{w}_0^2)(\mathsf{x}\cos\mathfrak{w}_{\rm R}   \tau -\mathsf{y} \sin \mathfrak{w}_{\rm R}  \tau)\sin\mathfrak{w}_{\rm R}   \tau\right\},
\label{eq_y}
\\
\dot{\mathsf{p}}_{\mathsf{z}}&=&-{\partial{\cal H}\over \partial\mathsf{z}}=-{Gm_1m_2\over \mathsf{r}^3}\mathsf{z}
-{\mu\mathfrak{r}_g\over 2\mathfrak{r}_0^3}
(1+3\mathfrak{r}_0^2\mathfrak{w}_0^2)\mathsf{z}.
\label{eq_z}
\ena

\subsection{Motion with 0.5 PN correction term}

Now we consider 0.5 PN terms.
As discussed in \S. \ref{binary}, we can assume  $\vect{R}=0$  by introduction of the acceleration given by Eq. (\ref{0.5PN_acceleration}).
We first solve the relative coordinates $\vect{r}$, and then 
the motion of the observer (or the CM).

\subsubsection{\bf Equations of motion for relative coordinates}
The equation of motion for relative coordinates $\vect{r}$ of a binary is now
given by 
\beann
\tilde {\cal L}_{\rm rel}(\vect{r}, \dot{\vect{r}})&=&{\cal L}_{\rm rel}(\vect{r}, \dot{\vect{r}})+{\cal L}_{1/2\mathchar`-{\rm rel}}(\vect{r}, \dot{\vect{r}})
\,,
\enann
where 
${\cal L}_{\rm rel}$  is given by Eq. (\ref{EoM_relative_coordinate}), while 
\beann
{\cal L}_{1/2\mathchar`-{\rm rel}}(\vect{r}, \dot{\vect{r}})&=&-\mu {2(m_1-m_2)\over 3(m_1+m_2)}\left(
\bar{\cal R}_{\hat 0\hat x\hat y\hat x}x(x\dot y-y\dot x)
+\bar{\cal R}_{\hat 0\hat z\hat y\hat z}z(z\dot y-y\dot z)\right)
\\
&=&-\mu {\mathfrak{r}_g\over \mathfrak{r}_0^3}{m_1-m_2\over m_1+m_2}
\mathfrak{r}_0\mathfrak{w}_0\sqrt{1+\mathfrak{r}_0^2\mathfrak{w}_0^2}
\left(
-x(x\dot y-y\dot x)
+z(z\dot y-y\dot z)\right)
\,.
\enann

 In non-rotating Fermi-Walker coordinates, 
 we find
 ${\cal L}_{\rm rel}$ is given by Eq. (\ref{EoM_relative_non-rotating}),
 while 
 \beann
{\cal L}_{1/2\mathchar`-{\rm rel}}(\vect{\mathsf{r}}, \dot{\vect{\mathsf{r}}})&=&
\mu {\mathfrak{r}_g\over \mathfrak{r}_0^3}{m_1-m_2\over m_1+m_2}
\mathfrak{r}_0\mathfrak{w}_0\sqrt{1+\mathfrak{r}_0^2\mathfrak{w}_0^2}
\Big{\{}\cos \mathfrak{w}_{\rm R}   \tau \left[\mathsf{x}\left(\mathsf{x}\dot{\mathsf{y}}-\mathsf{y}\dot{\mathsf{x}}\right)+\mathsf{z}\left(\mathsf{y}\dot{\mathsf{z}}-\mathsf{z}\dot{\mathsf{y}})+\mathfrak{w}_{\rm R}\mathsf{x}(\mathsf{x}^2+\mathsf{y}^2 -\mathsf{z}^2\right)\right]
\\
&&
~~~~~~~~~~~~~~~~~~~~~~~~~~~~~~~~~~~~~~~~-\sin \mathfrak{w}_{\rm R}   \tau\left[\mathsf{y}\left(\mathsf{x}\dot{\mathsf{y}}-\mathsf{y}\dot{\mathsf{x}}\right)+\mathsf{z}\left(\mathsf{z}\dot{\mathsf{x}}-\mathsf{x}\dot{\mathsf{z}})+\mathfrak{w}_{\rm R}\mathsf{y}(\mathsf{x}^2+\mathsf{y}^2 -\mathsf{z}^2\right)\right]\Big{\}}.
\enann

The momentum is obtained from the Lagrangian $\tilde {\cal L}_{\rm rel}(\vect{\mathsf{r}}, \dot{\vect{\mathsf{r}}})$ as 
\beann
p_{\mathsf{x}}&=&\mu \dot{\mathsf{x}} +\mu \mathfrak{w}_{\rm dS}\mathsf{y}
+\mu {\mathfrak{r}_g\over \mathfrak{r}_0^3}{m_1-m_2\over m_1+m_2}
\mathfrak{r}_0\mathfrak{w}_0\sqrt{1+\mathfrak{r}_0^2\mathfrak{w}_0^2}
\left(-\mathsf{x}\mathsf{y}\cos \mathfrak{w}_{\rm R} \tau+(\mathsf{y}^2-\mathsf{z}^2) \sin \mathfrak{w}_{\rm R}\tau\right),
\\
p_{\mathsf{y}}&=&\mu \dot{\mathsf{y}} -\mu \mathfrak{w}_{\rm dS}\mathsf{x}
+\mu {\mathfrak{r}_g\over \mathfrak{r}_0^3}{m_1-m_2\over m_1+m_2}
\mathfrak{r}_0\mathfrak{w}_0\sqrt{1+\mathfrak{r}_0^2\mathfrak{w}_0^2}
\left(-\mathsf{x}\mathsf{y}\sin\mathfrak{w}_{\rm R} \tau+(\mathsf{x}^2-\mathsf{z}^2) \cos \mathfrak{w}_{\rm R}\tau\right),
\\
p_{\mathsf{z}}&=&\mu \dot{\mathsf{z}} 
+\mu {\mathfrak{r}_g\over \mathfrak{r}_0^3}{m_1-m_2\over m_1+m_2}
\mathfrak{r}_0\mathfrak{w}_0\sqrt{1+\mathfrak{r}_0^2\mathfrak{w}_0^2}
\mathsf{z}\left(\mathsf{y}\cos \mathfrak{w}_{\rm R} \tau+\mathsf{x} \sin \mathfrak{w}_{\rm R}\tau\right).
\enann
The Hamiltonian is given by
\beann
\tilde {\cal H}_{\rm rel}(\vect{\mathsf{r}}, \vect{\mathsf{p}})&=&{\cal H}_{\rm rel}(\vect{\mathsf{r}}, \vect{\mathsf{p}})+{\cal H}_{1/2\mathchar`-{\rm rel}}(\vect{\mathsf{r}}, \vect{\mathsf{p}})
\,,
\enann
where 
${\cal H}_{\rm rel}(\vect{\mathsf{r}}, \vect{\mathsf{p}})$  is given by Eq. (\ref{Hamiltonian_non-rotating}), while 
\beann
{\cal H}_{1/2\mathchar`-{\rm rel}}(\vect{\mathsf{r}}, \vect{\mathsf{p}})&=&
-{1\over 2}\mu \left({\mathfrak{r}_g\over \mathfrak{r}_0^3}\right)^2{(m_1-m_2)^2\over (m_1+m_2)^2}
\mathfrak{r}_0^2\mathfrak{w}_0^2(1+\mathfrak{r}_0^2\mathfrak{w}_0^2)\Big[\left(-\mathsf{x}\mathsf{y}\cos \mathfrak{w}_{\rm R} \tau+(\mathsf{y}^2-\mathsf{z}^2) \sin \mathfrak{w}_{\rm R}\tau\right)^2
\\
&&
~~~~~~~~~~~~
+\left(-\mathsf{x}\mathsf{y}\sin\mathfrak{w}_{\rm R} \tau+(\mathsf{x}^2-\mathsf{z}^2) \cos \mathfrak{w}_{\rm R}\tau\right)^2+\mathsf{z}^2\left(\mathsf{y}\cos \mathfrak{w}_{\rm R} \tau+\mathsf{x} \sin \mathfrak{w}_{\rm R}\tau\right)^2\Big]
\,.
\enann
This Hamiltonian is very complicated, but it should be ignored because it is beyond quadrupole approximation, although momentum is modified. For an equal mass binary ($m_1=m_2$), the 0.5PN correction term vanishes and the momentum is also the same as the Newtonian one. As a result, the Newtonian solution is also a solution.

\subsubsection{\bf Motion of the CM of a binary and its stability}

In order to study stability of the CM of  a binary system, we analyze Eq. (\ref{eq_CM}).
Since $\vect{R}$ is measured by the circular observer at $\mathfrak{r}=\mathfrak{r}_0$, we can split the 4-velocity $u^\mu$ 
as
\beann
u^\mu&=&u^\mu_{(0)}+u^\mu_{(1)},
\enann
where
\beann
u^\mu_{(0)}&=&{d\mathfrak{x}^\mu_{(0)}\over d\tau}=(u^0_{(0)}\,,0\,,0\,,u^3_{(0)})=\left(\sqrt{1+\mathfrak{r}_0^2\mathfrak{w}_0^2\over f(\mathfrak{r}_0)}\,,0\,,0\,,\mathfrak{w}_0\right),
\\
u^\mu_{(1)}&=&{d\mathfrak{x}^\mu_{(1)}\over d\tau},
\enann
with 
\beann
\mathfrak{x}^\mu_{(0)}&=&\left(\sqrt{1+\mathfrak{r}_0^2\mathfrak{w}_0^2\over f(\mathfrak{r}_0)}\tau\,,\mathfrak{r}_0\,,{\pi\over 2}\,,\mathfrak{w}_0\tau\right),
\\
\mathfrak{x}^\mu_{(1)}&\equiv& e^{\mu}_{~\hat \ell}R^{\hat \ell}.
\enann
The acceleration $a^\mu$ is given by the motion of a binary $x^{\hat \mu}(\tau)$ 
in a rotating frame as
\beann
a^\mu=-{3\mu\over m_1+m_2}\mathfrak{w}_0\sqrt{1+\mathfrak{r}_0^2\mathfrak{w}_0^2}{\mathfrak{r}_g\over \mathfrak{r}_0^2}
\left[\left(\delta^\mu_0{\mathfrak{r}_0\mathfrak{w}_0\over \sqrt{f(\mathfrak{r}_0)}}+\delta^\mu_3{\sqrt{1+\mathfrak{r}_0^2\mathfrak{w}_0^2}\over \mathfrak{r}_0}\right)(\dot x x-\dot z z)\,
-\delta^\mu_1 \sqrt{f(\mathfrak{r}_0)}\dot y x\,-\delta^\mu_2{1\over \mathfrak{r}_0}\dot y z \right].
\enann

Here we assume that the deviation from a circular orbit is small, i.e.,  $\mathfrak{x}^\mu_{(1)}$ and $u^\mu_{(1)}$ are small perturbations.
Ignoring non-linear deviation terms in the equations of motion 
$\displaystyle{{Du^\mu\over d\tau}=a^\mu}$, because the circular orbit $\mathfrak{x}^\mu_{(0)}(\tau)$ is a geodesic,
we obtain
 a linear differential equation as
\beann
{du^\mu_{(1)}\over d\tau}
+2\Gamma^{\mu}_{~\rho\sigma}(\mathfrak{r}_0)u^\rho_{(0)} u^\sigma_{(1)}+{\partial \Gamma^{\mu}_{~\rho\sigma}
\over \partial \mathfrak{x}^\alpha}(\mathfrak{r}_0)\, \mathfrak{x}_{(1)}^\alpha 
\, u^\rho_{(0)} u^\sigma_{(0)}
=a^\mu,
\enann
where $a^\mu$ acts as an external force.
Describing the deviation as 
\beann
\mathfrak{x}^\mu_{(1)}=(\mathfrak{t}_{(1)},\mathfrak{r}_{(1)},\mathfrak{\theta}_{(1)},\mathfrak{\varphi}_{(1)})
\,,
\enann
we find
\bea
&&
{d^2\mathfrak{t}_{(1)}\over d\tau^2}+{\mathfrak{r}_g\over \mathfrak{r}_0^2f(\mathfrak{r}_0)}\sqrt{1+\mathfrak{r}_0^2\mathfrak{w}^2\over f(\mathfrak{r}_0)}{d\mathfrak{r}_{(1)}\over d\tau}=a^0=-{3\mu\over m_1+m_2}
{\mathfrak{w}_0^2 \mathfrak{r}_g\over \mathfrak{r}_0}{\sqrt{1+\mathfrak{r}_0^2\mathfrak{w}_0^2 \over  f(\mathfrak{r}_0)}}
(\dot x x-\dot z z),
\label{eq_t1}
\\
&&
{d^2\mathfrak{r}_{(1)}\over d\tau^2}
-{3\mathfrak{r}_g\over 2\mathfrak{r}_0^3} \left(1+\mathfrak{r}_0^2\mathfrak{w}_0^2\right)\mathfrak{r}_{(1)}
+ {\mathfrak{r}_g\over \mathfrak{r}_0^2}f(\mathfrak{r}_0)
\sqrt{1+\mathfrak{r}_0^2\mathfrak{w}^2\over f(\mathfrak{r}_0)}
{d\mathfrak{t}_{(1)}\over d\tau}-2\mathfrak{r}_0\mathfrak{w}_0 f(\mathfrak{r}_0){d\mathfrak{\varphi}_{(1)}\over d\tau}
\nonumber \\
&&~~~~~~~~~~
=a^1={3\mu\over m_1+m_2}{\mathfrak{w}_0\mathfrak{r}_g\over \mathfrak{r}_0^2}\sqrt{f(\mathfrak{r}_0)(1+\mathfrak{r}_0^2\mathfrak{w}_0^2)}
\dot y x,
~~~~~~
\label{eq_r1}
\\
&&
{d^2\mathfrak{\theta}_{(1)}\over d\tau^2}+\mathfrak{w}_0^2\mathfrak{\theta}_{(1)}=a^2={3\mu\over m_1+m_2}{\mathfrak{w}_0 \mathfrak{r}_g\over \mathfrak{r}_0^3}
\sqrt{1+\mathfrak{r}_0^2\mathfrak{w}_0^2}\,\dot y z,
\label{eq_th1}
\\
&&
{d^2\mathfrak{\varphi}_{(1)}\over d\tau^2}+2{\mathfrak{w}_0\over \mathfrak{r}_0}{d\mathfrak{r}_{(1)}\over d\tau}=a^3=-{3\mu\over m_1+m_2}{\mathfrak{w}_0 \mathfrak{r}_g\over \mathfrak{r}_0^3}
\left(1+\mathfrak{r}_0^2\mathfrak{w}_0^2\right)(\dot x x-\dot z z),
\label{eq_ph1}
\ena

Integrating Eqs. (\ref{eq_t1}) and (\ref{eq_ph1}), we obtain 
\bea
&&{d\mathfrak{t}_{(1)}\over d\tau}
=-{\mathfrak{r}_g\over \mathfrak{r}_0^2f(\mathfrak{r}_0)}\sqrt{1+\mathfrak{r}_0^2\mathfrak{w}_0^2\over f(\mathfrak{r}_0)}
\,\mathfrak{r}_{(1)}
-{3\mu\over 2(m_1+m_2)}
{\mathfrak{w}_0^2 \mathfrak{r}_g\over \mathfrak{r}_0}{\sqrt{1+\mathfrak{r}_0^2\mathfrak{w}_0^2 \over  f(\mathfrak{r}_0)}}
\left(x^2-z^2\right),
\label{eq_t2}
\\
&&
{d\mathfrak{\varphi}_{(1)}\over d\tau}
=-2{\mathfrak{w}_0\over \mathfrak{r}_0}\, \mathfrak{r}_{(1)}
-{3\mu\over 2(m_1+m_2)}{\mathfrak{w}_0 \mathfrak{r}_g\over \mathfrak{r}_0^3}
\left(1+\mathfrak{r}_0^2\mathfrak{w}_0^2\right)\left(x^2-z^2\right)
\,.
\label{eq_ph2}
\ena
where we set the integration constants are zero.
Plugging Eqs. (\ref{eq_t2}) and (\ref{eq_ph2}) into Eq. (\ref{eq_r1})
 with  Eq. (\ref{omega_0}), we obtain the perturbation equation for the radial coordinates 
$\mathfrak{r}_{(1)}$ as
\bea
{d^2\mathfrak{r}_{(1)}\over d\tau^2}
+ k^2\mathfrak{r}_{(1)}+A\left(x^2-z^2\right)+B\dot y x=0
\,,
\label{eq:radial_perturbation}
\ena
where
\bea
k^2&=&{\mathfrak{r}_g(\mathfrak{r}_0-3\mathfrak{r}_g)\over 
\mathfrak{r}_0^3(2\mathfrak{r}_0-3\mathfrak{r}_g)},
\label{coeff_k2}
\\
A&=&{3\mu\over m_1+m_2}{\mathfrak{r}_g^2(\mathfrak{r}_0-\mathfrak{r}_g)\over \mathfrak{r}_0^5 (2\mathfrak{r}_0-3\mathfrak{r}_g)}
\label{coeff_A}
\\
B&=&-{3\mu\over m_1+m_2}\sqrt{2\mathfrak{r}_g^3\over \mathfrak{r}_0^7}
{(\mathfrak{r}_0-\mathfrak{r}_g)\over (2\mathfrak{r}_0-3\mathfrak{r}_g)}.
\label{coeff_B}
\ena
We find that $k^2>0$ when $ \mathfrak{r}_0>3 \mathfrak{r}_g$, while 
$A>0$ and $B<0$ when 
$ \mathfrak{r}_0>3 \mathfrak{r}_g/2$.
The condition for $k^2>0$ is consistent with the fact that 
the radius of innermost stable circular orbit (ISCO) is $3 \mathfrak{r}_g$.

If the binary orbit is bounded ($x(\tau), y(\tau)$ and $z(\tau)$ are finite), 
the orbit of the center of mass is also bounded  
because $k^2>0$.
We expect that when $ \mathfrak{r}_0>3 \mathfrak{r}_g$ (ISCO radius),
a binary system near SMBH is linearly stable 
unless a binary is broken.

\end{widetext}
 
\section{Numerical Analysis}
\label{numerical_analysis}

\subsection{Validity and Stability}
\label{validity}
Before showing our numerical results, 
we discuss validity of the present approach.
The minimum curvature radius at the radius $\mathfrak{r}_0$ is evaluated as 
\beann
 \ell_{\bar{\cal R}} &=& {\rm min} \left[|\bar{\cal R}_{\hat \mu\hat \nu\hat \rho\hat \sigma}|^{-{1\over 2}}, |\bar{\cal R}_{\hat \mu\hat \nu\hat \rho\hat \sigma ; \hat \alpha}|^{-{1\over 3}}, |\bar{\cal R}_{\hat \mu\hat \nu\hat \rho\hat \sigma ; \hat \alpha;\hat \beta}|^{-{1\over 4}}
\right]
\\
&\sim&
 {\rm min} \left[\left({\mathfrak{r}_g\over \mathfrak{r}_0^3}\right)^{-{1\over 2}}\,,\left({\mathfrak{r}_g\over \mathfrak{r}_0^4}\right)^{-{1\over 3}}\,,
\left( {\mathfrak{r}_g\over \mathfrak{r}_0^5}\right)^{-{1\over 4}}
\right]
\\
&\sim &\mathfrak{r}_0 \left( \mathfrak{r}_0 \over {\mathfrak{r}_g}\right)^{1/4}
\geq 3\sqrt[4]{3}\mathfrak{r}_g\sim 8{\rm AU}\left({M\over 10^8 M_\odot}\right)
\,.
\enann
The equality is held at the ISCO radius $\mathfrak{r}_0 =3 \mathfrak{r}_g$.

When we put a binary at $\mathfrak{r}=\mathfrak{r}_0$,
 the  binary size $\ell_{\rm binary}$ should satisfy
\beann
\ell_{\rm binary}\ll  \ell_{\bar{\cal R}} 
\enann

As for stability of a binary, the mutual gravitational interaction between a binary 
should be much larger than the tidal force by a third body.
The condition is given by
\beann
{Gm_1m_2\over r^2}\gg {\mu \mathfrak{r}_g  \over \mathfrak{r}_0^3}\,r
\,,
\enann
which gives the constraint on a binary size $\ell_{\rm binary}$ as
\bea
\ell_{\rm binary} 
&\ll &
\left({m_1+m_2\over 2M}\right)^{1\over 3}\,\mathfrak{r}_0
\label{stability_tidal}
\\
&&\hskip -1.5cm
\approx 
4.64\times 10^{-3} \left({(m_1+m_2)\over20 M_\odot }\right)^{1\over 3}\left({M\over 10^8 M_\odot}\right)^{-{1\over 3}}\,\mathfrak{r}_0.
\nonumber
\ena

When we are interested in the orbit near the ISCO radius $\mathfrak{r}_0 =3 \mathfrak{r}_g$, we find 
\beann
\ell_{\rm binary} 
\ll 
3\times 10^{-2} {\rm AU} \left({(m_1+m_2)\over20 M_\odot }\right)^{1\over 3}\left({M\over 10^8 M_\odot}\right)^{{2\over 3}}.
\enann

As for the relativistic effect in a binary becomes important when
\beann
\ell_{\rm binary} &\leq& {G(m_1+m_2)\over c^2}
\\
 &\approx&
  2\times 10^{-7}{\rm AU}\left({(m_1+m_2)\over20 M_\odot }\right)
\,.
\enann

Hence,
for a binary with the size of
\beann
 &&
 2\times 10^{-7}{\rm AU}\left({(m_1+m_2)\over20 M_\odot }\right) \ll \ell_{\rm binary} 
 \\
 &&~~~~
 \ll 3 \times 10^{-2} {\rm AU}\left({(m_1+m_2)\over20 M_\odot }\right)^{1\over 3}\left({M\over 10^8 M_\odot}\right)^{{2\over 3}}\,,
 \enann
 we may apply the present Newtonian approach to the ISCO radius.

\subsection{Normalization and Initial Data}
We show some numerical examples for an equal-mass binary ($m_1=m_2$).
Hence, we have to solve Eqs. (\ref{px}) -(\ref{eq_z}).
In order to solve these basic equations, we shall introduce dimensionless variables as follows: The length scale of a binary is normalized by an initial semi-major axis $a_0$, while 
time scale is normalized by an initial binary mean motion $n_0$, which is defined by
\beann
n_0\equiv \bigg({G(m_1+m_2)\over a_0^3}\bigg)^{1/2}
\,.
\enann
Note that the initial binary period is given by $P_{\rm in}=2\pi/n_0$.
Introducing 
\beann
&&
\tilde \tau=n_0\tau,
\\
&&
\tilde{\mathsf{x}}={\mathsf{x}\over a_0}\,,~
\tilde{\mathsf{y}}={\mathsf{y}\over a_0}\,,~
\tilde{\mathsf{z}}={\mathsf{z}\over a_0}\,,~
\tilde{\mathsf{r}}={\mathsf{r}\over a_0},
\\
&&
\tilde{p}_\mathsf{x}={p_\mathsf{x}\over \mu a_0 n_0}\,,~
\tilde{p}_\mathsf{y}={p_\mathsf{y}\over \mu a_0 n_0}\,,~
\tilde{p}_\mathsf{z}={p_\mathsf{z}\over \mu a_0 n_0 }
\,,
\enann
we find the dimensionless equations of motion as
\bea
{d\tilde{\mathsf{x}}\over d\tilde \tau}&=&\tilde{p}_\mathsf{x}-\tilde{\mathfrak{w}}_{\rm dS}\tilde{\mathsf{y}},
\label{npx}
\\
{d\tilde{\mathsf{y}}\over d\tilde \tau}&=&\tilde{p}_\mathsf{y}+\tilde{\mathfrak{w}}_{\rm dS}\tilde{\mathsf{x}},
\label{npy}
\\
{d\tilde{\mathsf{z}}\over d\tilde \tau}&=&\tilde{p}_\mathsf{z},
\label{npz}
\ena
\begin{widetext}
and 
\bea
{d\tilde{\mathsf{p}}_{\mathsf{x}}\over d\tilde \tau}&=&
-{\tilde{\mathsf{x}}\over \tilde{\mathsf{r}}^3}-\tilde{\mathfrak{w}}_{\rm dS}\tilde{\mathsf{p}}_{\mathsf{y}}
-{\mathfrak{r}_g\over 2\mathfrak{r}_0}\epsilon^2
\left\{\tilde{\mathsf{x}}-3\left(1+\mathfrak{r}_0^2\mathfrak{w}_0^2\right)(\tilde{\mathsf{x}}\cos\tilde{\mathfrak{w}}_{\rm R}\tilde{\tau} -\tilde{\mathsf{y}} \sin \tilde{\mathfrak{w}}_{\rm R}\tilde{\tau})\cos \tilde{\mathfrak{w}}_{\rm R}\tilde{\tau}\right\}
\label{neq_x},
\\
{d\tilde{\mathsf{p}}_{\mathsf{y}}\over d\tilde \tau}&=&
-{\tilde{\mathsf{y}}\over \tilde{\mathsf{r}}^3}
+\tilde{\mathfrak{w}}_{\rm dS}\tilde{\mathsf{p}}_{\mathsf{x}}
-{\mathfrak{r}_g\over 2\mathfrak{r}_0}\epsilon^2
\left\{\tilde{\mathsf{y}}+3\left(1+\mathfrak{r}_0^2\mathfrak{w}_0^2\right)(\tilde{\mathsf{x}}\cos\tilde{\mathfrak{w}}_{\rm R} \tilde{\tau} -\tilde{\mathsf{y}} \sin \tilde{\mathfrak{w}}_{\rm R} \tilde{\tau})\sin\tilde{\mathfrak{w}}_{\rm R} \tilde{\tau}\right\},
\label{neq_y}
\\
{d\tilde{\mathsf{p}}_{\mathsf{z}}\over d\tilde \tau}&=&
-{\tilde{\mathsf{z}}\over \tilde{\mathsf{r}}^3}
-{\mathfrak{r}_g\over 2\mathfrak{r}_0}\epsilon^2
\left(1+3\mathfrak{r}_0^2\mathfrak{w}_0^2\right)\tilde{\mathsf{z}}\,,
\label{neq_z}
\ena
\end{widetext}
where
\beann
\tilde{\mathfrak{w}}_{\rm dS} &\equiv &{\mathfrak{w}_{\rm dS}\over n_0}
\,,~~
\tilde{\mathfrak{w}}_{\rm R} \,\equiv
\, {\mathfrak{w}_{\rm R}\over n_0},
\\
\mathfrak{r}_0^2\mathfrak{w}_0^2&=&\left({2\mathfrak{r}_0\over \mathfrak{r}_g}-3\right)^{-1},
\enann
and
\bea
\epsilon^2\equiv {1\over \mathfrak{r}_0^2 n_0^2}
=\left({a_0\over \mathfrak{r}_0}\right)^3 \left({\mathfrak{r}_0\over \mathfrak{r}_g}\right)
\left({2M\over m_1+m_2}\right).
\label{def:epsilon}
\ena

This $\epsilon$ corresponds to the initial semi-major axis $a_0$ as
\beann
a_0=\epsilon^{2\over 3}\, \mathfrak{r}_0\,\left({ m_1+m_2\over 2M}\right)^{1/3}
\left({\mathfrak{r}_0\over \mathfrak{r}_g}\right)^{-1/3}\,.
\enann
Using $\epsilon$, we find
\beann
\tilde{\mathfrak{w}}_{\rm dS} &=&\epsilon\, \left[\left({\mathfrak{r}_g\over 2\mathfrak{r}_0-3\mathfrak{r}_g}\right)^{1\over 2}-\left({\mathfrak{r}_g\over 2\mathfrak{r}_0}\right)^{1\over 2}\right],\\
\\
\tilde{\mathfrak{w}}_{\rm R} &=&\epsilon\, \left({\mathfrak{r}_g\over 2\mathfrak{r}_0}\right)^{1\over 2}.\\
\enann

As we discuss in \S. \ref{validity}, the necessary conditions for a stable ``Newtonian" binary is 
given by the length scale of a binary. If we set $\ell_{\rm binary}\sim a_0$, 
we find the condition for $\epsilon$
as
\beann{m_1+m_2\over 2M}\left({\mathfrak{r}_0\over \mathfrak{r}_g}\right)^{-1}\ll\epsilon\ll\sqrt{\mathfrak{r}_0\over \mathfrak{r}_g}.
\enann
When a binary locates near the ISCO radius, this condition is 
\beann
O(10^{-7})\ll\epsilon\ll O(1),
\enann
for $m_1=m_2=10M_\odot, M=10^8M_\odot$.

In order to solve Eqs. (\ref{px}) -(\ref{eq_z}), we first give masses $m_1,m_2, M$, the radius of the 
circular orbit $\mathfrak{r}_0$, and $\epsilon$, which corresponds to the initial semi-major axis of a binary $a_0$.
We then provide the initial data of a binary, i.e., $\tilde{\mathsf{x}}(0)\,,\tilde{\mathsf{y}}(0)\,,\tilde{\mathsf{z}}(0)$ and $\tilde{\mathsf{p}}_{\mathsf{x}}(0)\,,\tilde{\mathsf{p}}_{\mathsf{y}}(0)\,,\tilde{\mathsf{p}}_{\mathsf{z}}(0)$.

Since the motion of a binary can be approximated by an elliptic orbit, we shall fix the initial values 
by assuming an elliptic orbit given by
\beann
\mathsf{r}={a(1-e^2)\over 1+e\cos f}
\,,
\enann
where $a$ is a semi-major axis, $e$ is the eccentricity, and $f$ is true anomaly.
Since the orbital plane is not, in general, $\mathsf{z}=0$, we have to introduce 
three angular variables; the argument of periapsis $\omega$, the ascending node $\Omega$ and the inclination angle $I$.
We have the relations between the position $\bf{\mathsf{r}}=(\mathsf{x},\mathsf{y},\mathsf{z})$ of the component of a binary and the orbital parameters $(\omega\,,\Omega\,,  a\,, e\,, I\,, f)$ as
\begin{widetext}
\bea
\begin{pmatrix}
\mathsf{x} \\
\mathsf{y} \\
\mathsf{z} \\
\end{pmatrix}
&=&
\mathsf{r}\begin{pmatrix}
\cos \Omega\cos(\omega+f)-\sin\Omega\sin(\omega+f)\cos I \\
\sin \Omega\cos(\omega+f)+\cos\Omega\sin(\omega+f)\cos I\\
\sin(\omega+f)\sin I \\
\end{pmatrix}
\label{orbital_parameters}
\ena
 the initial position of a binary can be fixed by the orbital parameters.
As for the initial velocity, 
we have the relation between the mean anomaly $\mathfrak{l}$ and the true anomaly $f$ as
\bea
d\mathfrak{l}=(1-e^2)^{{3\over 2}}
\left(1+e\cos f \right)^{-2}df.
\label{relation_lf}
\ena
In Newtonian dynamics, $\mathfrak{l}=n(\tau-\tau_0)$, where $n$ is the mean motion.
Hence, in this approximation, the time derivative ($d/d\tau$) is given by the derivative with respect to 
the true anomaly ($d/df$).
\\

Assuming $f=0$ at $\tau=0$, we find
\beann
\tilde{\mathsf{x}}(0)
&=&(1-e_0)\cos \Omega_0\cos\omega_0-\sin\Omega_0\sin\omega_0\cos I_0,
\\
\tilde{\mathsf{y}}(0)
&=&(1-e_0)\sin \Omega_0\cos\omega_0+\cos\Omega_0\sin\omega_0\cos I_0,
\\
\tilde{\mathsf{z}}(0)
&=&(1-e_0)\sin\omega_0\sin I_0,
\enann
and
\beann
{d\tilde{\mathsf{x}}\over d\tilde \tau}(0)
&=&-\sqrt{1+e_0\over 1-e_0}\,\left[\cos \Omega_0\sin\omega_0+\sin\Omega_0\cos\omega_0\cos I_0\right],
\\
{d\tilde{\mathsf{y}}\over d\tilde \tau}(0)
&=&\sqrt{1+e_0\over 1-e_0}\,\left[-\sin \Omega_0\sin\omega_0+\cos\Omega_0\cos\omega_0\cos I_0\right],
\\
{d\tilde{\mathsf{z}}\over d\tilde \tau}(0)
&=&\sqrt{1+e_0\over 1-e_0}\,\cos\omega_0\sin I_0.  
\enann
Hence, when we prepare the initial orbital parameters ($e_0\,, I_0\,, \omega_0\,, \Omega_0$), 
we can provide the initial data for the normalized evolution equations (\ref{px}) -(\ref{eq_z}). 

\end{widetext}
\subsection{Binary motion near ISCO radius}
In a hierarchical triple system, there are several
important features. One is the so-called Kozai-Lidov (KL) oscillations.
If the system is inclined more than some critical angle, there appears an oscillation between the eccentricity and inclination angle.
The second interesting feature is an orbital flip, which  may appear when 
the inclination angle evolves into near $90^\circ$.
The last one which we show is a chaotic feature in the long-time evolution.

Here, we show our numerical results. In order to discuss the properties of 
a binary orbit, it is more convenient to use the orbital parameters assuming that the binary motion is close to an elliptic orbit.

In order to extract the orbital parameters 
from the orbit given by the Cartesian 
coordinates, one can use the osculating orbit when the orbit is close to an ellipse. However, one must be careful
with the definitions of orbital elements when using the osculating method. 
For instance, the magnitude of the 
normalized Laplace-Runge-Lenz vector,
which is defined by
\bea
\vect{e}\equiv \tilde{\vect{\mathsf{p}}}\times (\tilde{\vect{\mathsf{r}}}
\times \tilde{\vect{\mathsf{p}}})-{\tilde{\vect{\mathsf{r}}}\over \tilde{\mathsf{r}}}\,,
\label{LRL_vector}
\ena
is commonly used for a measure of orbital eccentricity, but 
it is
not always appropriate. 
It may show an ``apparent" rise in eccentricity or unphysical rapid oscillations especially when the eccentricity is very small~\cite{Will_2019}. 
We take caution and extract the elements' information from physical orbit. In that case, it may be better to 
define the eccentricity by the averaged one over one cycle
as
\beann
\langle e \rangle \equiv {\mathsf{r}_{\rm max}-\mathsf{r}_{\rm min}\over 
\mathsf{r}_{\rm max}+\mathsf{r}_{\rm min}}.
\enann
where $\mathsf{r}_{\rm max}$ and $\mathsf{r}_{\rm min}$  correspond to orbital separation at adjacent turning points of an eccentric orbit.  

The inclination angle $I$ is defined as mutual inclination between angular momenta of the inner and outer binary. In the present case, since the outer binary is just a circular motion on the equatorial plane, the inclination is given by 
\beann
I=\cos^{-1}\left({L_{\mathsf{z}} \over |\vect{L}|}\right)
\,,
\enann
where $\vect{L}\equiv \vect{\mathsf{r}}\times \vect{\mathsf{p}}$ is the angular momentum of a binary.

The other two essential angles $\Omega$ and $\omega$ governing the orientation of the orbital plane and the orbit are also computed in the post-process. The line that marks the intersection of the orbital plane with the reference plane (the equatorial plane in the present case) is called the node line, and the point on the node line where the orbit passes above the reference plane from below is called the ascending node. The angle between the reference axis (say $\mathsf{x}$-axis) and node line vector $\vect{N}$ is the longitude of ascending node $\Omega$.  First, node line is defined as,
\beann
\vect{N} = \vect{e}_{\mathsf{z}}\times \vect{h} \,.
\enann
where $\vect{e}_{\mathsf{z}}$ is normal to the reference plane (the unit vector in the $\mathsf{z}$ direction) and $\vect{h}=\vect{L}/\mu$ is specific angular momentum vector of a binary. Thus, $\Omega$ is computed as,
\beann
\Omega = \cos^{-1} (N_{\mathsf{x}}/N) \,.
\enann
The argument of periapsis $\omega$ is the angle between node line and periapsis measured in the direction of motion. Therefore,
\beann
\omega = \cos^{-1} \bigg(\frac{\vect{N} \cdot \vect{e}}{N\,e}\bigg) \,.
\enann
When the orbit can be approximated well by the osculating one, 
$\vect{e}$ is given by the normalized Laplace-Runge-Lenz vector (\ref{LRL_vector}).
Otherwise, we define the averaged eccentricity vector by 
\beann
\langle \vect{e}\rangle  \equiv -{\left(\vect{\mathsf{r}}_{\rm min}+\vect{\mathsf{r}}_{\rm max}\right)\over 
\left(\mathsf{r}_{\rm min}+\mathsf{r}_{\rm max}\right)}
\enann
pointing towards the periapsis,
where  $\vect{\mathsf{r}}_{\rm max}$ and $\vect{\mathsf{r}}_{\rm min}$
are numerical data of position vector.
We have used both definitions and found that most results agree well.

\subsubsection{Firmness of a binary and stability}
\label{stability}
As we discuss in \S. \ref{validity}, a binary near SMBH may be broken when the tidal force by SMBH is stronger than 
the mutual gravitational attractive force of a binary. We introduce a firmness parameter of a binary $\mathfrak{f}$
defined by
\beann
\mathfrak{f}&\equiv& {{\rm gravitational~force}\over {\rm tidal~force~by~SMBH}}={{Gm_1m_2/\ell_{\rm binary}^2}\over {\mu \mathfrak{r}_g \,\ell_{\rm binary} / \mathfrak{r}_0^3}}
\\&=&\left({ m_1+m_2\over 2M}\right)\left({\mathfrak{r}_0\over \ell_{\rm binary}}\right)^3
\approx {1\over \epsilon^2}\left({\mathfrak{r}_0\over \mathfrak{r}_g}\right)
\,.
\enann

If the firmness parameter $\mathfrak{f}$ is smaller than $O(1)$, we expect that the tidal force will break a binary.
Hence, $\mathfrak{f}>1$ is a necessary condition for stability of a binary. When we are interested in a binary motion near SMBH, 
this stability condition gives $\epsilon \leq O(1)$. We can confirm numerically  that our present system is really unstable 
when  $\epsilon \geq 0.5$. In what follows, we numerically analyze a binary system under the conditions of  $\epsilon \leq 0.4$.

\subsubsection{Kozai-Lidov oscillation}
\label{KL_oscillation}
Here, we show some numerical examples, which shows 
the KL oscillations in the long-time evolution.

We expect the KL oscillations to occur when the inclination angle is larger than the critical value.
Under the quadrupole approximation in a Newtonian  hierarchical system, the critical inclination angle is given by 
$I_{\rm crit}^{\rm (N)}=\sin^{-1}\sqrt{2/5}\approx 39.23^\circ$.
In the present model, it can be obtained by the double-averaging 
analysis of the Lagrange planetary equations, which shows 
that the critical value is slightly larger 
as the radius $\mathfrak{r}_0$ gets smaller, and  
it increases to $41.6^\circ$ near the ISCO radius (See Appendix \ref{DA_KL_oscillations}).
It seems hard to obtain the exact critical value 
by numerical simulations, although we 
have found consistent results.

The typical time scale of the KL oscillations is given by~\cite{Haruka2019,Antognini2015,Shevchenko17} 
\bea
T_{\rm KL}\sim {P_{\rm out}^2\over P_{\rm in}}
\,,
\label{KL_time_N}
\ena
where $P_{\rm in}$ and $P_{\rm out}$ are the periods of an inner binary and of an outer binary, respectively. 
For a hierarchical triple system, since $P_{\rm in}\ll P_{\rm out}$,
$T_{\rm KL}\gg P_{\rm out}$, which means that the KL oscillation is a secular effect.

\begin{figure}[h]
\begin{center}
\includegraphics[width=5.5cm]{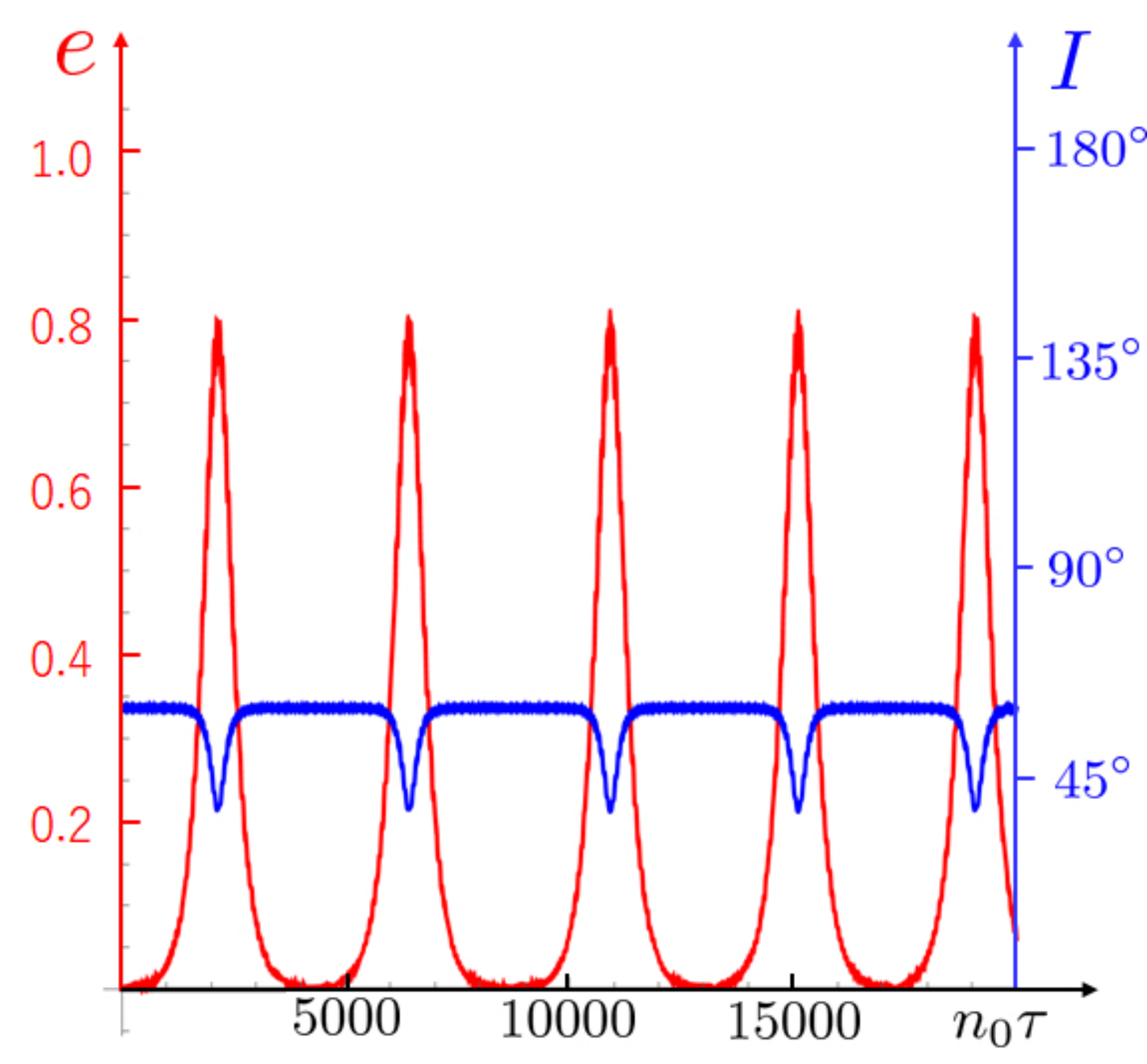}
\caption{This figure shows Kozai-Lidov oscillation between orbital eccentricity $e$
(the red curve) and relative inclination $I$ (the blue curve). For this particular model (Model Ic), we choose $\mathfrak{r}_0=3.5 \mathfrak{r}_g$ and $\epsilon=0.1$. 
The initial data is $e_0=0.01$, $I_0=60^\circ$, $\omega_0=60^\circ$ and $\Omega_0=30^\circ$. }
\label{KL_Ic}
\end{center}
\end{figure}

\begin{figure}[h]
\begin{center}
\includegraphics[width=6cm]{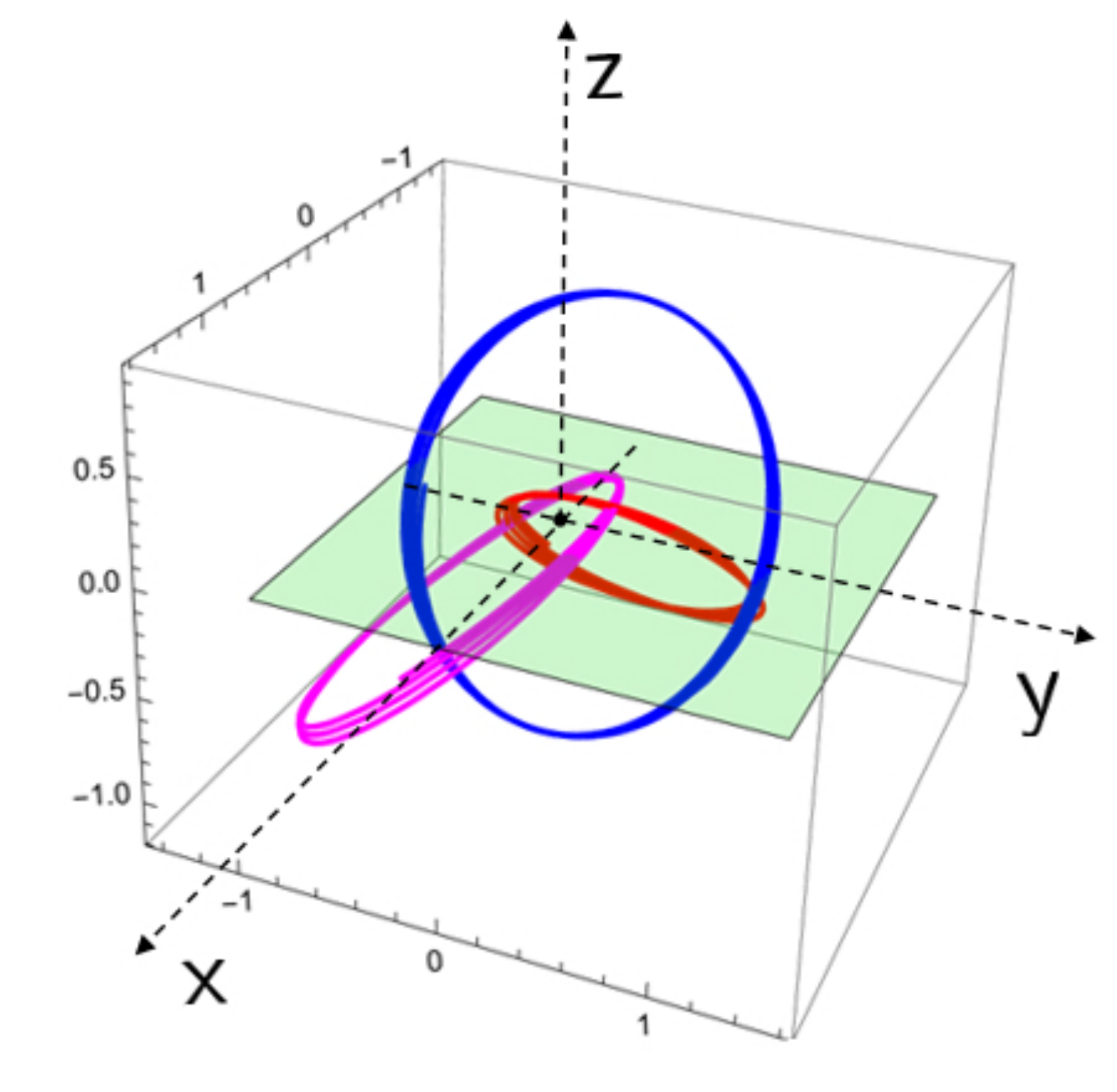}
\caption{Typical orbits in the same model as Fig. \ref{KL_Ic}.
 The red, blue  and magenta curves denote 
a few cycles near $n_0\tau=2120$, $4150$, and $6385$ respectively.
The inclination angles of red, blue  and magenta curves are $I\approx 38.5^\circ$, $61.1^\circ$, and $38.3^\circ$, while the eccentricities of those curves are 
$e\approx 0.80$, $0.0017$ and $0.80$, respectively.}
\label{KL_60_012}
\end{center}
\end{figure}

In the first model (Model Ic), we choose 
$ \mathfrak{r}_0=3.5  \mathfrak{r}_g$ and $\epsilon=0.1$.
It corresponds to 
$a_0=0.0023  \mathfrak{r}_g\approx 0.0045 {\rm AU}$
for $m_1=m_2=10M_\odot$ and $M=10^8 M_\odot$.
Since $G(m_1+m_2)/c^2\approx 30{\rm km} \ll a_0$, 
the binary motion can be described by a Newtonian orbit in a local inertial frame. We then adopt the initial conditions as 
the eccentricity $e_0=0.01$ and inclination $I_0=60^\circ$.
The numerical results are given in Figs. 
\ref{KL_Ic} and \ref{KL_60_012}.


The KL oscillation period is obtained numerically as 
$n_0T_{\rm KL} \sim 4000$.
Since $n_0 P_{\rm in}\sim 2\pi$ and $ \omega_0 P_{\rm out}\sim 2\pi $, 
we can evaluate it by Eq.~\eqref{KL_time_N} 
as $n_0 {P_{\rm out}^2/ P_{\rm in}}\sim 2\pi (n_0/\omega_0)^2\sim 2500$, 
which is consistent with the above numerical value.

\subsubsection{Chaotic feature}
\label{chaotic_feature}
One of the well-known features of three body system is 
a chaotic behaviour in a binary motion.
The model given in the previous subsection shows a stable KL oscillation.
The KL oscillation period is regular. The chaotic feature is not seen.
This is just because a binary is  extremely compact and very hard.
The firmness parameter is $\mathfrak{f}\sim 350$.

\begin{figure}[h]
\begin{center}
\includegraphics[width=5.5cm]{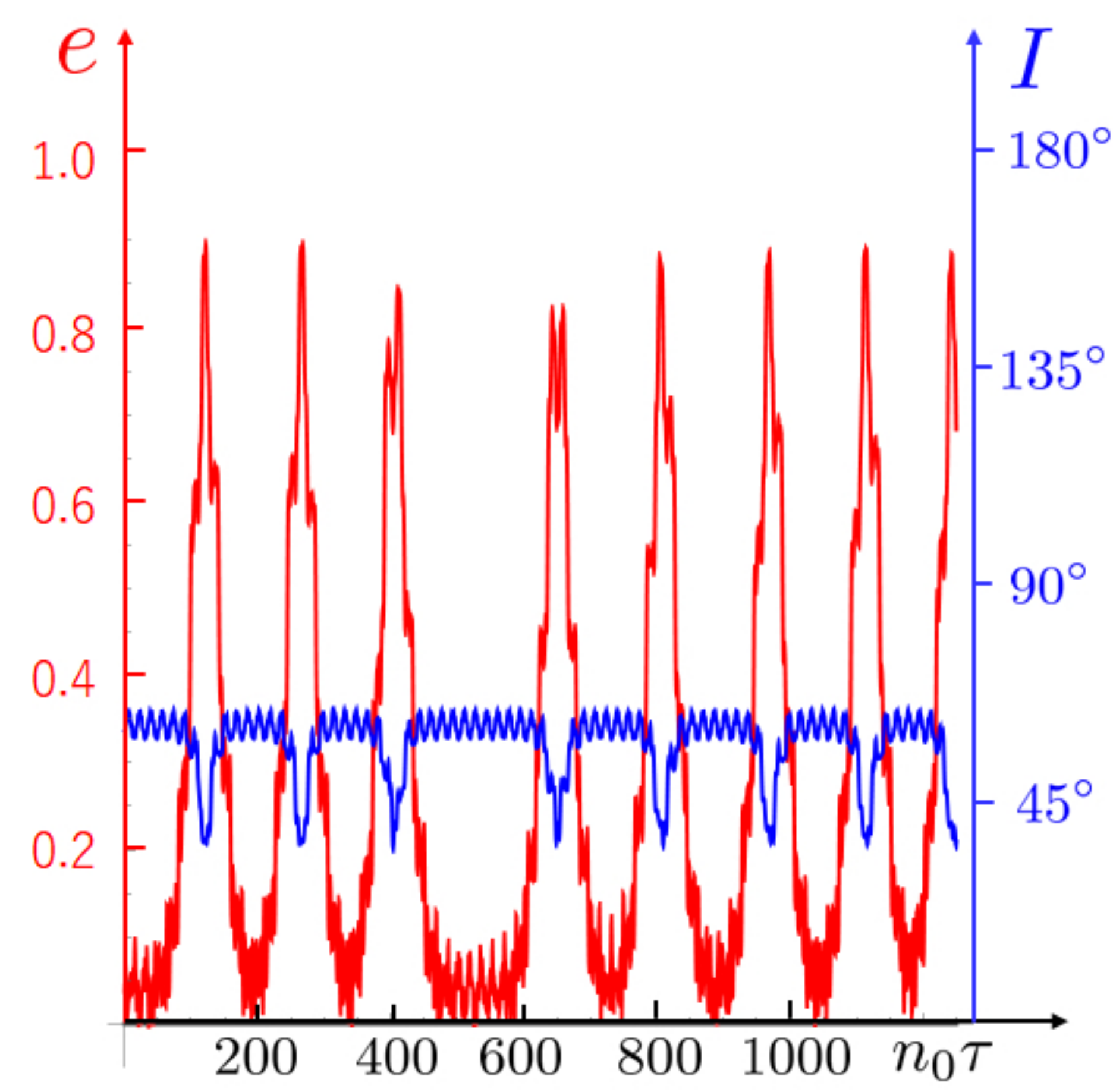}
\caption{Chaotic Kozai-Lidov oscillation. The red and blue curves denote the eccentricity and the inclination. For this model (Model IIc),
we choose $\mathfrak{r}_0=3.5 \mathfrak{r}_g$ and $\epsilon=0.4$.
The initial data is $e_0=0.01$, $I_0=60^\circ$, $\omega_0=60^\circ$ and $\Omega_0=30^\circ$. }
\label{KL_IIc}
\end{center}
\end{figure}

However, if a binary is not so highly compact, we find some chaotic features.
In Fig. \ref{KL_IIc}, we show one example (Model IIc). 
The model parameters are given by $\epsilon=0.4$ and $\mathfrak{r}_0=3.5 \mathfrak{r}_g$,
and the initial parameters are chosen as $e_0=0.01$, $I_0=60^\circ$, $\omega_0=60^\circ$ and $\Omega_0=30^\circ$. The larger value of $\epsilon$ corresponds to a larger-scale binary, i.e., the initial semi-major axis is $a_0=0.0058\mathfrak{r}_g\approx 0.0115 {\rm AU}$.
We can see clearly the KL oscillation, but the period is not strictly regular.
Since the firmness parameter is $\mathfrak{f}\sim 22>O(1)$, 
the system is still stable, but shows some 
irregular behaviours in the KL oscillations.
The maximum values of the eccentricity is also random as shown in 
Fig. \ref{KL_IIc}.

\subsubsection{Orbital flip}
\label{orbital_flip}
 Another interesting feature is an orbital flip, i.e., 
 an inclination angle goes beyond $90^\circ$. It may occur when the initial inclination angle is near $90^\circ$.
 \begin{figure}[h]
\begin{center}
\includegraphics[width=5.5cm]{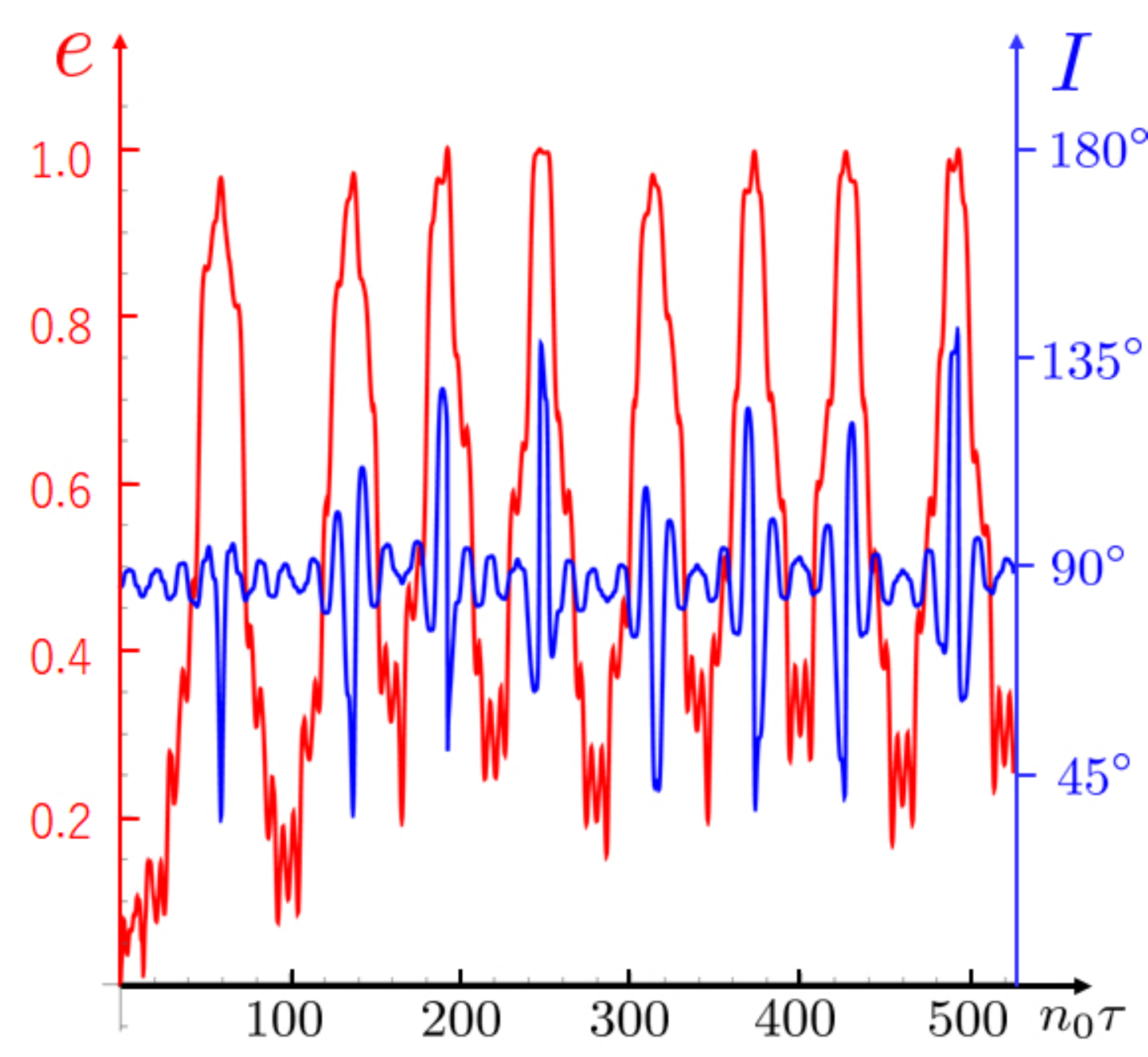}
\caption{Orbital flip in KL oscillation (Model IIa). 
We choose $\mathfrak{r}_0=3.5 \mathfrak{r}_g$ and $\epsilon=0.4$.
The initial data is $e_0=0.01$, $I_0=85^\circ$, $\omega_0=60^\circ$ and $\Omega_0=30^\circ$. The red and blue curves denote the eccentricity and the inclination.
The inclination angle evolves beyond $90^\circ$ several times.}
\label{KL_IIa}
\end{center}
\end{figure}

 In Fig. \ref{KL_IIa}, we find that
 the orbital flip accompanying KL oscillations occurs periodically. 
The model parameters are given by $\epsilon=0.4$ and $\mathfrak{r}_0=3.5 \mathfrak{r}_g$,
and the initial parameters are chosen as $e_0=0.01$, $I_0=85^\circ$, $\omega_0=60^\circ$ and $\Omega_0=30^\circ$ (Model IIa). 
The inclination blue curve in  Fig. \ref{KL_IIa}
evolves beyond $90^\circ$ several times, but the time period is irregular.

One interesting observation is 
there exists a strong correlation between an orbital flip and large eccentricity.
When an orbital flip occurs, 
the eccentricity becomes very close to unity.

\begin{figure}[h]
\begin{center}
\includegraphics[width=6cm]{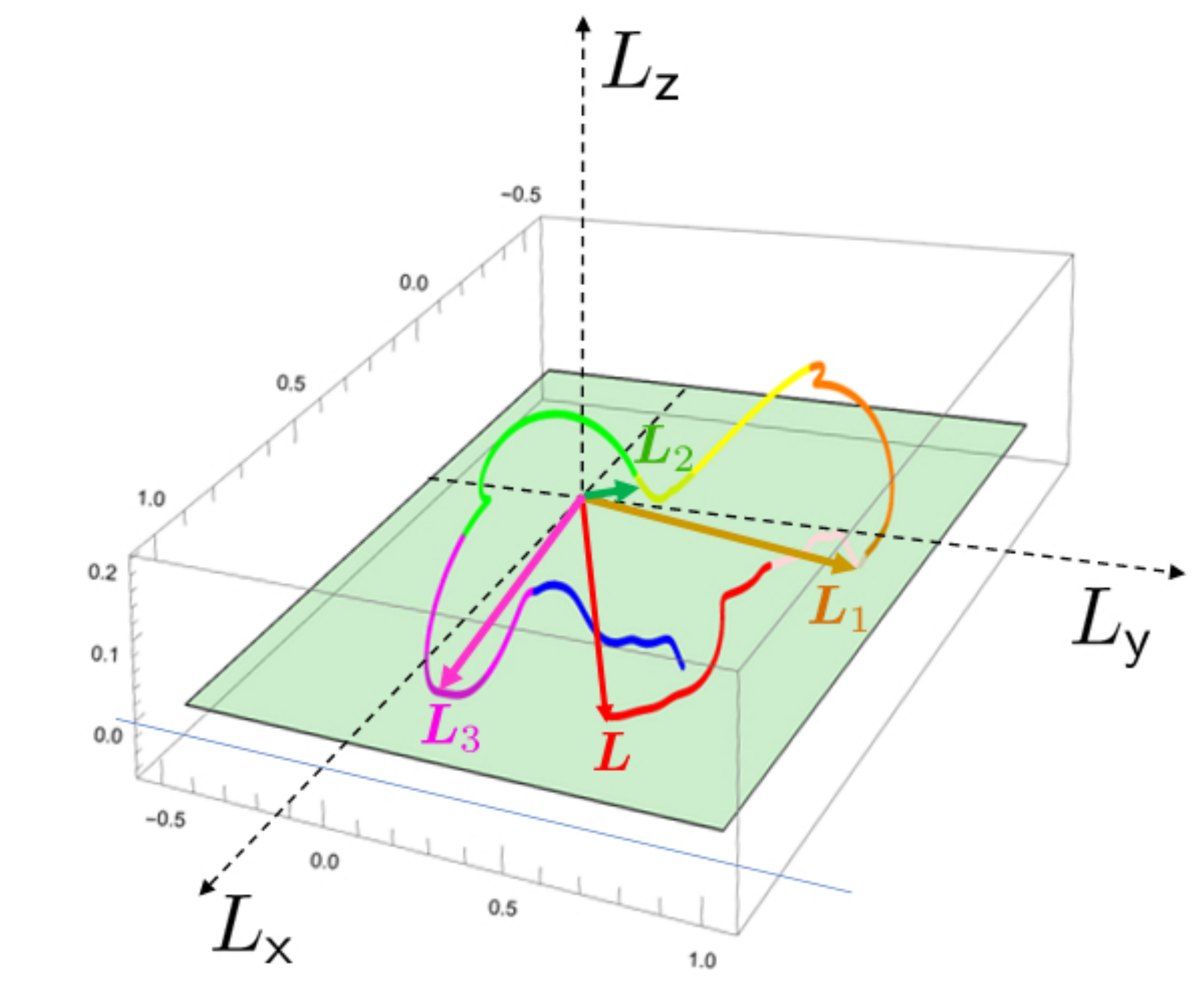}
\caption{Time evolution of angular momentum. The 
parameters and initial conditions are the same as those 
in Fig. \ref{KL_IIa}. The red, light-red, orange, 
yellow, green, magenta, and blue curves denote the 
angular momentum vectors in the periods $n_0\tau=40-
50,50-60, 60-70, 70-80, 80-90, 90-100$ and $100-110$, 
respectively.
The $z$-components of $\vect{L}_1\,,\vect{L}_2$\,, and 
$\vect{L}_3$ become negative.}
\label{angmom_orbitalflip}
\end{center}
\end{figure}

In Fig. \ref{angmom_orbitalflip}, we show the time evolution 
of the angular momentum vector, which is defined by 
$\vect{L}=\vect{\mathsf{r}}\times\vect{\mathsf{p}}$.
The $z$-component of the angular momentum becomes 
negative near $n_0\tau\approx 60-62\,,73-80\,,$ and $92-97$. The corresponding vectors are shown by the colored arrows $\vect{L}_1\,,\vect{L}_2\,,$ and $\vect{L}_3$,
respectively.

\subsubsection{Summary of various models}
\label{Model:summary}
We summarize our numerical results in Table~\ref{Table:summary}.
We have simulated three types of models (I, II and III).
The parameters of Model I, Model II and Model III are 
($\mathfrak{r}_0 =3.5 \mathfrak{r}_g$,  
$\epsilon=0.1$), ($\mathfrak{r}_0 =3.5 \mathfrak{r}_g$,  
$\epsilon=0.4$), and ($\mathfrak{r}_0 =7 \mathfrak{r}_g$,  
$\epsilon=0.1$), respectively. 
For each model, we choose the initial data as
(a) $e_0=0.01\,,I_0=85^\circ$, 
(b) $e_0=0.9\,,I_0=85^\circ$, 
(c) $e_0=0.01\,,I_0=60^\circ$, and 
(d) $e_0=0.9\,,I_0=60^\circ$.
The initial argument of periapsis and ascending node are chosen
as $\omega_0=60^\circ$ and $\Omega_0=30^\circ$ for all models.
We also performed the simulation with different values of those two parameters, the results do not change so much.

In Model I, since a binary is very compact (the firmness parameter $\mathfrak{f}\sim 350$), it is very stable. 
As shown in Fig. \ref{KL_Ic}, we find the regular KL oscillation, although the oscillation period $T_{\rm KL}$ is not strictly constant but slightly disperse within $1-9\%$.
When the initial eccentricity is large (Model Ib and Id),
the oscillation period gets smaller and minimum eccentricity 
becomes larger.
Except for Model Ic, the KL oscillation type is the so-called libration,
which shows the oscillation of the argument of periapsis around $90^\circ$.
In Model Ic, the KL oscillation seems to be 
the rotation type,
which denotes the argument of periapsis increases monotonically. However in  the present model, it does increase on average but not monotonically (sometimes going back and forth).

\begin{widetext}

\begin{table}[h]
\begin{center}
\scalebox{0.8}[0.9]{
  \begin{tabular}{|c||c|c|c||c|c|c|c||c|c|c|c|c|}
\hline 
 && & & &&&&& &&&\\[-.5em]
Model &$\mathfrak{r}_0 $&$\epsilon$&$a_0$&$e_0$ &$I_0$ &$\omega_0$&$\Omega_0$&KL oscillation
type&
$n_0T_{\rm KL}$&$e_{\rm min}/e_{\rm max}$&$I_{\rm min}/I_{\rm max}$&orbital flip
\\[.5em]
\hline
\hline
&& & & & &&&&& &&\\[-.5em]
Ia&  & &  &$0.01$ &$85^\circ$ &$60^\circ$ &$30^\circ$   &$\mathbb{L}$ [$36^\circ\leq \omega\leq 144^\circ$]&$2396-2430$&$0.01/0.995$ &$39.1^\circ/85.5^\circ$& No
\\[.5em]
\cline{1-1}\cline{5-13}
&& & & &&&&& &&&\\[-.5em]
Ib&  & & &$0.9$ &$85^\circ$ &$60^\circ$ &$30^\circ$    &$\mathbb{L}$ [$55^\circ\leq \omega\leq 125^\circ$]&$580-600$& $0.672/0.998$&$54.8^\circ/87.5^\circ$&No
\\[.5em]
\cline{1-1}\cline{5-13}
&&&& &&&& &&&&\\[-.5em]
Ic& $3.5\mathfrak{r}_g $&$0.1$&$0.0046\mathfrak{r}_g $ &$0.01$ &$60^\circ$ &$60^\circ$ &$30^\circ$    &$\mathbb{R}^*$&$3600-4300$& $0.003/0.8$&$38.5^\circ/60.5^\circ$&No
\\[.5em]
\cline{1-1}\cline{5-13}
&& & & &&& &&&&&\\[-.5em]
Id&  & & &$0.9$ &$60^\circ$ &$60^\circ$ &$30^\circ$    &$\mathbb{L}$ [$56^\circ\leq \omega\leq 124^\circ$]&$753-770$& $0.49/0.95$&$47.3^\circ/74.3^\circ$&No
\\[.5em]
\hline
\hline
&& & & & &&&&&& &\\[-.5em]
IIa&  & &  &$0.01$ &$85^\circ$ &$60^\circ$ &$30^\circ$   &$\mathbb{R}^*$&$50-78$& ($0.15-0.34$)/& ($33.6^\circ-42.0^\circ$)/& Yes
\\[.5em]
&  & &    &  & &  &   && & $(0.97-1.00)$&
$(137.9^\circ-143.9^\circ)$ &  (irregular)
\\[.5em]
\cline{1-1}\cline{5-13}
&& & & &&&&&& &&\\[-.5em]
IIb&  & & &$0.9$ &$85^\circ$ &$60^\circ$ &$30^\circ$  &$\mathbb{L}$ [$(50^\circ-53^\circ)\leq $&$39-44$& $(0.533-0.589)/$&$(50.2^\circ-53.7^\circ)/$
 & Yes 
\\[.5em]
&  & & &  & &     & &$\omega\leq (126^\circ-131^\circ)$]&& $(0.978-0.999)$&$(126.4^\circ-128.6^\circ)$
 & (irregular)
\\[.5em]
\cline{1-1}\cline{5-13}
&& & & &&&&&& &&\\[-.5em]
IIc&  $3.5\mathfrak{r}_g $&$0.4$&$0.0116\mathfrak{r}_g $ &$0.01$ &$60^\circ$ &$60^\circ$ &$30^\circ$    & $\mathbb{R}^*$ &$116-250$&$(0.04-0.08)/$ &$(36.5^\circ-37.7^\circ)/$ & No
\\[.5em]
& && &  &  &    &   &   & &$(0.82-0.90)$ &
$(59^\circ-65^\circ)$ &
\\[.5em]
\cline{1-1}\cline{5-13}
&&&& &&&&&& &&\\[-.5em]
IId&  & & &$0.9$ &$60^\circ$ &$60^\circ$ &$30^\circ$    & $\mathbb{L}$ [$(50^\circ-56^\circ)\leq$&$54-60$& $(0.41-0.49)/$&$(43^\circ-46^\circ)/$ & No
\\[.5em]
&  & & &  &  &   &  & $ \omega\leq (124^\circ-130^\circ)$]&
& $(0.92-0.97)$&$(69^\circ-75^\circ)$ & 
\\[.5em]
\hline
\hline
&& & & & &&&&&& &\\[-.5em]
IIIa& && &$0.01$ &$85^\circ$ &$60^\circ$ &$30^\circ$    &$\mathbb{L}$ [$38^\circ\leq \omega\leq 142^\circ$]
&$6776-6928$& $0.009/0.997$
&$39.1^\circ/85.2^\circ$&No
\\[.5em]
\cline{1-1}\cline{5-13}
&& & & &&&&&& &&\\[-.5em]
IIIb&  & & &$0.9$ &$85^\circ$ &$60^\circ$ &$30^\circ$   & $\mathbb{L}$ [$56^\circ\leq \omega\leq  124^\circ$] &$1585-1628$& $0.675/0.998$&$55.1^\circ / 87.2^\circ$&No
\\[.5em]
\cline{1-1}\cline{5-13}
&& & &&&&&&& &&\\[-.5em]
IIIc& $7\mathfrak{r}_g $&$0.1$&$0.0073\mathfrak{r}_g $ &$0.01$ &$60^\circ$ &$60^\circ$ &$30^\circ$     &$\mathbb{L}$ [$44^\circ\leq \omega \leq 136^\circ$]& $9450-10600$&$0.004/0.792$&$38.3^\circ/60.3^\circ$&No
\\[.5em]
\cline{1-1}\cline{5-13}
&& & & &&&&& &&&\\[-.5em]
IIId&  & & &$0.9$ &$60^\circ$ &$60^\circ$ &$30^\circ$    &$\mathbb{L}$ [$56^\circ\leq \omega\leq 124^\circ$]&$2045-2074$&$0.493/0.949$&$47.5/74.5$& No
\\[.5em]
\hline
\hline 
 \end{tabular}
 }
\caption{The properties of the KL oscillations. The oscillations in 
Models I and III are regular, although there are small amount of dispersion in the KL
oscillation time scale. For Model II, the oscillations are irregular and 
there appears dispersion in the maximum and minimum values of the eccentricity and inclination, whose values are shown in the bracket ().
$\mathbb{L}$ and $\mathbb{R}$ denote libration and rotation type, respectively.
$\mathbb{R}^*$ means that the argument of periapsis does increase on average 
but not monotonically, and sometimes going back and forth. }
\label{Table:summary}
\end{center}
\end{table}

For Model II, a binary is slightly less compact (the firmness parameter $\mathfrak{f}\sim 22$). It is still stable but becomes irregular 
both in the oscillation period and in the amplitude as shown in Fig. \ref{KL_IIc}.
There appear dispersion in the maximum and minimum values of the eccentricity and inclination, whose values are shown in the bracket ().
When the initial inclination angle is large (Models IIa and IIb), we find 
irregular orbital flips.
The KL oscillation is either libration type (Models IIb and IId),
in which the maximum and minimum angles of the argument of the periapsis 
disperse as shown by the brackets (), 
or irregular rotation type (Models IIa and IIc), in which the inclination angle increases on average but not monotonically (sometimes going back and forth).

Model III is the case with a hard binary as Model I, but its location 
is a little far from SMBH ($\mathfrak{r}_0=7\mathfrak{r}_g$).
Hence the relativistic effect as well as the tidal force are smaller than 
those in Model I. As a result, the KL oscillation becomes more regular and stable.
The KL oscillation timescale is larger than that in Model I because the outer orbital period $P_{\rm out}$ 
is larger.

\subsection{Motion of the CM of a binary}
Next we show the motion of the CM of a binary.
In order to know how much the motion deviates from a circular motion with the radius $\mathfrak{r}_0$, we 
solve the radial perturbation equation (\ref{eq:radial_perturbation}).
Using the previous numerical solutions, we integrate 
Eq. (\ref{eq:radial_perturbation})
 with the initial conditions such that $\mathfrak{r}_{(1)}(0)=0$ and $\dot{\mathfrak{r}}_{(1)}(0)=0$. Here, we discuss some typical cases.
 
\subsubsection{\rm Model Ia}
In this case, we find stable and regular oscillations with the period $T_{\rm osc}\equiv {2\pi/ k}\approx 2.65 P_{\rm out}$, where $k$ is defined by Eq. (\ref{coeff_k2}) and $P_{\rm out}={2\pi/ \mathfrak{w}_0}$ is the period of the circular motion.
The oscillation center is given by $\langle \mathfrak{r}_{1}\rangle \approx 1.60\times 10^{-4}\mathfrak{r}_0$ with the amplitude $\Delta\mathfrak{r}_{1}\approx 1.71\times 10^{-4}\mathfrak{r}_0$, but the center increases to 
 $2.18\times 10^{-4}\mathfrak{r}_0$ when the eccentricity becomes close unity, keeping 
 the oscillation amplitude $\Delta\mathfrak{r}_{1}$ to be constant
 (See Fig. \ref{radial_CM} (left)).
 
 There exists a good correlation between the shift of 
 the oscillation center and the eccentricity.
 Since the oscillation amplitude is very small compared with the circular radius $\mathfrak{r}_0$,
 the deviations from the circular motion can be treated as perturbations, which confirms our approach.

\begin{figure}[h]
\begin{center}
\includegraphics[width=6cm]{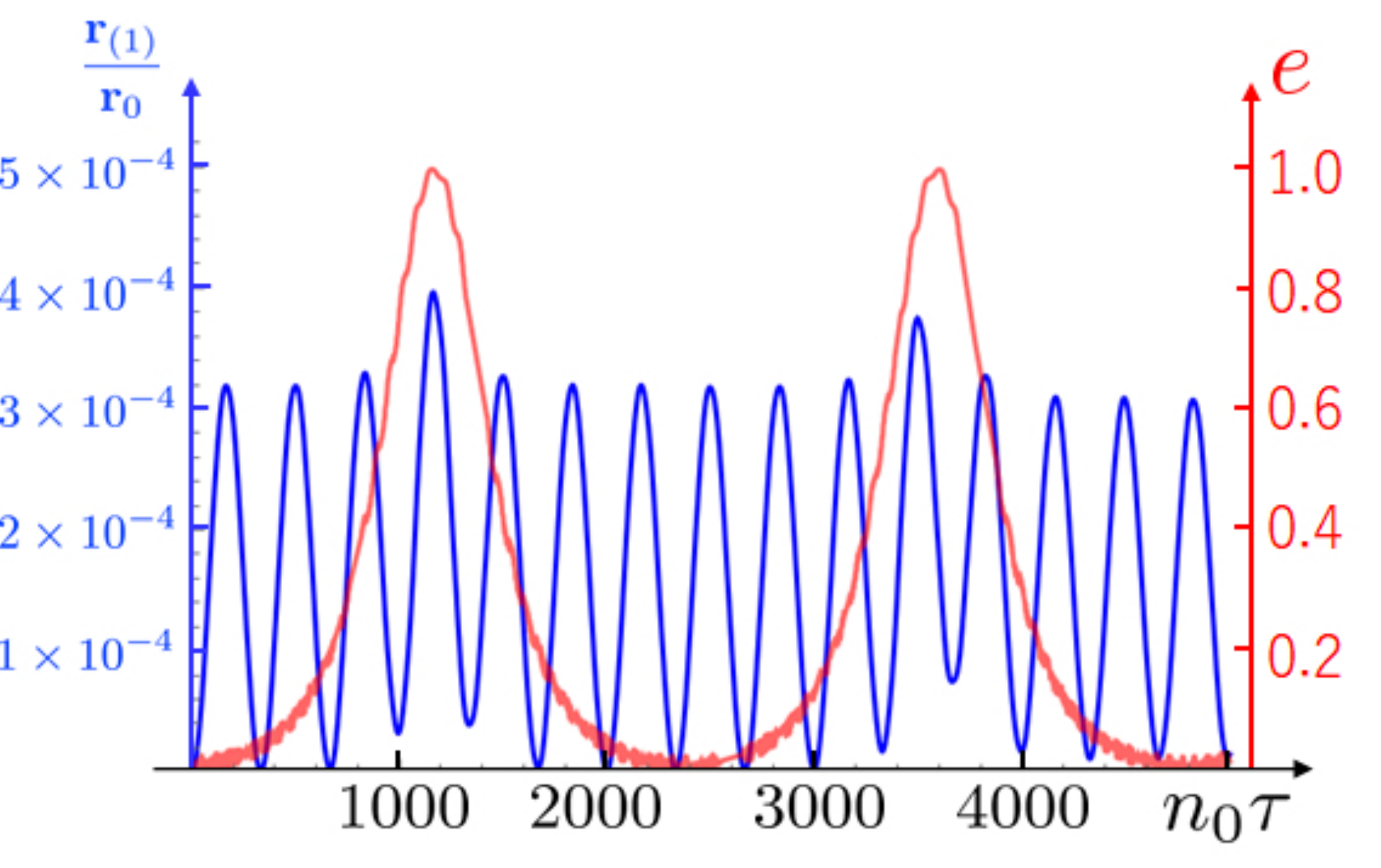}
~~~~
\includegraphics[width=6cm]{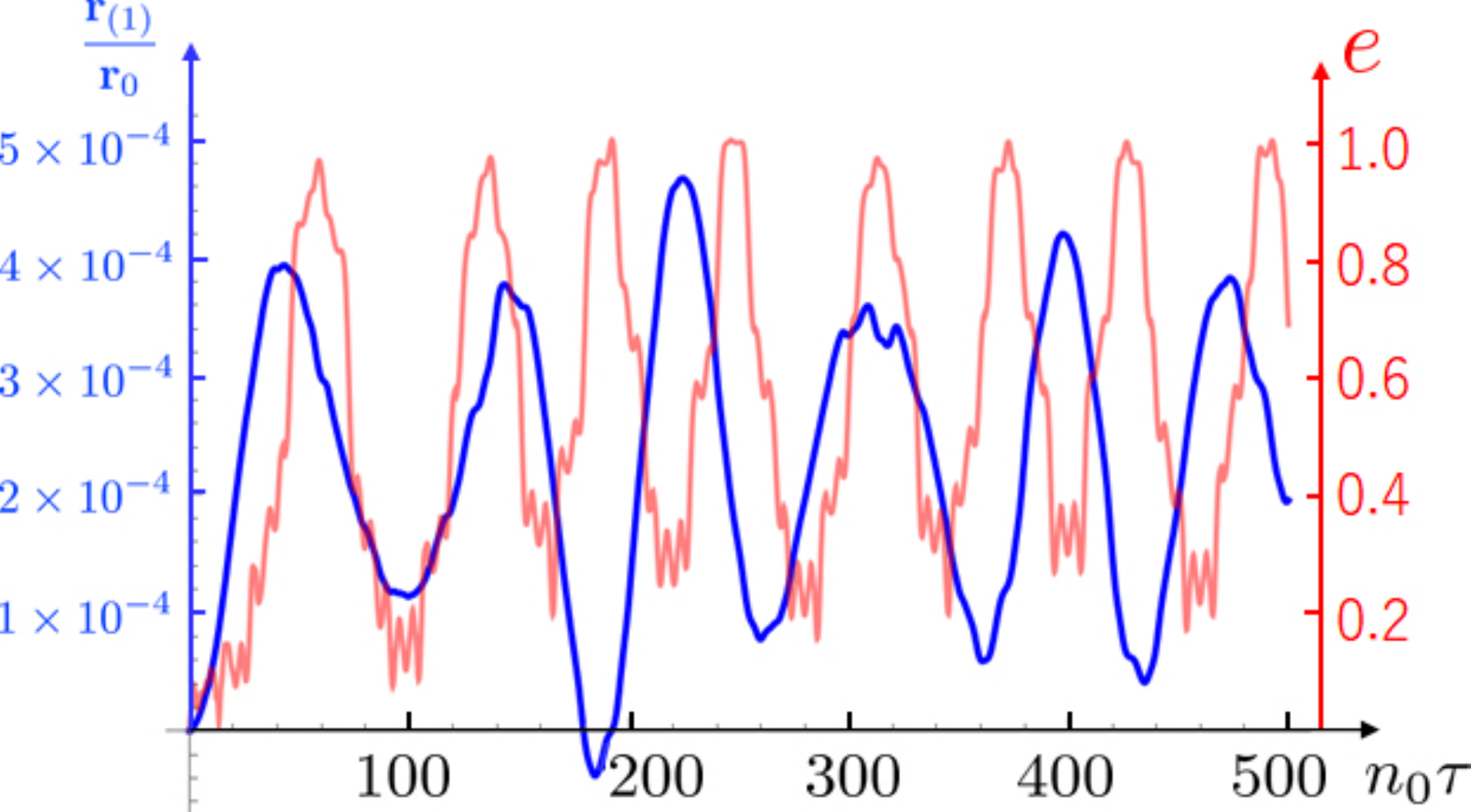}
\caption{The radial deviation $\mathfrak{r}_{(1)}$ normalized by 
the radius of the circular orbit $\mathfrak{r}_0$ are shown by blue curves 
for Model Ia (left) and Model IIa (right). The deviation from a circular orbit 
in Model Ia 
is regular oscillations around $\langle \mathfrak{r}_{1}\rangle$, which increases when the eccentricity increases (the red curve), while that in Model IIa 
becomes irregular oscillations without correlation with the eccentricity (the red curve).
}
\label{radial_CM}
\end{center}
\end{figure}

\subsubsection{\rm Model IIa}

In this case, the KL oscillation is stable but irregular. As a result, 
we find irregular oscillations of the CM as shown by a blue curve in Fig \ref{radial_CM} (right).
There is no correlation with the eccentricity $e$.
Although the oscillations are irregular, the amplitude is very small and then the deviation from the circular motion can be treated as stable perturbations.

\subsubsection{\rm Other Models}
We summarize the results in Table \ref{Table:deviation} for the  Models
given in Table \ref{Table:summary}.
We find that Models (I) and (III) give 
stable and regular oscillations of the CM.
It is just because those binary motions  also show stable and regular KL oscillations. 
On the other hand, Model (II) shows irregular oscillations
because of the irregular KL oscillations in those binary motions.

\begin{table}[h]
\begin{center}
\scalebox{0.8}[0.9]
{
  \begin{tabular}{|c||c|c|c|c||c|c||c|c|c||}
\hline 
&&&&&&&&&\\[-.5em]
Model &$\mathfrak{r}_0 ({\rm AU}) $&$a_0 ({\rm AU})$&$\epsilon$&$\mathfrak{f}$&$e_0$ &$I_0$ &
$[\langle \mathfrak{r}_{1}\rangle \pm \Delta\mathfrak{r}_{1}]/\mathfrak{r}_0$&
$T_{\rm ocs}/P_{\rm out}$&oscillation property
\\[.5em]
\hline
\hline
&&&&&&&&&\\[-.5em]
Ia&  & &  &&$0.01$ &$85^\circ$ &
$[(1.60-2.18)\pm 1.71]\times 10^{-4}$& &regular (good correlation with  $e$)  
\\[.5em]
\cline{1-1}\cline{6-8}\cline{10-10}
&&&&&&&&&\\[-.5em]
Ib&  & & &&$0.9$ &$85^\circ$ &
$[0.18\pm 3.51]\times 10^{-3}$& &regular
\\[.5em]
\cline{1-1}\cline{6-8}\cline{10-10}
&&&&&&&&&\\[-.5em]
Ic& $6.91$&$0.0091$ &$0.1$&$350$&$0.01$
&$60^\circ$ &
$[(7.01-7.37)\pm 7.04]\times 10^{-4}$&$2.65$&regular (good correlation with $e$)  
\\[.5em]
\cline{1-1}\cline{6-8}\cline{10-10}
&&&&&&&&&\\[-.5em]
Id&  & & &&$0.9$ &$60^\circ$ &
$[(4.09-4.40)\pm 4.51]\times 10^{-4}$& &regular (good correlation with $e$) 
\\[.5em]
\hline
\hline
&&&&&&&&&\\[-.5em]
IIa&  & &  &&$0.01$ &$85^\circ$ &
$[(1.60 - 2.91)\pm 1.88]\times 10^{-4}$& &irregular (no correlation with  $e$)  
\\[.5em]
\cline{1-1}\cline{6-8}\cline{10-10}
&&&&&&&&&\\[-.5em]
IIb&  & & &&$0.9$ &$85^\circ$ &
$[(2.89-3.18)\pm 2.85]\times 10^{-4}$& &regular (no correlation with  $e$)   
\\[.5em]
\cline{1-1}\cline{6-8}\cline{10-10}
&&&&&&&&&\\[-.5em]
IIc&  $6.91$&$0.0229 $ &$0.4$&$21.9$&$0.01$ 
&$60^\circ$ &
$[(4.30-5.77)\pm 6.09]\times 10^{-4}$&$2.65$&irregular (no correlation with $e$)  
\\[.5em]
\cline{1-1}\cline{6-8}\cline{10-10}
&&&&&&&&&\\[-.5em]
IId&  & & &&$0.9$ &$60^\circ$ &
$[(4.05-4.31)\pm 4.41]\times 10^{-4}$& &irregular (no correlation with  $e$)  
\\[.5em]
\hline
\hline
&&&&&&&&&\\[-.5em]
IIIa& && &&$0.01$ &$85^\circ$ &
$[(1.23-1.56)\pm 1.28]\times 10^{-5}$& &regular (good correlation with  $e$)  
\\[.5em]
\cline{1-1}\cline{6-8}\cline{10-10}
&&&&&&&&&\\[-.5em]
IIIb&  & & &&$0.9$ &$85^\circ$ &
$[(1.22-1.43)\pm 1.35]\times 10^{-5}$& &regular (good correlation with  $e$)  
\\[.5em]
\cline{1-1}\cline{6-8}\cline{10-10}
&&&&&&&&&\\[-.5em]
IIIc& $13.8$&$0.0144 $ 
    &$0.1$&$700$&$0.01$ &$60^\circ$ &
$[(5.78-5.98)\pm 5.83]\times 10^{-5}$&$1.32$&regular (good correlation with  $e$) 
\\[.5em]
\cline{1-1}\cline{6-8}\cline{10-10}
&&&&&&&&&\\[-.5em]
IIId&  & & &&$0.9$ &$60^\circ$ &
$[(3.14-3.33)\pm 3.38]\times 10^{-5}$& &regular (good correlation with  $e$) 
\\[.5em]
\hline
\hline 
 \end{tabular}
 }
\caption{The oscillations of the CM, which are the radial deviations from a circular geodesic motion. $\langle \mathfrak{r}_{1}\rangle$ denotes the center of the oscillations, while $ \Delta\mathfrak{r}_{1}$ gives the amplitude of the oscillations
(or dispersion for the irregular oscillations (Model II)).
The typical oscillation period is given by $T_{\rm osc}\equiv 2\pi/k$, where $k$ is defined by Eq. (\ref{coeff_k2}).}
\label{Table:deviation}
\end{center}
\end{table}

\subsection{Comparison with double-averaging approach}
In Appendix \ref{planetary equation}, we present 
one of standard approaches on hierarchical triple system, which is the Lagrange planetary equations
for the orbital parameters. 
Since we are interested in the long-time behaviour 
such as the Kozai-Lidov mechanism, 
taking averages of the Hamiltonian over two periods
of the inner and outer binaries, 
we can analyze the simplified doubly-averaged 
planetary equations.

Here, we compare our numerical results with those obtained in the double-averaging (DA) approach. We show the evolution of the eccentricity for Models Ia, IIa and IIIa in 
the left, center and right panels of Fig. \ref{DAvsDI_ec}, respectively.

For Model Ia, two results agree very well although the oscillation period of the DA approach is slightly longer than that of the direct integration (DI) method. For Model IIa, the DA approach gives 
a regular periodic oscillations, but the DI method shows 
irregular chaotic oscillations. Two results do not agree well, although the maximum values of the eccentricity are almost the same.
For Model IIIa, two results agree almost completely.
For other models, we also find the similar results.

We conclude that the DA approach for Models I and III 
may give good results although the period of oscillations deviates slightly. 
On the other hand, for Model II, which shows chaotic feature in the KL oscillations, the DA approach 
does not give correct results.
\vskip 5cm

\begin{figure}[h]
\begin{center}
\includegraphics[width=5cm]{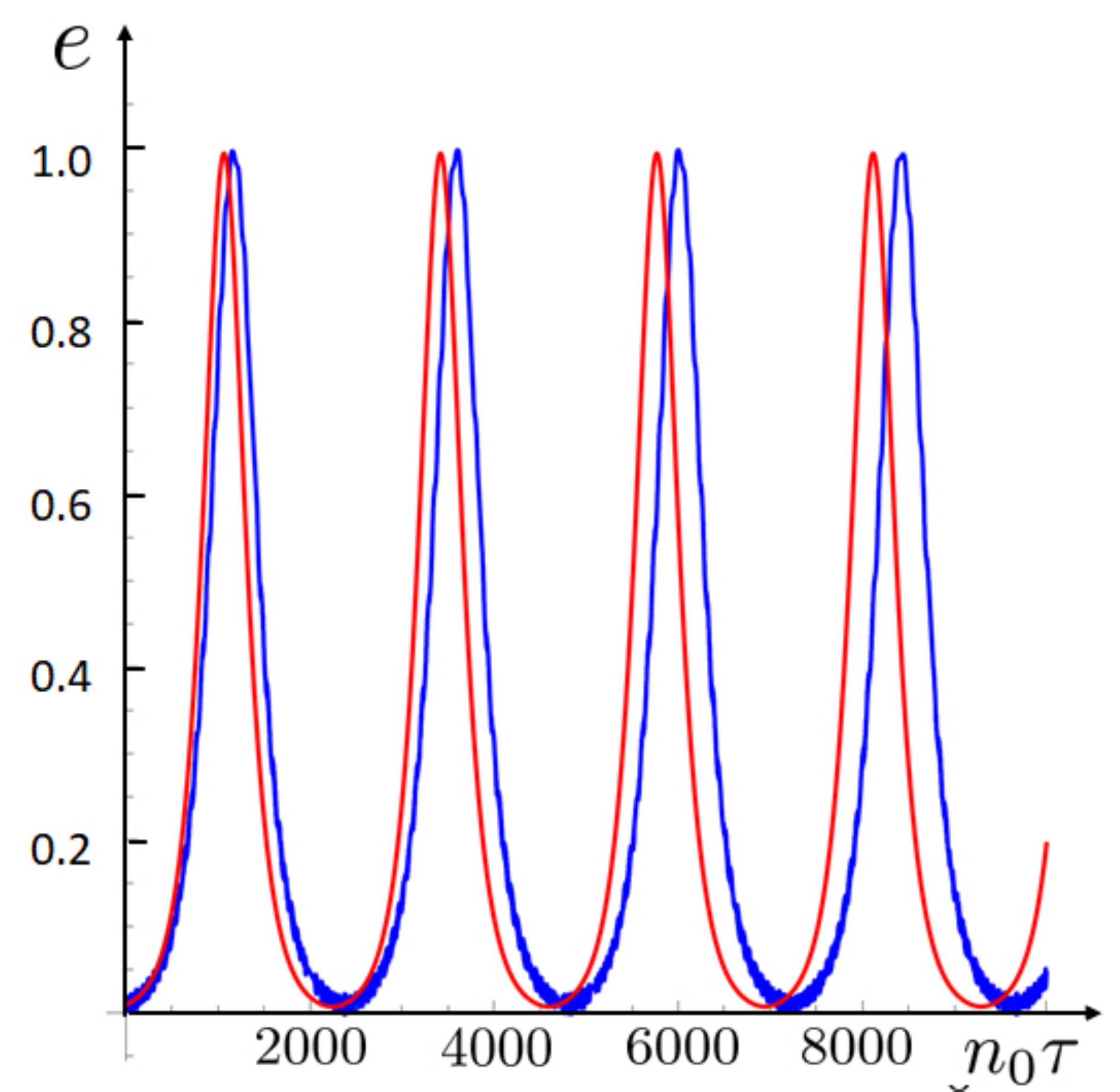}~~~~\includegraphics[width=5cm]{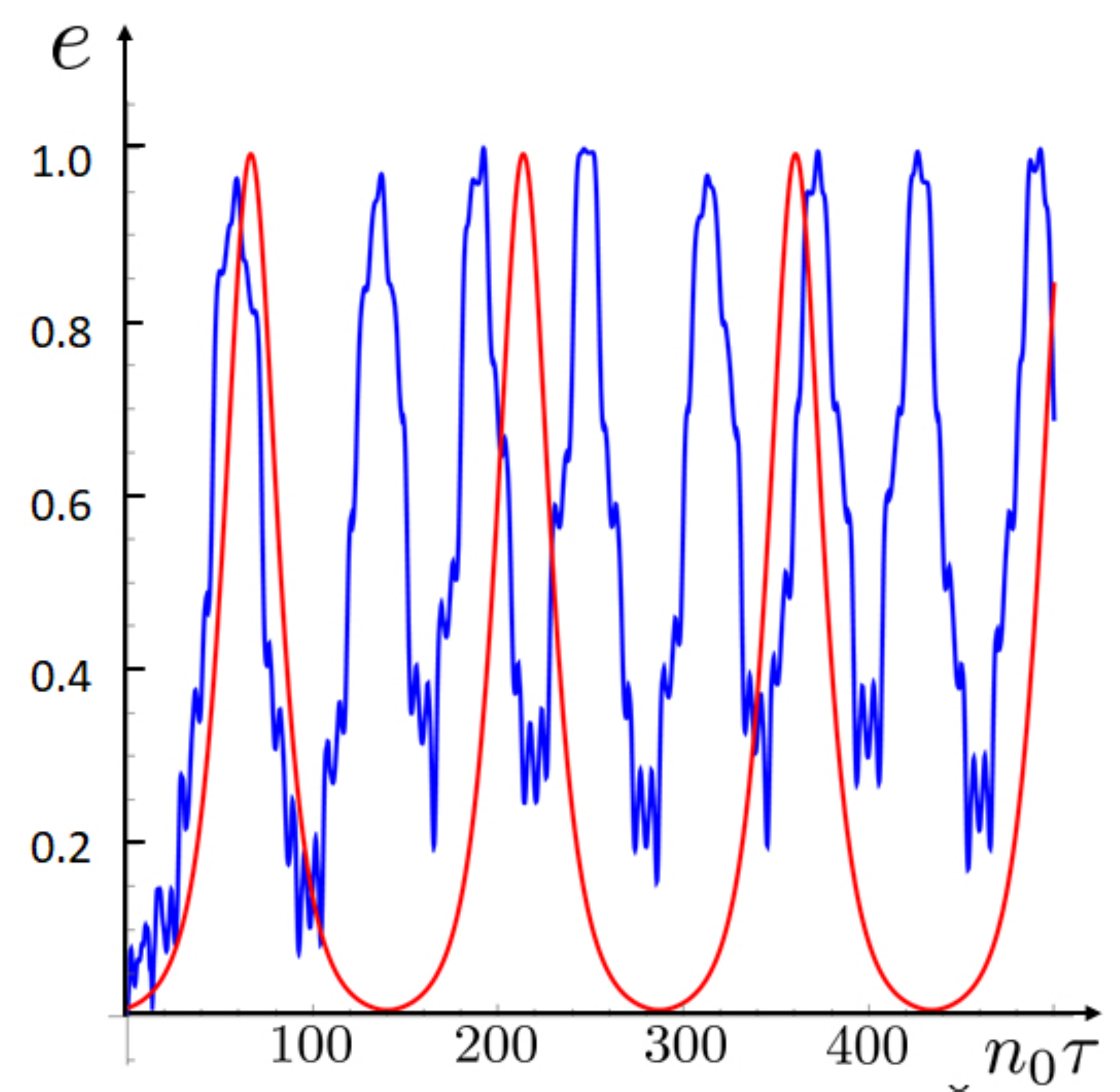}~~~~\includegraphics[width=5cm]{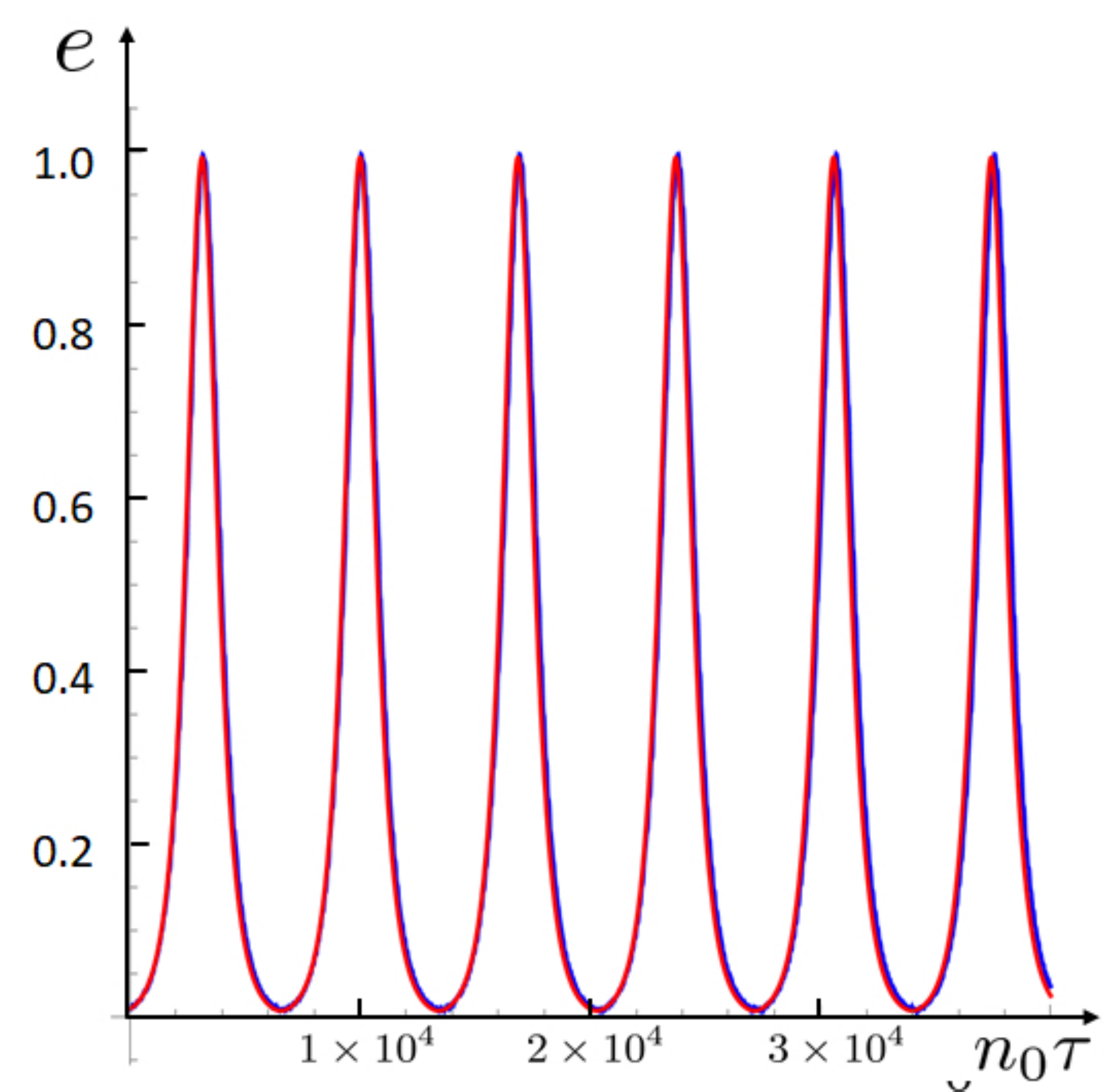}
\caption{The evolution of the 
eccentricity for Models Ia (left), IIa (center) and IIIa (right).
The red curves denote the results by the double-averaging (DA) approach, while the blue ones by the direct integration (DI) method.
For Models Ia and IIIa, two results agree well, but for Model IIa, they do not agree well.}
\label{DAvsDI_ec}
\end{center}
\end{figure}
\end{widetext}
\section{Concluding Remarks}
In this paper, we discuss motion of a binary system near SMBH.
Using Fermi-normal coordinates, we set up Newtonian self-gravitating system in the local 
proper reference frame.
Assuming a circular geodesic observer around a Schwarzschild SMBH, 
we write down the equations of motion of a binary.
To remove the interaction terms between the CM of a binary and its relative coordinates, 
we introduce a small acceleration of the observer.
As a result, the CM follows the observer's orbit, but its motion deviates from an exact 
geodesic.
Since the relative motion is decoupled from the system, we first solve it, and  then 
find the motion of the CM by the perturbation equations with the small acceleration, which is given by the relative motion. 

We show that there appears the KL oscillations when a binary is compact enough and 
the inclination angle is larger than the critical value.
If the firmness parameter $\mathfrak{f}$ is larger than a few hundreds,
the oscillations are regular.
However when  $\mathfrak{f}$ is around a few tens, 
although we find the stable KL oscillations, but the oscillations 
become irregular both in the oscillation period and in the amplitude.
Especially, if the initial inclination is large, we find an orbital flip.

One of the most interesting and important subjects of a binary system near SMBH is
the gravitational waves. When the KL oscillations is found in a binary motion, 
we expect a large amount of the GW emission because the eccentricity becomes large.
The large eccentricity also provides much higher frequencies than that from a circular binary~\cite{Lisa2019,Hoang2019,Gupta_2020}.
Another interesting point on the GWs from the KL oscillations is 
that the large amount of the GW emission repeats periodically
with the KL oscillation time scale.
It is a good advantage in the observations because we have 
a certain preparation time for next observations.

There are two GW sources in a hierarchical triple system: One is the GWs from an inner binary and the other is those from the outer binary.
The time scale of the emission of gravitational waves from a circular binary is evaluated as~\cite{MMbook}
 \beann
 \tau_{\rm GW}={5\over 256}{c^5 R_0^4\over G^3 
 m^2\mu}
 \,,
  \enann
  where $m=m_1+m_2$ and $\mu=m_1m_2/(m_1+m_2)$ and $R_0$ is the initial distance.
  Hence, the ratio of the time scale of an outer binary to that of an inner binary  
  is
  \beann
  {\tau_{\rm outer}\over \tau_{\rm inner}}={m_1m_2\over M^2}
  \left({\mathfrak{r}_0\over \ell_{\rm binary}}\right)^4.
  \enann
In the present model, as we discuss in \S. \ref{validity}, 
we have some constraints. From stability condition of a binary (\ref{stability_tidal}), we have 
\beann
{\ell_{\rm binary}\over \mathfrak{r}_0} \lsim \left({m_1+m_2\over 2M}\right)^{1/3}
\,,
\enann
while from the validity of Newtonian dynamics, we find
\beann
{\ell_{\rm binary}\over \mathfrak{r}_0}\gsim\left({m_1+m_2\over 2M}\right){\mathfrak{r}_g\over \mathfrak{r}_0}
\,.
\enann

If  a binary 
exists near the ISCO radius ($\mathfrak{r}_0\sim 3 \mathfrak{r}_g$), 
we find 
   \beann
2\times 10^{-5} \lsim {\tau_{\rm outer}\over \tau_{\rm inner}}\lsim 8\times 10^{16}
\,.
  \enann

Hence, in most cases,  gravitational waves from the outer orbit are less effective compared with those from the inner binary.
However when the binary is close to instability range, it is not the case.
In fact, in our examples discussed in this paper, 
if we assume a circular binary, we find 
\beann
  {\tau_{\rm outer}\over \tau_{\rm inner}}\sim 
  \left\{
  \begin{array}{cl}
 3\times 10^{-3} &{\rm (Model ~I)}
  \\
 8\times 10^{-5} &{\rm (Model ~II)}
  \\
 8\times 10^{-3} &{\rm (Model ~III)}
  \\
  \end{array}
  \right.
 \enann
The GWs from the outer binary become much larger than those from the inner binary. However if there exists the Kozai-Lidov oscillation,
the emission time scale is reduced by the factor~\cite{Peters-Mathews} 
\beann
F(e_{\rm in})\approx  {768\over  429 }\left(1-e_{\rm in}^2\right)^{7/2},
\enann
when $e_{\rm in}\approx 1$.
As a result, the GWs from the inner binary may become larger that those from the outer binary.

In recent years, three-body systems and the emission of GWs from them have received significant attention~\cite{Amaro-Seoane2010, Antonini2012,hoang18,Antonini2016,Meiron2017,Robson2018,Lisa2018,Lisa2019,Hoang2019,Loeb2019,Gupta_2020}. Our future work will involve evaluating the GWs from the present hierarchical triple setting using the black hole perturbation approach, since near the ISCO radius the quadrupole formula may not be valid~\cite{PhysRevD.103.L081501}.

In this paper, we assume that 
the CM of a binary moves along a circular orbit, but 
an eccentric orbit is interesting to be studied since the Kozai-Lidov oscillation may be modulated on a longer timescale\cite{Lithwick:2011hh,Katz:2011hn,Naoz:2011mb,Li2014ApJ,Liu2015MNRAS}.
However, for such a highly eccentric orbit, 
the present proper reference frame expanded 
up to the second order of the spatial coordinates $x^{\hat a}$
may not be sufficient.
We may need higher-order terms in the metric,
where the derivatives of the Riemann curvature appear\cite{Nesterov_1999,Delva:2011abw}.
Although the basic equations are very complicated, 
such an extension is straightforward.

Another natural direction would be an extension to a rotating SMBH that may allow us to study the precession of the binary orbit around the Kerr black hole. Such systems can reveal the impact of spin on GWs emitted from a nearby binary. Recent research \cite{camilloni2023tidal} considers the secular dynamics of the binary system distorted by a much larger Kerr black hole's tidal forces. This is done by deriving the magnetic and electric tidal moment at quadrupole orders.


\begin{acknowledgments}
We thank  Luc Blanchet, Vitor Cardoso, Eric Gourgoulhon and Haruka Suzuki for
useful discussions. This work was supported in part by JSPS KAKENHI
Grant Numbers  JP17H06359, JP19K03857 (K.M.), and by
JP20K03953 (H.O.). P.G. acknowledges support from C.V. Raman fellowship grant. K.M.
would like to acknowledges the Yukawa Institute for Theoretical
Physics at Kyoto University, where the present work was begun during
the Visitors Program of FY2022. 
He would also thank Niels Bohr Institute/Niels Bohr International Academy, The Max Planck Institute for Gravitational Physics (Albert Einstein Institute), The laboratory AstroParticle and Cosmology (APC), and CENTRA/Instituto Superior T\'ecnico, where most of this work was performed.

\end{acknowledgments}

\begin{widetext}
\appendix
\section{Coplanar Binary}
In this Appendix, we analyze motion of a coplanar binary, that is $\mathsf{z}=0\,, {\mathsf{p}}_z=0$.
It is an exact solution for Eqs. (\ref{pz}) and (\ref{eq_z}).
\subsection{Coplanar motion: Numerical results}
In the case of coplanar motion of a binary, the relative inclination angle $I$ is always zero. We then have the coupled equations (\ref{px}), (\ref{py}), (\ref{eq_x}), and  (\ref{eq_y}) for $\mathsf{x}$ and $\mathsf{y}$.
We first show numerical results for Model I ($\mathfrak{r}_0=3.5 \mathfrak{r}_g\,, \epsilon=0.1$) and II ($\mathfrak{r}_0=3.5 \mathfrak{r}_g\,, \epsilon=0.4$) 
in Fig. \ref{coplanar_orbit}.
We choose the initial conditions as $e_0=0.9$ and $\omega_0=60^\circ$. For Model I, as shown in Fig. \ref{coplanar_orbit} (left), the orbit 
is approximately elliptic, but the periapsis is rotating because of relativistic effect and the shift is quite regular.
On the other hand, for Model II, we find some irregular behaviours
in the orbit as shown in Fig.  \ref{coplanar_orbit}(right).
As discussed in the text, this model is not tightly bounded (the firmness parameter $\mathfrak{f}\sim 22$)
and the effects of the tidal force by SMBH is not so small. As a result, the system 
shows some chaotic feature. In the present coplanar case, 
the orbital shape is deformed from an ellipse.

\begin{figure}[h]
\begin{center}
\includegraphics[width=5.cm]{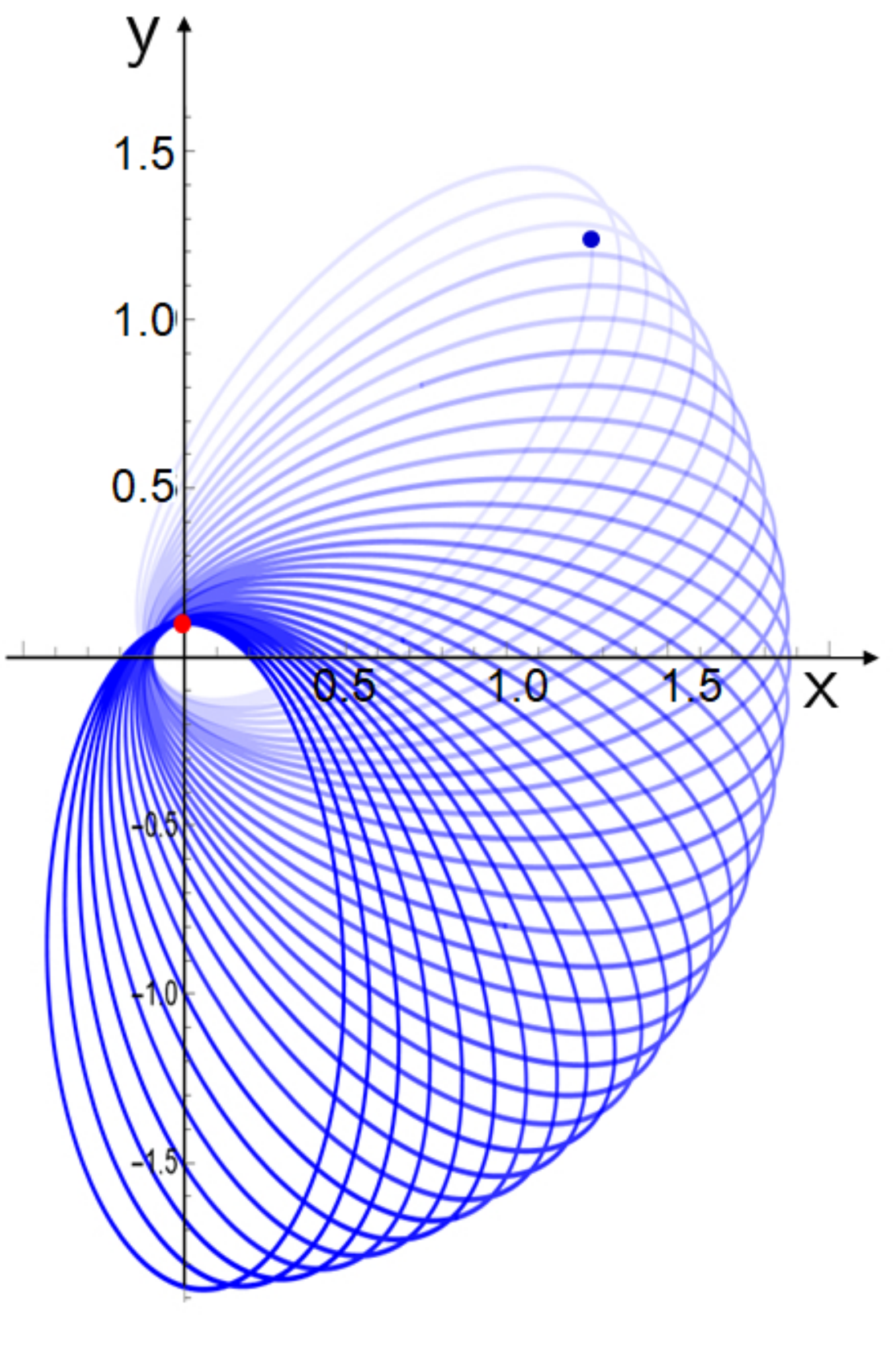}
~~~~~~~~~~~~~~~
\includegraphics[width=5.cm]{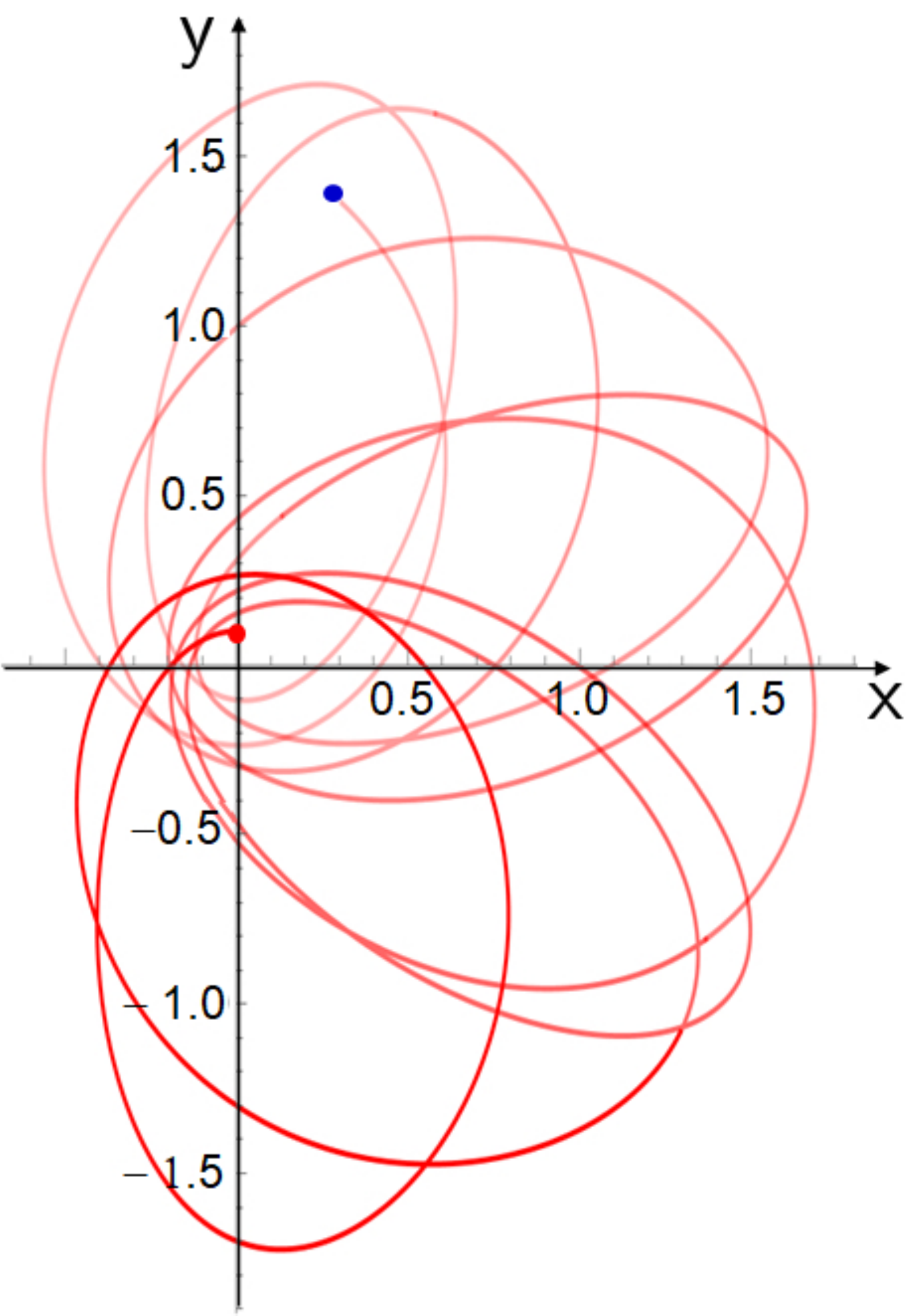}
\caption{The coplanar orbits for Model I (left) and for Model II (right).
We choose the initial eccentricity $e_0=0.9$.
The orbits start from the red point, and end at the blue point
[($n_0\tau=200$ (Model I) and $n_0\tau=50$ (Model II)].
}
\label{coplanar_orbit}
\end{center}
\end{figure}

It is further confirmed by the time evolution of the eccentricity, 
which is given  in Fig.~\ref{coplanar_eccentricity}.
 For Model I, the eccentricity is almost constant ($e\sim 0.9$).
On the other hand,  for Model II,  it oscillates irregularly 
between $e\sim 0.7$ and $0.9$.
Although the orbital shape is not well-approximated
by an ellipse, we evaluate the eccentricity by the osculating orbit. 
If we use the averages eccentricity over one cycle, the 
eccentricity oscillates between $\langle e \rangle\sim 0.75$ and $0.87$
, which is slightly different from the values in Fig. \ref{coplanar_eccentricity}.

\begin{figure}[h]
\begin{center}
\includegraphics[width=6cm]{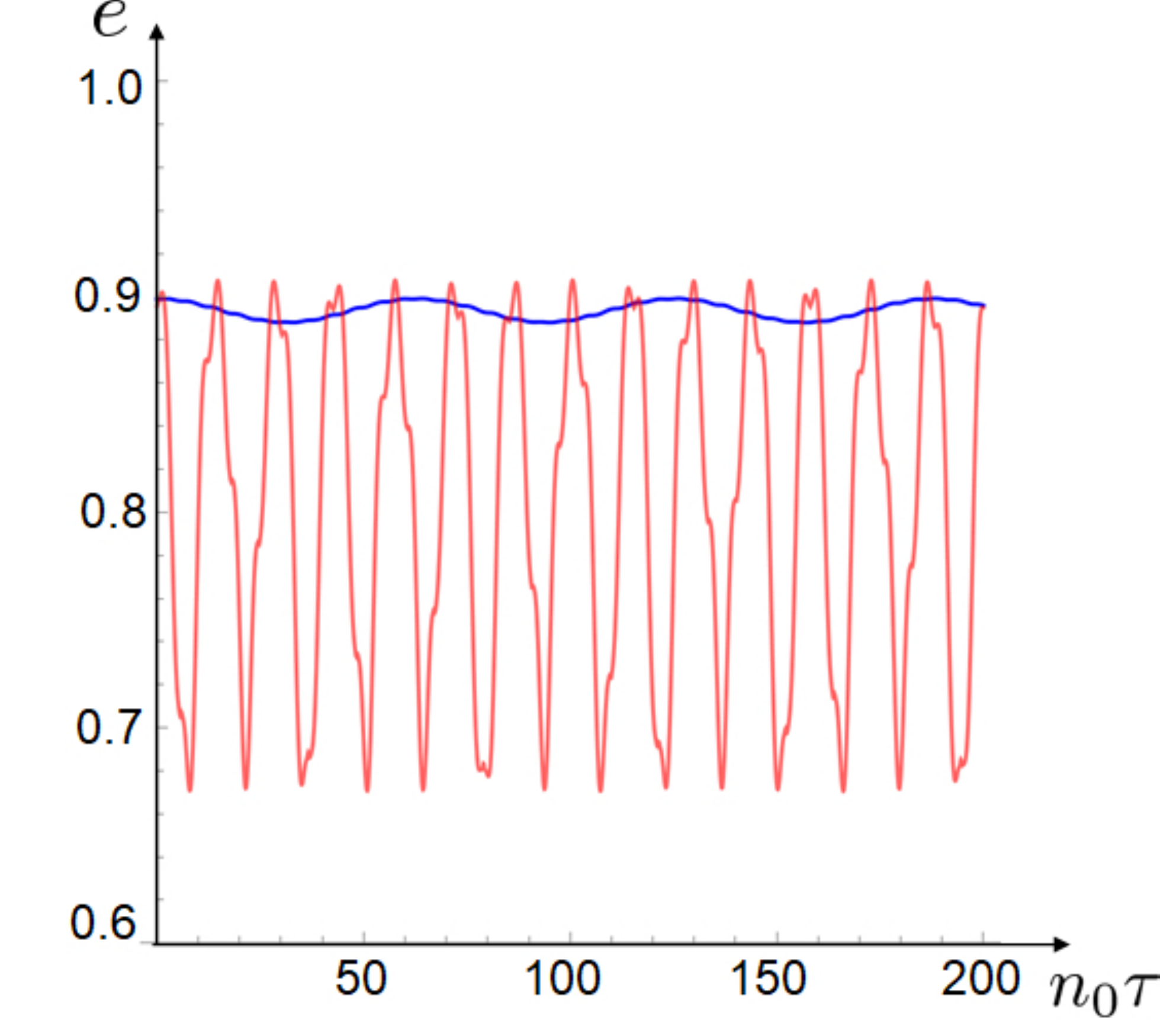}
\caption{Time evolution of the eccentricity for Model I (the blue curve)
and for Model II (the red curve). The eccentricity for Model I is almost constant ($e\approx 0.9$), while that for Model II oscillates irregularly between $e\sim 0.7$ and $0.9$.}
\label{coplanar_eccentricity}
\end{center}
\end{figure}

\subsection{Circular motion}
There exists an exact circular motion of a binary as follows: Assuming a circular solution as 
\beann
\zeta\equiv \mathsf{x}+i \mathsf{y}=\rho_0 \exp\left[
i\theta(\tau)\right],
~~~~{\rm with}~~\theta(\tau)=\bar \theta(\tau)+\mathfrak{w}_{\rm dS}\tau,
\enann
where the radius $\rho_0$ is constant, 
we find two equations  for $\bar \theta$ as
\bea
&&
-\dot{\bar\theta}^2+{G(m_1+m_2)\over \rho_0^3}-{\mathfrak{r}_g\over 4\mathfrak{r}_0^3}
(1+3\mathfrak{r}_0^2\mathfrak{w}_0^2)-{3\mathfrak{r}_g\over 4\mathfrak{r}_0^3}
(1+\mathfrak{r}_0^2\mathfrak{w}_0^2)\cos(2(\bar\theta+2\mathfrak{w}_{\rm R}\tau))=0
\label{eq:R},
\\
&&\ddot{\bar\theta}+{3\mathfrak{r}_g\over 4\mathfrak{r}_0^3}
(1+\mathfrak{r}_0^2\mathfrak{w}_0^2)\sin(2(\bar\theta+2\mathfrak{w}_{\rm R}\tau))=0
\label{eq:theta}.
\ena

Since the derivative of Eq. (\ref{eq:R}) with respect to $\tau$ gives Eq. (\ref{eq:theta})
unless $\bar \theta=0$, we 
first solve Eq. (\ref{eq:theta}). 
Setting $\eta\equiv 2(\bar\theta+2\mathfrak{w}_{\rm R}\tau)$, we find
\beann
\ddot{\eta}+{3\mathfrak{r}_g\over 2\mathfrak{r}_0^3}
(1+\mathfrak{r}_0^2\mathfrak{w}_0^2)\sin\eta=0
\,,
\enann
which can be integrated as
\beann
\dot{\eta}^2-{3\mathfrak{r}_g\over \mathfrak{r}_0^3}
(1+\mathfrak{r}_0^2\mathfrak{w}_0^2)\cos\eta={\rm constant}
\,.
\enann
By use of this equation, we can eliminate the term with $\cos\eta$ in Eq.(\ref{eq:R}), resulting in $\dot\eta$ and $\dot{\bar\theta}$ to be constant.
It follows that $\cos\eta=\cos(2(\bar\theta+2\mathfrak{w}_R\tau))$ is constant.
We then obtain a solution such that 
\beann
\bar \theta+2\mathfrak{w}_{\rm R}\tau={n\over 2}\pi 
\,,
\enann
where $n$ is an integer.
Eq. (\ref{eq:R}) becomes 
\beann
-4\mathfrak{w}_{\rm R}^2=-{G(m_1+m_2)\over \rho_0^3}+{\mathfrak{r}_g\over 4\mathfrak{r}_0^3}
(1+3\mathfrak{r}_0^2\mathfrak{w}_0^2)+{3\mathfrak{r}_g\over 4\mathfrak{r}_0^3}
(1+\mathfrak{r}_0^2\mathfrak{w}_0^2) (-1)^n
\,.
\enann
Since $\mathfrak{w}_{\rm R}=\sqrt{\mathfrak{r}_g/2\mathfrak{r}_0^3}$,
we obtain two analytic solutions as follows:\\
When $n$ is even,
\bea
\rho_0&=&\rho_0^{(+)}\equiv \sqrt[3]{(m_1+m_2)\over 2M(1+\mathfrak{r}_0^2\mathfrak{w}_0^2)}~\mathfrak{r}_0,
\label{sol:Re}\\
\theta&=&\theta^{(+)}\equiv \bar\theta^{(+)}+\mathfrak{w}_{\rm dS}\tau=
(\mathfrak{w}_{\rm dS}-2\mathfrak{w}_{\rm R})\tau+m\pi~~(m\in \mathbb{Z})
\label{sol:thetae}\,,
\ena
while when $n$ is odd,
\bea
\rho_0&=&\rho_0^{(-)}\equiv \sqrt[3]{(m_1+m_2)\over M(1+\mathfrak{r}_0^2\mathfrak{w}_0^2)}~\mathfrak{r}_0,
\label{sol:Ro}\\
\theta&=&\theta^{(-)}\equiv \bar\theta^{(-)}+\mathfrak{w}_{\rm dS}\tau=(\mathfrak{w}_{\rm dS}-2\mathfrak{w}_{\rm R})\tau 
+\left(m+{1\over 2}\right)\pi~~(m\in \mathbb{Z}).
\label{sol:thetao}
\ena
Using these solutions, we can find the analytic solution for 
the motion of the CM of a binary.
The relative coordinate $(x, y,z) $ of a binary in the rotating proper frame are given by
\beann
x&=&\mathsf{x}\cos(\mathfrak{w}_{\rm R}\tau)-\mathsf{y}\sin(\mathfrak{w}_{\rm R}\tau)=\rho_0^{(\pm)}
\cos[(\mathfrak{w}_{\rm dS}-\mathfrak{w}_{\rm R})\tau+\phi_0^{(\pm)}],
\\
y&=&\mathsf{x}\sin(\mathfrak{w}_{\rm R}\tau)+\mathsf{y}\cos(\mathfrak{w}_{\rm R}\tau)=\rho_0^{(\pm)}
\sin[(\mathfrak{w}_{\rm dS}-\mathfrak{w}_{\rm R})\tau+\phi_0^{(\pm)}],
\\
z&=&\mathsf{z}=0,
\enann
where $\rho_0^{(\pm)}$ are given by the previous solutions, and $\phi_0^{(+)}=m\pi$ and $\phi_0^{(-)}=\left(m+{1\over 2}\right)\pi$.
We then find the perturbation equation for $\mathfrak{r}_{(1)}$ as
\bea
{d^2\mathfrak{r}_{(1)}\over d\tau^2}
+ k\mathfrak{r}_{(1)}+C^{(\pm)}\cos^2[(\mathfrak{w}_{\rm dS}-\mathfrak{w}_{\rm R})\tau+\phi_0^{(+)}]=0
\label{eq_r3}
\,,
\ena
where
\beann
C^{(\pm)}\equiv \left[A+(\mathfrak{w}_{\rm dS}-\mathfrak{w}_{\rm R}) B\right]\left(\rho_0^{(+)}\right)^2
\,.
\enann

Since two independent solutions of the homogeneous equation are $\sin k\tau$ and 
$\cos k\tau$,  introducing  two unknown functions $u(\tau)$ and $v(\tau)$,
we set 
\beann
\mathfrak{r}_{(1)}=u(\tau)\sin k\tau+v(\tau)\cos k\tau
\,.
\enann
Inserting this into Eq. (\ref{eq_r3}), we find
\beann
\ddot u \sin k\tau+\ddot v \cos k\tau+2 k\left(\dot u\cos k\tau
-\dot v \sin k\tau \right)+C^{(\pm)}\cos^2[(\mathfrak{w}_{\rm dS}-\mathfrak{w}_{\rm R})\tau+\phi_0^{(\pm)}]=0
\,.
\enann
We assume one constraint equation such that 
\bea
\dot u\sin k\tau
+\dot v \cos k\tau =0
\label{constraint1}
\,,
\ena
which yields
\beann
\ddot u\sin k\tau
+\ddot v \cos k\tau
+ k\left(\dot u\cos k\tau
-\dot v \sin k\tau\right)=0.
\enann
We then find
\bea
k \left(\dot u\cos k\tau
-\dot v \sin k\tau \right)+C^{(\pm)}\cos^2[(\mathfrak{w}_{\rm dS}-\mathfrak{w}_{\rm R})\tau+\phi_0^{(\pm)}]=0.
\label{eq_r4}
\ena
From Eqs. (\ref{constraint1}) and (\ref{eq_r4}), we obtain 
\beann
\dot u&=&-{C^{(\pm)}\over k}\cos k\tau \cos^2[(\mathfrak{w}_{\rm dS}-\mathfrak{w}_{\rm R})\tau+\phi_0^{(\pm)}],
\\
\dot v&=&{C^{(\pm)}\over k}\sin k\tau \cos^2[(\mathfrak{w}_{\rm dS}-\mathfrak{w}_{\rm R})\tau+\phi_0^{(\pm)}].
\enann
We can integrate these equations as
\beann
u&=&u_0-{C^{(\pm)}\over 2k}\left[{1\over k}\sin k\tau +{1\over 2}\left(
{\sin[(k+2(\mathfrak{w}_{\rm dS}-\mathfrak{w}_{\rm R}))\tau+2\phi_0^{(\pm)}]\over k+2(\mathfrak{w}_{\rm dS}-\mathfrak{w}_{\rm R})}+{\sin[(k-2(\mathfrak{w}_{\rm dS}-\mathfrak{w}_{\rm R}))\tau-2\phi_0^{(\pm)}]
\over k-2(\mathfrak{w}_{\rm dS}-\mathfrak{w}_{\rm R})}\right)\right],
\\
v&=&v_0-{C^{(\pm)}\over 2k}\left[{1\over k}\cos k\tau +{1\over 2}\left(
{\cos[(k+2(\mathfrak{w}_{\rm dS}-\mathfrak{w}_{\rm R}))\tau+2\phi_0^{(\pm)}]\over k+2(\mathfrak{w}_{\rm dS}-\mathfrak{w}_{\rm R})}+{\cos[(k-2(\mathfrak{w}_{\rm dS}-\mathfrak{w}_{\rm R}))\tau-2\phi_0^{(\pm)}]
\over k-2(\mathfrak{w}_{\rm dS}-\mathfrak{w}_{\rm R})}\right)\right],
\enann
where $u_0$ and $v_0$ are integration constants.
As a result, we obtain general solution as
\beann
\mathfrak{r}_{(1)}=-{C^{(\pm)}\over 2k^2}+u_0\sin k\tau+v_0\cos k\tau
-{C^{(\pm)}\over 2[k^2-4(\mathfrak{w}_{\rm dS}-\mathfrak{w}_{\rm R})^2]}
\cos\left[2(\mathfrak{w}_{\rm dS}-\mathfrak{w}_{\rm R})\tau+2\phi_0^{(\pm)}\right].
\enann
The initial conditions determines the integration constants $u_0$ and $v_0$.
For example, if we assume $\mathfrak{r}_{(1)}(0)=0$ and $\dot{\mathfrak{r}}_{(1)}(0)=0$,
we find
\bea
\mathfrak{r}_{(1)}=-{C^{(\pm)}\over k^2}\sin^2{k\tau\over 2}
+ {(-1)^n C^{(\pm)}\over k^2-4(\mathfrak{w}_{\rm dS}-\mathfrak{w}_{\rm R})^2}\sin \left[\left(\mathfrak{w}_{\rm dS}-\mathfrak{w}_{\rm R}+{k\over 2}\right)\tau \right]
\sin \left[\left(\mathfrak{w}_{\rm dS}-\mathfrak{w}_{\rm R}-{k\over 2}\right)\tau\right],
\,
\label{sol:r1}
\ena
where $n=2m$ or $2m+1$.
As a result, $\mathfrak{r}_{(1)}$ oscillates around zero.
As for the other variables $\mathfrak{t}_{(1)}\,,\mathfrak{\theta}_{(1)}\,,
\mathfrak{\varphi}_{(1)}$,
although some of them may diverge as $\tau\rightarrow\infty$,
no singularity appear in the evolution equations. Hence, we conclude that the coplanar circular orbit is linearly stable.


\section{Planetary equations for a binary system in Newtonian limit}
\label{planetary equation}

In order to understand our numerical results, it may be better to introduce 
the Lagrange planetary equations, which give time evolution of the orbital parameters
such as the semi-major axis, eccentricity and inclination. To derive the planetary equations, we treat 
 the proper Hamiltonian with unit mass $\mu=1$,
which is given by
\beann
\bar{{\cal H}}=\bar{{\cal H}}_0+\bar{{\cal H}}_1,
\enann
where
\beann
\bar{{\cal H}}_0
&=&{1\over 2} \bar{\vect{\mathsf{p}}}^2- {G (m_1+m_2)\over \mathsf{r}},
\\
\bar{{\cal H}}_1
&=&\bar{{\cal H}}_{1\mathchar`-{\rm dS}}+\bar{{\cal H}}_{1\mathchar`-\bar{\cal R}},
\enann
with
\beann
\bar{{\cal H}}_{1\mathchar`-{\rm dS}}&=&
\mathfrak{w}_{\rm dS}\left(\bar{\mathsf{p}}_{\mathsf{y}}\mathsf{x}-\bar{\mathsf{p}}_{\mathsf{x}}\mathsf{y}\right),
\\
\bar{{\cal H}}_{1\mathchar`-\bar{\cal R}}&=&{\mathfrak{r}_g\over 4\mathfrak{r}_0^3} 
\left[\mathsf{x}^2+\mathsf{y}^2+\mathsf{z}^2-3\left(1+\mathfrak{r}_0^2\mathfrak{w}_0^2\right)\left(\mathsf{x}\cos\mathfrak{w}_{\rm R}   \tau -\mathsf{y} \sin \mathfrak{w}_{\rm R}  \tau\right)^2
+3\mathfrak{r}_0^2\mathfrak{w}_0^2 \mathsf{z}^2 \right].
\enann
The position $\bf{\mathsf{r}}=(\mathsf{x},\mathsf{y},\mathsf{z})$ 
of a binary 
should be described in the non-rotating proper reference frame.
\\


For the unperturbed Hamiltonian $\bar{{\cal H}}_0$, it is just the same as that of a binary in Newtonian dynamics.
We find an elliptic orbit, which is described by
\bea
\mathsf{r}= {a(1-e^2)\over 1+e\cos f}
\label{elliptic_r}
\,,
\ena
where $\mathsf{r}, a, e$ and $f$ are the radial distance from the center of mass, the semi-major axis, the eccentricity, and the true anomaly.  This orbital plane is inclined
with the inclination angle $I$ from the equatorial plane in the proper reference frame.


Hence, the relative position $\bf{\mathsf{r}}=(\mathsf{x},\mathsf{y},\mathsf{z})$ of a binary is given by
the orbital parameters $(\omega\,,\Omega\,,  a\,, e\,, I\,, f)$ as Eq. (\ref{orbital_parameters}) with Eq. (\ref{elliptic_r}). Introducing the Delaunay variables
as
\beann
\left\{
\begin{array}{l}
\mathfrak{l}=n(t-t_0)
\\
\mathfrak{g}=\omega
\\
\mathfrak{h}=\Omega
\\
\end{array}
\right.
~~ {\rm and} \hskip .5cm
\left\{
\begin{array}{l}
\mathfrak{L}=\sqrt{G(m_1+m_2)a}
\\
\mathfrak{G}=\sqrt{G(m_1+m_2)a(1-e^2)}
\\
\mathfrak{H}=\sqrt{G(m_1+m_2)a(1-e^2)}\cos I
\\
\end{array}
\right.
\,,
\enann
where
\beann
n\equiv {2\pi \over  P}=\sqrt{{G(m_1+m_2)\over a^3}},
\enann
is the mean motion,
we find new unperturbed Hamiltonian as
\beann
\widetilde{\bar{\cal H}}_0=-{G^2(m_1+m_2)^2\over 2\mathfrak{L}^2}.
\enann
\\[-1em]

Including the perturbations $\bar{\cal H}_1$,
we obtain the Hamiltonian the Delaunay variables as
\beann
\widetilde{\bar{\cal H}}=\widetilde{\bar{\cal H}}_0+\bar{\cal H}_1
\,.
\enann

After some calculations, the Hamilton equations are reduced to 
\bea
\dot{a}&=&-{2\over na}{\partial \bar{\cal H}_1\over \partial \mathfrak{l}},
\label{LPeq_a}
\\
\dot{e}&=&{\sqrt{1-e^2}\over na^2 e}{\partial \bar{\cal H}_1\over \partial \omega}-{1-e^2\over na^2 e}{\partial \bar{\cal H}_1\over \partial \mathfrak{l}},
\label{LPeq_e}
\\
\dot I&=&{1\over na^2\sin I\sqrt{1-e^2}}{\partial \bar{\cal H}_1\over \partial \Omega}-{\cos I\over na^2\sin I\sqrt{1-e^2}}
{\partial \bar{\cal H}_1\over \partial \omega},
\label{LPeq_I}
\ena
\bea
\dot{\mathfrak{l}}&=&n+{2\over na}{\partial \bar{\cal H}_1\over \partial a}
+{1-e^2\over na^2 e}{\partial \bar{\cal H}_1\over \partial e},
\label{LPeq_ell}
\\
\dot \omega&=&-{\sqrt{1-e^2}\over na^2 e}{\partial \bar{\cal H}_1\over \partial e}+{\cos I\over na^2\sin I\sqrt{1-e^2}}{\partial \bar{\cal H}_1\over \partial I},
\label{LPeq_om}
\\
\dot \Omega&=&-{1\over na^2\sin I\sqrt{1-e^2}}{\partial \bar{\cal H}_1\over \partial I}.
\label{LPeq_Om}
\ena
The partial derivative $\partial/\partial \mathfrak{l}$ can be replaced by that of the true anomaly as
\beann
{\partial \over \partial \mathfrak{l}}=(1-e^2)^{-{3\over 2}}
\left(1+e\cos f \right)^2{\partial \over \partial f}.
\enann
Hence, once we find the perturbation Hamiltoninan $\bar{\cal H}_1$  in terms of the orbital parameters, we obtain the planetary equations.

The proper Hamiltonian is described by the orbital parameters by inserting the relation given in Eq.(\ref{orbital_parameters}) with Eq. (\ref{elliptic_r}).
We then find the perturbed Hamiltonian as
\beann
\bar{\cal H}_1=\bar{{\cal H}}_{1\mathchar`-{\rm dS}}+\bar{{\cal H}}_{1\mathchar`-\bar{\cal R}}
\,,
\enann
where
\bea
\bar{{\cal H}}_{1\mathchar`-{\rm dS}}&=&\mathfrak{w}_{\rm dS}\mathsf{r}^2(a, e, f)
\left\{n\cos I (1-e^2)^{-3/2}(1+e\cos f)^2 -\mathfrak{w}_{\rm dS}\left(\cos^2(\omega+f)+\sin^2(\omega+f)\cos^2 I \right)\right\}~~~~~~~
\label{H1_dS}\\
\bar{{\cal H}}_{1\mathchar`-\bar{\cal R}}&=&{\mathfrak{r}_g\over 4\mathfrak{r}_0^3}
\mathsf{r}^2(a, e, f)\Big\{1-3(1+ \mathfrak{r}_0^2\mathfrak{w}_0^2 )\left[\cos (\Omega+\mathfrak{w}_{\rm R} \tau)
\cos(\omega+f)-\sin(\Omega+\mathfrak{w}_{\rm R} \tau)\sin(\omega+f)\cos I 
\right]^2
\nn
&&~~~~~~~~~~~~~~~~~~~~~~~
+3 \mathfrak{r}_0^2\mathfrak{w}_0^2 \sin^2(\omega+f)\sin^2 I\Big\}
\,.
\label{H1_R}
\ena
We then obtain the planetary equations for the present hierarchical triple system from Eqs. (\ref{LPeq_a}) - (\ref{LPeq_Om}).

\subsection{Double-averaging approach}
Here, instead of solving the Lagrange planetary equations themselves, 
which is equivalent to our numerical methods in the text,
we take average of the perturbed Hamiltonian over two periods,
the inner and outer orbital periods, 
and then analyze the simplified equations, because we are interested in the long-time behaviour of the present system
such as the Kozai-Lidov mechanism.

The double-averaged Hamiltonian is defined by
\beann
\langle\langle\bar{\cal H}_1 \rangle\rangle\equiv {1\over 2\pi}\int_0^{2\pi}
d\mathfrak{l}_{\rm out}\left( {1\over 2\pi}\int_0^{2\pi}
d\mathfrak{l}\, \bar{\cal H}_1\right)
\enann
Since the outer orbit is circular, we find that $\mathfrak{l}_{\rm out}=f_{\rm out}=\mathfrak{w}_0\tau$.
We also have
\beann
d\mathfrak{l}={1\over \sqrt{1-e^2}}\left({\mathsf{r}\over a}\right)^2 df
\,.
\enann
Inserting Eqs. (\ref{H1_dS}) and (\ref{H1_R}) into the above integrals, 
we find the doubly-averaged Hamiltonian as
\bea
\langle\langle\bar{\cal H}_1\rangle\rangle
&=&\mathfrak{w}_{\rm dS}na^2 \sqrt{1-e^2}\cos I
-{a^2\over 8}\Big\{
(2+3e^2)\Big[\mathfrak{w}_{\rm dS}^2(3+\cos 2I)+
{\mathfrak{r}_g\over 8\mathfrak{r}_0^3}(1+3\mathfrak{r}_0^2\mathfrak{w}_0^2)(1+3\cos 2I)\Big]
\nn
&&
~~~~~~~~~~~~~~~~~~~~~~~~~~~~~~~~~
+10e^2\sin^2 I\cos 2\omega\left[
\mathfrak{w}_{\rm dS}^2+{3\mathfrak{r}_g\over 8\mathfrak{r}_0^3}(1+3\mathfrak{r}_0^2\mathfrak{w}_0^2)\right]
\Big\}.
\label{DA_perturbed_H1}
\ena
Using the doubly-averaged Hamiltonian Eq.(\ref{DA_perturbed_H1}), we obtain the Lagrange planetary equations as
\bea
\dot e&=&{5\over 4}  \left(\mathfrak{w}_{\rm dS}^2+{3\mathfrak{r}_g\over 8\mathfrak{r}_0^3}(1+3\mathfrak{r}_0^2\mathfrak{w}_0^2)
\right){e\sqrt{1-e^2}\over n}\left(1-\cos 2I\right)\sin (2\omega),
\label{PE_Sch_e}
\\
\dot I&=&-{5\over 4}  \left(\mathfrak{w}_{\rm dS}^2+{3\mathfrak{r}_g\over 8\mathfrak{r}_0^3}(1+3\mathfrak{r}_0^2\mathfrak{w}_0^2)
\right){e^2\over n\sqrt{1-e^2}}\sin 2I \sin(2\omega),
\label{PE_Sch_I}
\\
\dot \omega&=&
{1\over 4n}\left(\mathfrak{w}_{\rm dS}^2+{3\mathfrak{r}_g\over 8\mathfrak{r}_0^3}(1+3\mathfrak{r}_0^2\mathfrak{w}_0^2)
\right)
\Big{[}\sqrt{1-e^2} [3+5\cos 2I +5\left(1-\cos 2I \right)\cos 2\omega]
\nonumber
\\
&&+{5e^2\over \sqrt{1-e^2} }(1+\cos 2I)(1-\cos2\omega)\Big{]}
+{2\sqrt{1-e^2} \over n}\mathfrak{w}_{\rm dS}^2,
\label{PE_Sch_omega}
\\
\dot \Omega&=&\mathfrak{w}_{\rm dS}+
{\cos I\over 2n\sqrt{1-e^2}}\left(\mathfrak{w}_{\rm dS}^2+{3\over 8}{\mathfrak{r}_g\over \mathfrak{r}_0^3}(1+3\mathfrak{r}_0^2\mathfrak{w}_0^2)
\right)
\left[-(2+3e^2)
+5e^2\cos(2\omega)\right].
\label{PE_Sch_Omega}
\ena
The semi-major axis $a$ is constant in the present approximation.
Also, we can easily check from Eqs. (\ref{PE_Sch_e}) and (\ref{PE_Sch_I}) that
\beann
{d\over d\tau}\left(\sqrt{1-e^2}\cos I\right)=0
\,,
\enann
which corresponds to conservation of  the $z$-component of the angular momentum.

\subsection{KL oscillations}
\label{DA_KL_oscillations}

Introducing a ``potential" by
$V_S\equiv - \langle\langle\bar{\cal H}_1\rangle\rangle$,
we rewrite the above planetary equations as
\bea
\dot{e}&=&-{\sqrt{1-e^2}\over na^2 e}{\partial {V_S}\over \partial \omega},
\label{LPeq_e1}
\\
\dot I&=&{\cos I\over na^2\sin I\sqrt{1-e^2}}
{\partial {V_S}\over \partial \omega},
\label{LPeq_I1}
\\
\dot \omega&=&{\sqrt{1-e^2}\over na^2 e}{\partial {V_S}\over \partial e}-{\cos I\over na^2\sin I\sqrt{1-e^2}}{\partial {V_S}\over \partial I},
\label{LPeq_om1}
\\
\dot \Omega&=&{1\over na^2\sin I\sqrt{1-e^2}}{\partial {V_S}\over \partial I}
\label{LPeq_Om1}
\,.
\ena
We obtain the closed form of a set of the differential equations 
for $e\,, I\,,$ and $\omega$ by Eqs. (\ref{LPeq_e1}), (\ref{LPeq_I1})
and (\ref{LPeq_om1}). 
It gives several properties of KL oscillations such as
the oscillation amplitude of the eccentricity 
and the oscillation time scale as analyzed in Newtonian and 1PN hierarchical triple system~\cite{Suzuki:2020zbg}. 

The potential is written by use of 
$\eta \equiv  \sqrt{1-e^2}$ and $\mu_I \equiv  \cos I$
as
\beann
V_S&\equiv &
-\langle\langle\bar{\cal H}_1\rangle\rangle
= {a^2 \mathfrak{r}_g (1+3\mathfrak{r}_0^2\mathfrak{w}_0^2)\over 32 \mathfrak{r}_0^3}v_S(\eta, \mu_I),
\enann
where
\beann
v_S(\eta, \mu_I)&\equiv & 2(-1+3\mu_I^2\eta^2)(1+\alpha_{\rm dS})+12 C_{KL}
+4\alpha_{\rm dS}\left(2-{3n\over\mathfrak{w}_{\rm dS}}\mu_I\eta\right),
\enann
with
\beann
\alpha_{\rm dS}&\equiv&{8\mathfrak{w}_{\rm dS}^2\mathfrak{r}_0^3\over 3\mathfrak{r}_g(1+3\mathfrak{r}_0^2\mathfrak{w}_0^2)},
\\
C_{\rm KL}&\equiv& (1-\eta^2)\left[(1+2\alpha_{\rm dS})-{5\over 2}(1+\alpha_{\rm dS})(1-\mu_I^2)\sin^2\omega \right].
\enann

Note that when $\alpha_{\rm dS}=0$, we find the same equations for Newtoninan hierarchical triple system with quadrupole approximation.
The terms with $\alpha_{\rm dS}$ give relativistic corrections.
\\

Introducing the normalized time $\tilde \tau$, which is defined by
\beann
\tilde \tau\equiv
{\tau \over \tau_{\rm KL}} ,
\enann
with the typical KL time scale 
\beann
\tau_{\rm KL}\equiv {32 n \mathfrak{r}_0^3\over \mathfrak{r}_g(1+3\mathfrak{r}_0^2\mathfrak{w}_0^2)}
\left(\sim {P_{\rm out}^2\over P_{\rm in}}\right)
\,,
\enann
the above planetary equation is rewritten as
\beann
{d\eta\over d\tilde \tau}&=&{\partial {v_S}\over \partial \omega},
\\
{1\over \mu_I}{d\mu_I\over d\tilde \tau}&=&-{1\over \eta}{\partial {v_S}\over \partial \omega},
\\
{d\omega\over d\tilde \tau}&=&
-{\partial {v_S}\over \partial \eta}+{\mu_I\over \eta}
{\partial {v_S}\over \partial \mu_I}.
\enann
From these equations, we can easily show that 
\beann
{d(\mu_I\eta) \over d\tilde \tau}=0
~\,,~~~
{dv_S\over d\tilde \tau}=0
\,,
\enann
which means there exist two conserved quantities $\vartheta\equiv \mu_I\eta$ and $C_{\rm KL}$ just as the Newtoninan and 1PN hierarchical triple system under dipole approximation.
Using these two conserved quantities, we obtain a single equation for $\eta$ as
\beann
{d\eta^2\over d\tilde \tau}=-24\sqrt{2}\sqrt{f(\eta^2)g(\eta^2)},
\enann
with
\beann
f(\eta^2)&\equiv&(1+2\alpha_{\rm dS})(1-\eta^2)-C_{\rm KL},
\\
g(\eta^2)&\equiv&-5(1+\alpha_{\rm dS})\vartheta^2
+\left[5(1+\alpha_{\rm dS})\vartheta^2+3+\alpha_{\rm dS}+2C_{\rm KL}\right]\eta^2
-(3+\alpha_{\rm dS})\eta^4.
\enann

Setting $\xi=\eta^2$, we find
\beann
{d\xi\over d\tilde \tau}=-24\sqrt{2(1+2\alpha_{\rm dS})(3+\alpha_{\rm dS})}\sqrt{(\xi-\xi_0)(\xi-\xi_+)(\xi-\xi_-)},
\enann
where
\beann
\xi_0&=&1-{C_{\rm KL}\over 1+2\alpha_{\rm dS}},
\\
\xi_\pm&=&{1\over 2}\Big[\left(1+{5(1+\alpha_{\rm dS})\over 3+\alpha_{\rm dS}}
\vartheta^2+{2\over 3+\alpha_{\rm dS}}C_{\rm KL}\right)
\pm \sqrt{\left(1+{5(1+\alpha_{\rm dS})\over 3+\alpha_{\rm dS}}
\vartheta^2+{2\over 3+\alpha_{\rm dS}}C_{\rm KL}\right)^2-{20(1+\alpha_{\rm dS})\over 3+\alpha_{\rm dS}}\vartheta^2
}\Big],
\enann
are the solutions of $f(\xi)=0$ and $g(\xi)=0$, respectively.
\\

We can find  the relativistic 
corrections with $\alpha_{\rm dS}$, which is evaluated as 
\beann
\alpha_{\rm dS}&=&{3\mathfrak{r}_g^2\over \mathfrak{r}_0^2\left(1+\sqrt{1-{3\mathfrak{r}_g\over 2\mathfrak{r}_0}}\right)^2}
\leq 
{1\over 3\left(1+{1\over \sqrt{2}}\right)^2} \approx 0.114382.
\enann
The equality is found in the case of the ISCO radius ($\mathfrak{r}_0=3\mathfrak{r}_g)$.
\\

Analyzing the above equation, we find that there exists KL oscillation in this system just the same as in Newtonian hierarchical triple system, and we can classify the KL solutions by the sign of $C_{\rm KL}$ into two cases: 
(1) $C_{\rm KL}>0$ (rotation) and (2) $C_{\rm KL}<0$ (libration).

\subsubsection{$C_{\rm KL}>0$ {\rm (rotation)}}
In this case, $0<\xi_-<1<\xi_+$ and $0<\xi_0<1$.
This is possible if 
\beann
0<C_{\rm KL}<1+2\alpha_{\rm dS}
\,.
\enann

Hence we find the maximum and minimum values of the eccentricity as
\beann
e_{\rm max}=\sqrt{1-\xi_-}
~~,~~
e_{\rm min}=\sqrt{1-\xi_0}.
\enann

The KL time scale is given by 
\bea
T_{\rm KL}=\tau_{\rm KL}\, \mathfrak{T}_{\rm KL}^{\rm (rot)},
\ena
where
\bea
&&\mathfrak{T}_{\rm KL}^{\rm (rot)}\equiv {1\over 12\sqrt{2(1+2\alpha_{\rm dS})(3+\alpha_{\rm dS})}}
\times \int_{\xi_-}^{\xi_0}d\xi {1\over\sqrt{(\xi-\xi_0)(\xi-\xi_+)(\xi-\xi_-)}}
\,.~~~~~~~
\label{TKL_rotation}
\ena
\\

\subsubsection{$C_{\rm KL}<0$ {\rm (libration)}}
Since $0<\xi_-<\xi_+<1$ and $\xi_0<0$ in this case, 
 we find 
\beann
e_{\rm max}=\sqrt{1-\xi_-}\,,~~~
e_{\rm min}=\sqrt{1-\xi_+}.
\enann
It occurs when 
\beann
 -{3+\alpha_{\rm dS}\over 2}<C_{\rm KL}<0\,,~~~{\rm and}~~~
\vartheta<{(\sqrt{3+\alpha_{\rm dS}}-\sqrt{-2C_{\rm KL}})\over \sqrt{5(1+\alpha_{\rm dS})}}
\,.
\enann

The KL time scale is given by 
\bea
T_{\rm KL}=\tau_{\rm KL}\, \mathfrak{T}_{\rm KL}^{\rm (lib)},
\ena
where
\bea
\mathfrak{T}_{\rm KL}^{\rm (lib)}\equiv {1\over 12\sqrt{2(1+2\alpha_{\rm dS})(3+\alpha_{\rm dS})}}
\times \int_{\xi_-}^{\xi_+}d\xi {1\over\sqrt{(\xi-\xi_0)(\xi-\xi_+)(\xi-\xi_-)}}
\,.~~~~~~~
\label{TKL_libration}
\ena
\end{widetext}

\begin{figure}[h]
\begin{center}
\includegraphics[width=7cm]{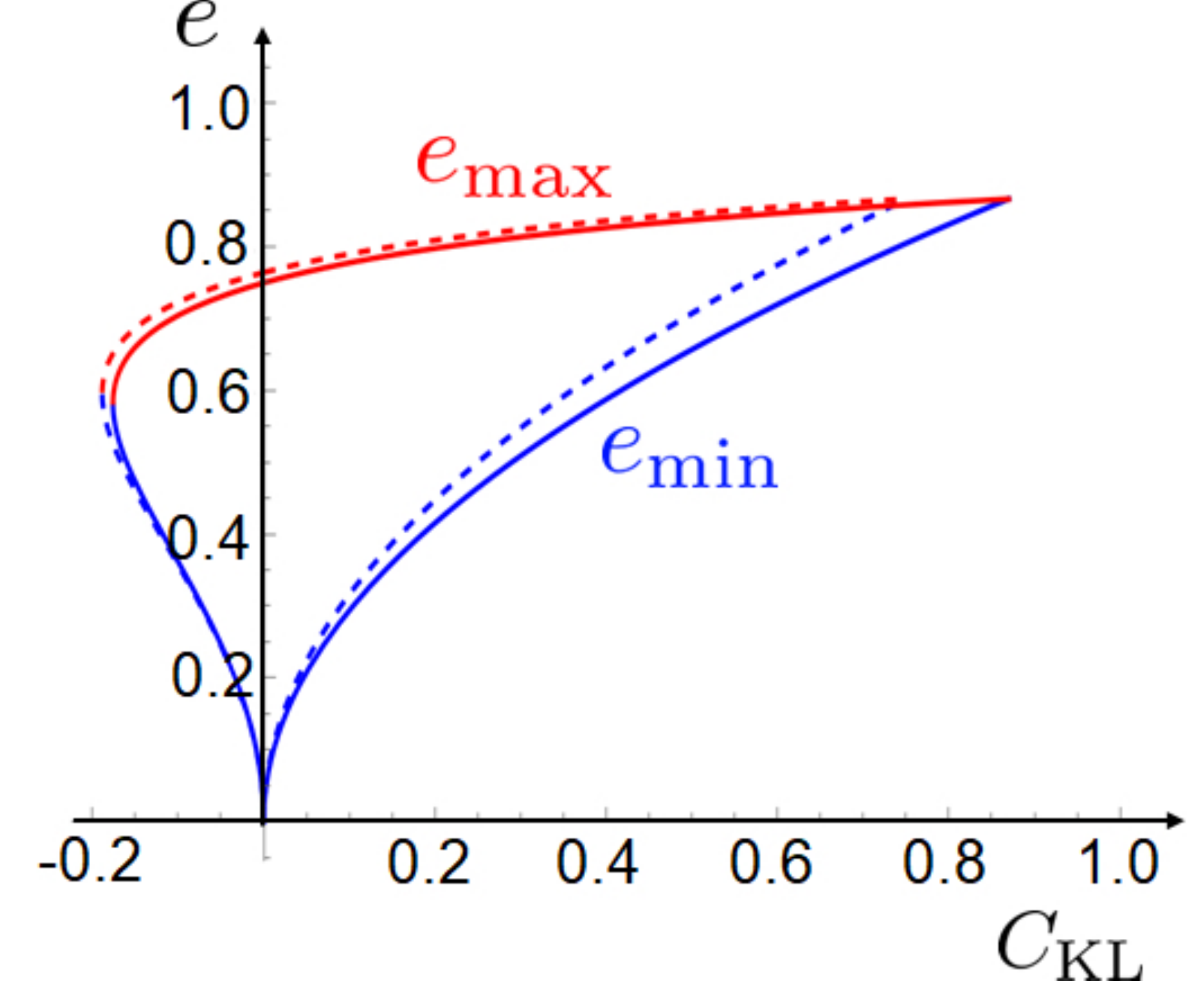}
\caption{The maximum and minimum values of the eccentricity in KL oscillations. 
We choose $\vartheta=0.5$. 
The red and blue curves depict the maximum and minimum values, 
respectively.
The solid curves denote the case of $\alpha_{\rm dS}=0.0794$ (Model I). We also show the case of $\alpha_{\rm dS}=0$ by the dotted curves.}
\label{fig:emaxemin}
\end{center}
\end{figure}

The maximum and minimum values of the eccentricity in the KL oscillations
are determined by two conserved parameters, $\vartheta$ and $C_{\rm KL}$.
We present one example in Fig. \ref{fig:emaxemin}. 
We choose $\vartheta=0.5$ and show the maximum value, $e_{\rm max}$
(the red curve),
 and the minimum value, $e_{\rm min}$ (the blue curve), in terms of  $C_{\rm KL}$. 
The solid curves denote the case of $\alpha_{\rm dS}=0.0794$
(Model I), while  the dotted curves is the case of $\alpha_{\rm dS}=0$. 
For Model Ic, we find $C_{\rm KL}=-0.0000359$, which is consistent with our result in Table I.
The relativistic effect with $\alpha_{\rm dS}$ is small.

\begin{figure}[h]
\begin{center}
\includegraphics[width=6.5cm]{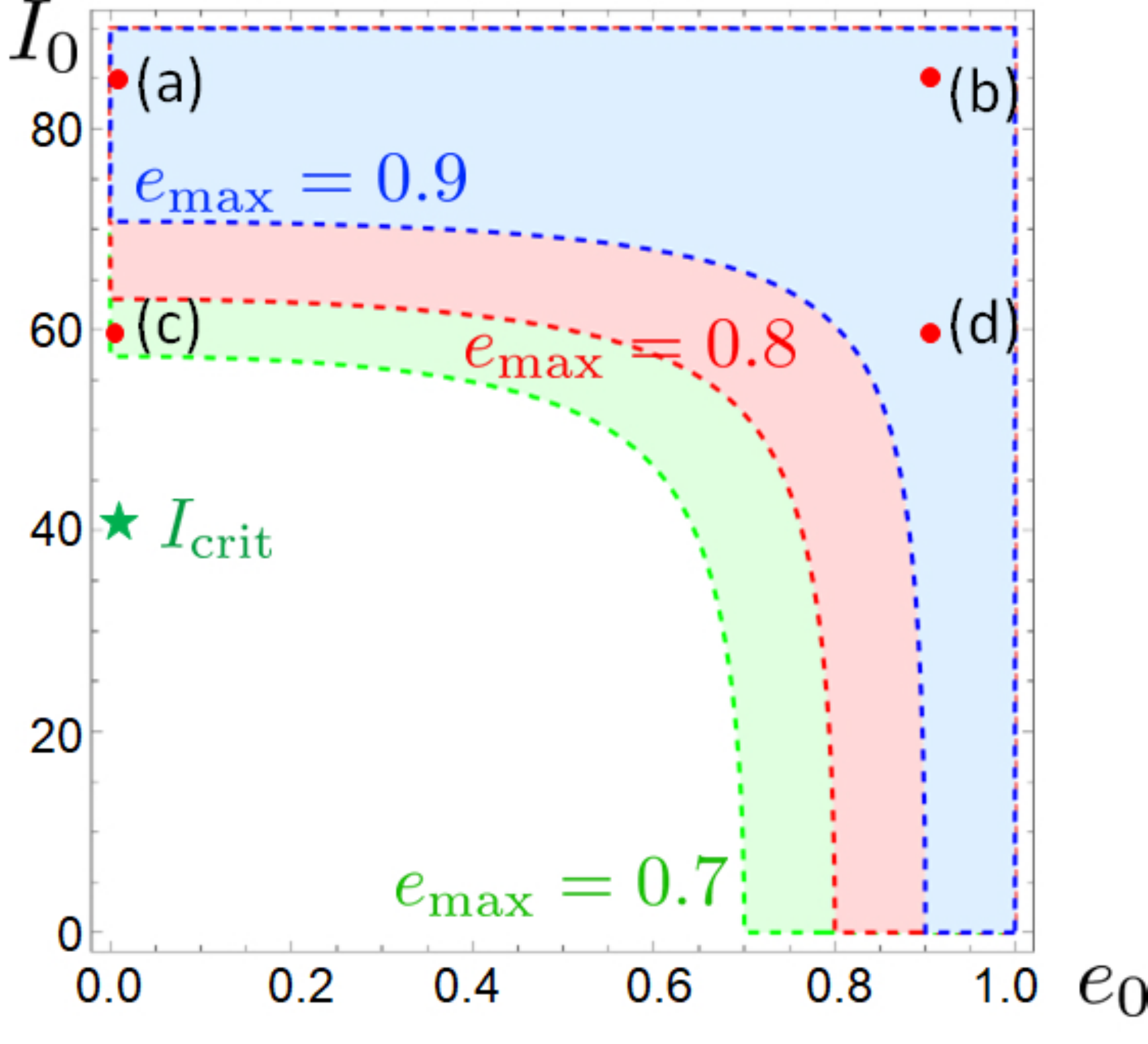}
\caption{The range of the initial conditions
($e_0$ and $I_0$) for the large 
maximum values of the eccentricity.
The light blue, light red and light green regions correspond to the regions of 
$0.9 \leq e_{\rm max}<1.0$, 
$0.8 \leq e_{\rm max}<0.9$, and $0.7 \leq e_{\rm max}<0.8$, respectively.
The red dots denote Models a, b, c, and d. 
The critical angle is also shown by the green star.}
\label{high_e_initial_conditions}
\end{center}
\end{figure}

The maximum eccentricity in KL oscillations is important, especially when we 
discuss emission of GWs.
We plot the range of the initial conditions ($e_0$ and $I_0$) 
which show the large 
maximum values of the eccentricity in Fig. \ref{high_e_initial_conditions}.
The light blue, light red and light green regions correspond to 
$0.9 \leq e_{\rm max}<1.0$, 
$0.8 \leq e_{\rm max}<0.9$, and $0.7 \leq e_{\rm max}<0.8$, respectively.
The red dots denote Models a, b, c, and d. Note that this figure is 
valid for all models I, II, and III.
Hence, when the initial inclination angle is large, the maximum eccentricity may grow close to unity.

The time scale of the KL oscillations is important for 
observation of the gravitational waves.
From Eqs (\ref{TKL_rotation}) and (\ref{TKL_libration}), assuming the integrals do not so much depend on the relativistic parameters, 
we may roughly evaluate the relativistic effect
(including de Sitter precession),
which is given by
\beann
{T_{\rm KL}\over T_{\rm KL}^{\rm (N)}}\approx
{1+3\mathfrak{r}_0^2\mathfrak{w}_0^2
\over \sqrt{(1+2\alpha_{\rm dS})(1+\alpha_{\rm dS}/3)}}
\,.
\enann
where $T_{\rm KL}^{\rm (N)}$ is the Newtonian value.
This ratio changes from $0.4427$ to 1 as $\mathfrak{r}_0$
increases from the ISCO radius to infinity.
Hence, the KL time scale near ISCO radius may become smaller 
than one-half of the Newtonian value.

We can also 
evaluate a critical inclination angle, 
beyond which the KL oscillation occurs even when the initial eccentricity is very small. It is given by the condition for a bifurcation point with 
$C_{\rm KL}=0$ with $\omega=90^\circ$. 
Setting 
\beann
(1+2\alpha_{\rm dS})-{5\over 2}(1+\alpha_{\rm dS})\sin^2 I_{\rm crit}=0
\,,
\enann
we obtain 
\beann
I_{\rm crit}=\sin^{-1}\sqrt{2(1+2\alpha_{\rm dS})\over 5(1+\alpha_{\rm dS})}
\,.
\enann
We find that the critical inclination angle 
changes from $41.6^\circ$ to the Newtonian value $I_{\rm crit}^{\rm (N)}=\sin^{-1}\sqrt{2/5}\approx 39.2^\circ$
as $\mathfrak{r}_0$ increases from the ISCO radius to infinity.

\newpage
\bibliography{refer}

\end{document}